\documentclass[11pt]{article}
\usepackage{latexsym}
\usepackage{amsmath}
\usepackage{amsfonts}
\usepackage{mathrsfs}
\usepackage{dsfont}
\usepackage{enumerate}
\usepackage{verbatim}
\usepackage{hyperref}
\usepackage{tikz}
\usepackage{mathtools}
\usepackage{simplewick}
\usepackage{float}
\usepackage{caption}
\usepackage{slashed}
\usepackage[title]{appendix}
\usepackage{graphicx}
\usepackage{subfig}
\usepackage{tikz}
\newcommand*{\myov}[1]{\overbracket[0.95pt][-1pt]{#1}}

 \csname
@addtoreset\endcsname{equation}{section}
\usepackage{amssymb}
\usepackage{color}
\usepackage{wasysym}
\begin{document}

\def\be{\begin{equation}}
\def\ee{\end{equation}}
\def\bea{\begin{eqnarray}}
\def\eea{\end{eqnarray}}
\def\nn{\nonumber}
\newcommand{\eq}[1]{(\ref{#1})}

\renewcommand{\thefootnote}{\roman{footnote}}

\begin{flushright}

\end{flushright}

\vspace{40pt}

\begin{center}

{\Large\sc Generalized Darmois-Israel junction conditions }

\vspace{50pt}

{\sc Chong-Sun Chu${}^{1,2}$ and Hai Siong Tan${}^1$ }

\vspace{15pt}
       {\sl\small
         ${}^1$ Physics Division, National Center for Theoretical Sciences, 
  National Tsing Hua University \\
 ${}^2$  Department of Physics, National Tsing-Hua
  University, Hsinchu 30013, Taiwan }

\vspace{15pt}

\vspace{70pt} {\sc\large Abstract}\end{center}

We present a general method to derive the appropriate Darmois-Israel
junction conditions for gravitational theories with higher-order
derivative terms by integrating the bulk equations of motion across
the singular hypersurface. In higher derivative theories, the field
equations can contain terms which are more singular than the Dirac
delta distribution. To handle them appropriately, we formulate a
regularization procedure based on representing the delta function as
the limit of a sequence of classical functions.  This procedure
involves imposing suitable constraints on the extrinsic curvature such
that the field equations are compatible with the singular source being
a delta distribution.  As explicit examples of our approach, we
demonstrate in detail how to obtain the generalized junction
conditions for quadratic gravity, $\mathcal{F}(R)$ theories, a 4D
low-energy effective action in string theory and action terms that are
Euler densities. Our results are novel, and refine the accuracy of
previously claimed results in $\mathcal{F} (R)$ theories and quadratic
gravity.  In particular, when the coupling constants of quadratic
gravity are those for the Gauss-Bonnet case, our junction conditions
reduce to the known ones for the latter obtained independently by
boundary variation of a surface term in the action.  Finally, we
briefly discuss a couple of applications to thin-shell wormholes and
stellar models.

\newpage

\tableofcontents

\section{Introduction}
\label{intro}

In general relativity, the Darmois-Israel junction conditions
prescribe the appropriate boundary conditions across a singular
hypersurface $( \Sigma )$ supported by a localized energy-momentum
source that contains a Dirac delta distribution.  Apart from metric
continuity at $\Sigma$, these `junction conditions' can be expressed
simply as the following equation \be
\label{DIJunction}
[K] h_{ij} - [K_{ij}] = 8 \pi S_{ij}, \qquad K \equiv K^m_m, \ee where
the square bracket indicates the jump discontinuity across $\Sigma$,
$h_{ij}$ is the induced metric, $S_{ij}$ is the singular Dirac delta
source localized within $\Sigma$, and all indices pertain to
coordinates of $\Sigma$. The Darmois-Israel junction conditions in
\eqref{DIJunction} were historically obtained in
\cite{Darmois,Israel,Lanczos} by integrating the field equations
across the infinitesimal width of $\Sigma$, and were later shown to
equivalently follow from the boundary variation of the
Gibbons-Hawking-York term in the action \cite{Gibbons,York}.

Since its discovery, \eqref{DIJunction} has found numerous
applications in a plethora of gravitational topics, being primarily
used to govern the construction of spacetime geometries obtained by a
`cut-and-paste' approach with $\Sigma$ as the locus of
identification. These geometries include thin-shell wormholes
\cite{Visser} (and their other variants such as gravastars
\cite{Mazur,Uchikata}), static and dynamical stellar models
\cite{Fayos,Fayos:1996} like the Oppenheimer-Snyder solution
\cite{Oppenheimer} describing gravitational collapse, as well as
domain wall configurations \cite{Blau} separating true and false vacua
in cosmology.

In recent years, there have been interesting explorations of
generalizing \eqref{DIJunction} to the wider context of gravitational
theories beyond Einstein's theory which we can consider to be a
long-distance, effective description connected to some presumably
UV-complete framework of quantum gravity such as string theory. When
interpreted as such, from the effective field theory perspective, it
is natural to consider adding higher-derivative terms to the
Einstein-Hilbert action.  Models that have attracted interest include
supergravity, $\mathcal{F}(R)$-theories, Gauss-Bonnet gravity, etc.
Accordingly, there has been a small number of different proposals for
how \eqref{DIJunction} should be modified in several contexts. For
example, junction conditions were proposed in
\cite{SenoFR,Sasaki,Olmo} for $\mathcal{F}(R)$ gravity, and in
\cite{Reina,Berezin} for quadratic gravity.
However, we find these results to be incomplete as they miss 
contributions which can only be seen with a proper and careful treatment of
terms involving products of the step function,
Dirac delta function, and their derivatives. For example, consider
the case that the extrinsic curvature is discontinuous across a hypersurface:
$K_{ij} =
K^-_{ij} + [K_{ij}] \Theta(n) $, where $n$ is the normal coordinate and
$\Theta$ is the step function \cite{SenoFR,Mars}
with $\Theta(0^-)=0$, $\Theta(0^+) =1$ and $K^\pm_{ij} \equiv K_{ij} (0^\pm) $.
The Gauss-Codazzi relation
$R = \hat{R} - K^2 - 2\partial_n K - K_{ab}K^{ab}$ implies that
\be
\label{T2one}
R = R^+ \Theta + R^- (1 - \Theta) - [K_{ab}][ K^{ab}] (\Theta^2 -
\Theta).
\ee
Naively, one may take 
\be \label{tht}
\Theta^2 = \Theta.
\ee
However
the field equations in higher derivative gravity theory imply
that generally, these terms are not isolated. Instead, they appear
together with delta functions and their derivatives (this will be discussed
extensively below in
Section \ref{GJC}).
As a result, \eq{tht} cannot be taken 
prematurely from the outset  as was assumed in
\cite{SenoFR,Mars}.
In fact, in any sensible regularization of the delta function,
we have $\delta(n) = \Theta'(n)$ and 
$
\int dn\,\,\, \Theta^n(n)  \delta(n) = \frac{1}{n+1}.
$
This implies that in the distributional sense,
$$ \delta(n) \left( \Theta^2(n)  - \Theta
(n) \right) \neq 0.$$
In Section \ref{411} below, we show how the correct treatment of the
$(\Theta^2 -\Theta)$-term
in \eq{T2one} leads to contributons
of the form $[K_{ij}][K_{ab}][K^{ab}]$ in the junction conditions of the
$R^2$-theory. Such contributions were missed in the literature
\cite{SenoFR,Sasaki,Reina}.
As  products of singular distributions is unavoidable in higher derivative
gravity and the handling of them are subtle,
one of the key motivations of this work is to provide a consistent
mathematical framework for the derivation of the junction conditions.

In this paper, we present a general method to derive the appropriate
junction conditions for gravitational theories with higher-order
derivative terms by integrating the bulk equations of motion across
the infinitesimal width of the singular surface $\Sigma$, in the same
spirit as was done in the foundational paper of Israel in
\cite{Israel}.  Fundamentally, asserting an equivalence between both
sides of the field equations in the sense of distribution translates
to ensuring the convergence of the integral which is now much more
complicated by virtue of the higher-order differential
operators. Completing the integral then yields the generalized
junction conditions compatible with a Dirac $\delta$-distribution
carried by the energy-momentum tensor.

We find that a combination of the following techniques furnishes an
effective toolbox for solving this problem.  First, we choose to work
in Gaussian normal coordinates describing a local neighborhood of
$\Sigma$ which is now defined by taking $n=0$, where $n$ parametrizes
the proper distance along a geodesic orthogonal to $\Sigma$. In this
chart, we invoke the Gauss-Codazzi relations extensively to express
various curvature terms in the equations of motion in terms of the
intrinsic geometric quantities of $\Sigma$, the extrinsic curvature
and their derivatives. The problem is now reduced to examining the
convergence of a one-dimensional integral. For this purpose, we find
that expressing the $\delta$-function as a limit of a sequence of
classical functions gives us a powerful language for organizing
various integrand terms according to their algebraic order of
singularity.\footnote{We note that in both the math and physics
  literature, the representation of the $\delta$-distribution by the
  limit of delta-convergent sequences has a long history (see
  e.g. \cite{Jones} for the context of the theory of distributions). A
  good, old example relevant for both communities is the
  Kramers-Kronig or Sokhotski-Plemelj equation which defines, in the
  distributional sense,
$
  \delta(x) = \mp \frac{1}{\pi i} \left( \lim_{\epsilon \rightarrow 0}
  \frac{1}{x\pm i \epsilon} - \text{p.v.} \left( \frac{1}{x} \right)
  \right) $.  Taking the sum of the two choices of sign leads to the
  $\delta$-distribution being a limit of a sequence of Cauchy
  distributions : $ \delta (x) = \lim_{\epsilon \rightarrow 0}
  \frac{1}{\pi} \frac{\epsilon}{x^2 + \epsilon^2}
$,
which is a concrete example of a `nascent delta function' that will
feature frequently in our narrative.  } This allows us to identify
what are the additional conditions on the embedding geometry that
naturally ensure the absence of singular terms that are incompatible
with a $\delta$-distribution source term.  We call these conditions
`\emph{regularity constraints}', and they are typically expressed as
the vanishing of some function of the extrinsic curvature components
on each side of $\Sigma$. As we shall elaborate later on, it turns out
that they are intimately related to the Hadamard regularization
procedure \cite{Hadamard} in the theory of distributions (see
e.g. \cite{Ram}). In the absence of the regularity constraints, the
generalized junction conditions are precisely the Hadamard-finite part
of the integral.  Our method can be readily adopted and applied to a
wide variety of gravitational theories, including matter and gauge
couplings, although in this paper, we focus almost exclusively on
action terms that are invariants constructed from the Riemann tensor
as illustrative examples.

Our exposition is structured as follows. In Section \ref{GJC}, we
present the background canvas of concepts underlying our general
approach of deriving the junction conditions via a bulk integration
across $\Sigma$, including the notion of delta-convergent sequences,
regularity constraints and how our method relates to Hadamard
regularization.  Section \ref{quadraticSec} contains a detailed
derivation of the junction conditions for quadratic gravity. This is
accompanied by Appendix \ref{AppA} containing some integral identities
that we developed for our purpose, and by Appendix \ref{GBsec}
containing details of how the junction conditions in quadratic gravity
reduce consistently to known ones for Gauss-Bonnet gravity.  Section
\ref{OtherSec} presents the junction conditions for three types of
theories: $\mathcal{F}(R)$ theories, a low-energy effective action
descending from string theory, and higher-dimensional Euler density
terms that appear in Lovelock gravity. In Section \ref{AppSec}, we
briefly discuss a couple of applications by examining how thin-shell
wormholes and stellar models are governed by junction conditions in
the context of $\mathcal{F}(R) = R + \beta R^2$ theory. Finally, we
end with some concluding remarks in Section \ref{Conclude}. For the
reader who is uninterested in the illustrative examples of
gravitational theories we have chosen, Section \ref{GJC} suffices as a
stand-alone explanation of the tapestry of techniques used in our
method, but in our opinion, reviewing a couple of concrete examples in
Section \ref{quadraticSec} or \ref{OtherSec} should still be useful
towards learning how to apply our method for other theories.

\emph{Conventions}: We work in natural units where
$\mathcal{G}=c=1$. Greek indices refer to all of spacetime, whereas
Latin indices pertain to coordinates of the singular hypersurface
$\Sigma$, with `$n$' reserved for the direction normal to
$\Sigma$. Our convention for the Riemann tensor is
${R^\rho}_{\sigma
  \mu \nu} = \partial_\mu \Gamma^\rho_{\nu \sigma} + \ldots$.
\vskip 0.3cm
\noindent {\bf \large Comparison with the existing results}
\vskip 0.1cm

Our results for $\mathcal{F}(R)$ theory and quadratic gravity refine
the accuracy of those reported earlier in literature
in a number of ways. For clarity, they are collected here:

{\it 1.} As mentioned 
above, we explain
in Section
\ref{411} explicitly the origin of a missing term in
\cite{SenoFR,Sasaki,Reina} for junctions conditions in the theory with
$R^2$ action term.

{\it 2.} In the papers \cite{SenoFR,Olmo,Reina,Berezin},
the authors argued that the naive
     appearance of products of delta function and delta derivatives
     in the equations of motion
leads to inconsistency, and that there is no unambiguous manner to sum
them up.
As a result it was assumed that $[K]=0$ so that $R$ is non-singular.
On this issue, we find that the use of delta-convergent
sequences clearly quantifies the growth property of each integrand
term, and provides an organizational principle for identifying
regularity constraints and a well-defined procedure to sum up the
residual finite terms, eventually leading us to the junction
conditions.
For us, singularities are handled by a careful procedure of regularization.
This is inspired by the common practice of
regularization in quantum field theory.
We will show below that, rather than insisting on the use of
distribution-valued curvature, our approach is more practical and
much less restrictive
and allows for more general solutions to the theory.

     {\it 3.} In \cite{Reina}, the authors posited that in
quadratic gravity, the energy-momentum tensor should contain the
distributional derivative of the $\delta$-function that they derived
from the higher-order differential operators acting on a discontinuous
Riemann tensor.\footnote{They dubbed this component as the `double
  layer' (see also \cite{Dl1,Dl2}). Also, they postulated the
  regularity of the on-shell action as a necessary condition, but
  here, we take the equations of motion as the physical starting point
  of our analysis.}  In this paper, we focus only on the explicit
form of the generalized junction conditions, noting that only the
delta-distribution component of $T_{\mu \nu}$
appears in final expression, since any regular component and those
containing higher-order derivatives
of the delta function do not survive the infinitesimal integral across
$\Sigma$.

     {\it 4.}
We should also mention that our method of derivation passes a
stringent consistency test that is noticeably absent in the past
literature, namely that when applied to quadratic gravity, i.e. taking
the Lagrangian $\mathcal{L} = \beta_1 R^2 + \beta_2 R^{\alpha \beta}
R_{\alpha \beta}+ \beta_3 R^{\alpha \beta \mu \nu} R_{\alpha \beta \mu
  \nu}$, our junction conditions reduce to those of Gauss-Bonnet
gravity $(\beta_2 = - 4\beta_3 = -4\beta_1)$ that were derived some
time ago in \cite{Davis} via a completely different approach --- the
boundary variation of a surface term \cite{Myers} that accompanies the
gravitational action. As already argued in \cite{Myers,Madsen} and
definitively shown in \cite{Smolic}, the equations of motion of a
generic higher-derivative gravitational theory do not descend from a
well-posed variational principle with Dirichlet conditions, and hence
the approach of obtaining their appropriate junction conditions by
boundary variation of surface terms is not always
applicable.\footnote{In \cite{Berezin}, an attempt was made to derive
  the junction conditions for a generic quadratic theory using a
  variational principle, but imposed a variation of the extrinsic
  curvature inconsistent with its fundamental definition. See further
  comments in Appendix \ref{GBsec}. }  An exception lies in the family
of theories defined by Euler density terms which are the linear
combinations of curvature invariants that generate only second-order
field equations, and thus, in principle, these exists appropriate
surface terms for them. The simplest example would be the Ricci scalar
being the 2D Euler density, with the Gibbons-Hawking-York term $
S_{GHY} = \frac{1}{8\pi } \int d^{d-1}x \,\, \sqrt{h} K $ being its
surface term, and of which variation with respect to the induced
metric yields \eqref{DIJunction}.  The junction condition derived in
this manner nicely furnishes a consistency check for the other
derivation route obtained by integrating across $\Sigma$. In Appendix
\ref{GBsec}, we will demonstrate how the complicated junction
conditions for a generic quadratic gravity theory reduce to the known
ones for Gauss-Bonnet gravity. The regularity constraints need not be
imposed precisely for $\beta_2 = - 4\beta_3 = -4\beta_1$, and this
also serves as a consistency check for the corresponding equations
used in determining them.  On this point, we note that previously in
\cite{Reina} where the authors proposed junction conditions for
quadratic gravity, they imposed $[K_{ij}]=0$ even in the Gauss-Bonnet
case which contradicts the known ones derived by a boundary variation
\cite{Davis,Gravanis}.

\section{Generalized Junction Conditions - a general approach}
\label{GJC}

\subsection{Some Preliminaries}
For a generic gravitational theory with higher-order derivatives of
the metric in the action, let us denote the field equations as
\be
\label{Gfield}
\tilde{G}_{\mu \nu} \equiv R_{\mu \nu} - \frac{1}{2} R g_{\mu \nu} +
\ldots = 8 \pi T_{\mu \nu},
\ee
where the ellipses represent the
additional terms descending from the modified action.  In this work,
we will mainly be studying gravitational theories where these terms
are products of various contractions of the Riemann tensor and
differential operators acting on them. On the RHS, we assume that the
energy-momentum tensor harbors a singular source localized on some
codimension one hypersurface $\Sigma$.  If we integrate across the
infinitesimal width of $\Sigma$, only singular terms on both sides of
\eqref{Gfield} survive and the resulting integral yields the
appropriate \emph{junction conditions} for the gravitational theory
considered.  The main purpose of this paper is to develop a conceptual
and computational basis for this integration procedure. Fundamentally,
apart from topological action terms which are not reflected in
\eqref{Gfield}, the bulk equations of motion suffice as the starting
point for the derivation of the junction conditions which can be
regarded as consistency conditions for the bulk dynamics induced on
$\Sigma$.

In the full generality, the induced metric on $\Sigma$ $h_{ij}$ can be
timelike, spacelike or null. For the rest of the paper, we will mainly
focus on the timelike case which has garnered the most interest in the
past literature.  When $\Sigma$ is endowed with a singular
energy-momentum tensor localized on it, it represents a singular thin
shell of matter. (For a spacelike $h_{ij}$, this would be a
gravitational instanton whereas a null $h_{ij}$ would represent a
singular light cone.)

In the most generic setting, $\Sigma$ is a codimension-one
hypersurface that is embedded in a bulk manifold ($\mathcal{M}$)
constructed from two distinct ones ($\mathcal{M}_1, \mathcal{M}_2$)
that is joined by identifying
$$ \Sigma_1 \sim \Sigma_2 \sim \Sigma,
$$ where $\Sigma_{1,2}$ are hypersurfaces in $\mathcal{M}_{1,2}$ that
can be related via a homeomorphism. In this cut-and-paste procedure,
the final manifold $\mathcal{M}$ is the union of $\Sigma$ and the
interiors of $\mathcal{M}_{1,2}$ previously bounded by $\Sigma_{1,2}$.
We can introduce a set of coordinates $(\zeta^i)$ intrinsic to
$\Sigma$ which is defined parametrically by
$$ f\left(x^\mu (\zeta^i ) \right) = 0,
$$
with the induced metric 
\be
h_{ij} = g_{\mu \nu} \frac{\partial x^\mu}{\partial \zeta^i}
\frac{\partial x^\nu}{\partial \zeta^j}.  \ee Denoting the sign of
$n^2$ by $\xi$, the unit normal vectors to $\Sigma$ read
$$ n_\mu = \xi N \frac{\partial f}{\partial x^\mu}, \qquad N = \left|
g^{\alpha \beta} \partial_\alpha f \partial_\beta f \right|^{-1/2},
\,\,\,\, n^2 = \xi,
$$
where $\xi = \{+1,-1,0 \}$ for the cases of $\Sigma$ being
timelike, spacelike and null respectively. The second fundamental form
or the extrinsic curvature enacts an essential role in our discussion
and can be defined as follows.
\be
K_{ab} = h^m_a h^l_b \nabla_m n_l,
\qquad n^a K_{ab} = 0.
\ee
For our purpose, we find it useful to work
in the Gaussian normal coordinate system in the local neighborhood of
$\Sigma$ constructed from the family of non-intersecting geodesics
orthogonal to $\Sigma$.
\be
\label{Gaussian}
ds^2 = \xi dn^2 + h_{ij} (n, x^k ) dx^i dx^j,
\ee
with $\Sigma$
defined by taking $n=0$ and $x^i$ are the intrinsic coordinates.
Having briefly described the geometric characterization of $\Sigma$,
let us now return to the field equations in \eqref{Gfield}. 
On the
RHS, we can express the energy-momentum tensor as
$$ T_{\mu \nu} = S_{\mu \nu} \delta(n) + \ldots,
$$ where we have denoted the delta function source by
$S_{\mu \nu} = S_{\mu \nu} (\vec{x})$. The
ellipses refer to regular components of the energy-momentum
tensor, and more generally some linear combination of derivatives of
the delta function.
\footnote{A known result in the theory of distribution (see e.g.
  \cite{Lighthill}) states that any distribution with point support is 
  a linear combination of derivatives of the delta distribution,
  hence such an ansatz is the most general
  one for a singular source localized on a codimension-one hypersurface
  expressed in Gaussian normal coordinates. For example, a source of
  the form $h(n) \delta^{(k)} (n)$ is equivalent to
$
\sum_{k=0}^{N_t} h_k (n)  \sum_{m=0}^k  h^{(m)} (0) {}^k C_m (-1)^m \delta^{(k-m)} (n)
= \left( h^{(k)}_k (0) (-1)^k \right) \delta(n) + \ldots$
in the sense of distribution. 
    }
Working in the local Gaussian normal chart, we can integrate
\eqref{Gfield} across the infinitesimal width of $\Sigma$ as follows.
\be
\label{integration}
\lim_{\epsilon \rightarrow 0^+} \int^\epsilon_{-\epsilon} dn \,\,
\tilde{G}_{\mu \nu} = 8 \pi S_{\mu \nu}.
\ee
The infinitesimal
integral domain picks up only the $\delta$-distribution part of the
energy-momentum tensor, since other components vanish in the
$\epsilon \rightarrow 0$ limit.  Formally, we can see this by writing
$$
\lim_{\epsilon \rightarrow 0^+} \int^\epsilon_{-\epsilon} dn =
\int^\infty_{-\infty} dn - \text{p.v.} \int^\infty_{-\infty} dn,
$$
where the second term on the RHS is the Cauchy principal value of
the first integral term. In local Gaussian normal coordinates, the
Christoffel symbols read
\be
\Gamma^n_{ij} = -\frac{1}{2} \partial_n
h_{ij} = - K_{ij}, \,\,\, \Gamma^i_{nj} = K^i_j = \frac{1}{2} h^{im}
\partial_n h_{jm}.
\ee
The discontinuity in $\partial_n h_{ij}$ or
the extrinsic curvature induces various singular terms in
$\tilde{G}_{\mu \nu}$ which is then related to the singular source
$S_{\mu \nu}$ in \eqref{integration}.

A main theme of this work is to seek the conditions characterizing the
embedding of $\Sigma$, under which the LHS of \eqref{integration} is
well-defined. This then leads to an appropriate set of junction
conditions on the LHS of \eqref{integration} upon performing the
integral. \emph{We take this to be the operational meaning of the
  delta function source in the energy-momentum tensor, since the field
  equations should be understood in the sense of distribution when the
  energy-momentum tensor carries a delta-singular source. }

Now in our derivation, the Gauss-Codazzi relations turn out to be
crucial as they express various curvature quantities in terms of the
intrinsic Riemann tensor and the extrinsic curvature tensor,
elucidating the form of various singular terms in $\tilde{G}_{\mu
  \nu}$ easily.  Recall that the Gauss-Codazzi equations read
\bea
\hat{R}^a{}_{bcd} &=& \mathcal{P} \left( {R^a}_{bcd} \right) + \xi
\left( K^a_c K_{bd} - K^a_d K_{bc} \right), \\ n^a R_{aqrs} h^q_b
h^r_c h^s_d &=& \xi \left( D_d K_{bc} - D_c K_{bd} \right),
\eea
where
$\mathcal{P}$ refers to the indices being projected with the induced
metric $h_{ab}$, $D_a$ is the projected covariant derivative defined
with the induced metric and hatted variables are geometrical
quantities intrinsic to $\Sigma$. Using Gaussian normal coordinates
and contracting some of the indices, we further obtain the following
useful set of equations.
\bea
    {R^n}_{s n v} &=& \xi \left( -
\partial_n K_{v s} + K_{a v} K^a_s \right),\,\,\, R_{nijk} = D_k
K_{ij} - D_j K_{ik}, \\ R_{ij} &=& \xi \left( - \partial_n K_{ij} +
2K^a_j K_{ai} - KK_{ij} \right) + \hat{R}_{ij}, \\ R_{nn} &=& -
\partial_n K - K^{ab} K_{ab} = - h^{ij} \partial_n K_{ij} + K^{ab}
K_{ab},\\ R_{in} &=& D^k K_{ik} - D_i K, \\ R &=& \hat{R} - \xi \left(
2 \partial_n K + K^2 + K^{ab} K_{ab} \right).
\eea


\subsection{On the standard Darmois-Israel junction conditions in GR}

It is instructive to first review how one can obtain the standard Darmois-Israel junction conditions in ordinary GR before proceeding to more complicated gravitational theories. In \eqref{integration},
we now simply take $\tilde{G}_{\mu \nu} = G_{\mu \nu}$ and integrate
\be
\label{integrationGR}
\lim_{\epsilon \rightarrow 0^+} \int^\epsilon_{-\epsilon} dn \,\, G_{\mu \nu} = 
8 \pi  S_{\mu \nu}.
\ee
On the LHS, only singular terms in the Einstein tensor survive in the limit $\epsilon \rightarrow 0$. The singularity of these
terms are of the Dirac delta type and can be traced to a discontinuity in the 
derivative of the metric in the direction normal to $\Sigma$ which, in the Gaussian normal chart,
is effectively the discontinuous extrinsic curvature. 
We keep track of these terms appearing in the Einstein tensor which can be
simplified to read
\bea
\label{Einsteinbulk}
&& \xi \left( \partial_n K^i_j - \delta^i_j \partial_n K \right) +\xi
\left( - K K^i_j + \ldots \right) =8 \pi   T^i_j, \cr
&&\xi \left( - \nabla_i K + \nabla_j K^j_i \right) =  8 \pi T^n_i,\,\,\,
-\frac{1}{2} \hat{ R} + \frac{1}{2} \xi \left( K^2 - \text{Tr} (K^2) \right) = 8\pi  T^n_n,  
\eea
where the ellipses are terms which do not involve $\partial_n$. 
Integrating over $\Sigma$ as in \eqref{integrationGR}
on both sides of the field equations then yields the junction conditions 
\be
\label{stanDI}
\xi\left( -[K_{ij} ] + [K] g_{ij} \right) = 8 \pi S_{ij},
\ee
which is the standard Darmois-Israel junction conditions in GR, and 
we have used the bracket $[...]$ to denote the jump or difference between
 the limiting values of the bracketed quantity
on each side of $\Sigma$, for example $[K] \equiv K\vert_{n=0^+} - K\vert_{n=0^-}$. 
We also note that the junction conditions in other directions are trivial, i.e. 
in \eqref{integration}, 
\be
\lim_{\epsilon \rightarrow 0^+} \int^\epsilon_{-\epsilon} dn \,\, G_{i n} = 
\lim_{\epsilon \rightarrow 0^+} \int^\epsilon_{-\epsilon} dn \,\, G_{n n} = 0.
\ee
Thus in GR, the localized singular source $S_{\mu \nu}$ cannot 
have components non-parallel to $\Sigma$. As we shall see later, this
is not generally true for other gravitational theories. Another feature characterizing the junction
condition \eqref{stanDI} that differentiates a generic higher-derivative theory from GR is that the junction
equation does not only involve bracketed terms but also acquires terms that represent averaging 
across $\Sigma$.

A crucial ingredient in our formulation is a procedure that takes into account singular terms more divergent than delta function terms in the LHS of the field equations \eqref{Gfield},
that may appear in the equations of motion upon assuming a discontinuous extrinsic curvature (note that this problem is absent in \eqref{Einsteinbulk}). 
We demonstrate how to impose suitable regularity constraints on the extrinsic curvature such that 
the integral in \eqref{integration} remains convergent, leading to generalized junction conditions
which are the appropriate boundary conditions at $\Sigma$. We will find that an essential tool is to regard the delta function
as a limit of a sequence of absolutely convergent functions, a delicate subject which we turn to next.

\subsection{Delta sequences and a double scaling limit}

Since the singularities in \eqref{Gfield} essentially arise from derivatives of the Heaviside step function
parametrizing the discontinuity in the extrinsic curvature, 
we first consider a suitable representation of it by defining
it as a limit $b\rightarrow 0$ of a sequence of classical functions $\Theta (n,b)$, with
\be
\label{classical}
\Theta (n) = \lim_{b\rightarrow 0} \Theta (n,b), 
\qquad 
\Theta (n,b) = \frac{1}{2} + \Theta_X \left( \frac{n}{b} \right), 
\ee
where we have chosen to parametrize it such that its derivative yields a representation of the delta function as follows. 
\be
\label{classical2}
\delta (n) = \lim_{b\rightarrow 0} \partial_n \Theta (n,b) = \lim_{b \rightarrow 0} 
\frac{1}{b} \Theta'_X  \left( X \right) \equiv \lim_{b \rightarrow 0} 
\frac{1}{b} F(X), \qquad X \equiv \frac{n}{b},
\ee
The function $F(X)$ is sometimes known as a nascent delta function, with $\Theta_X (X)$
being its antiderivative. 
As a generalized function or distribution, we require that 
\be
\label{normDelta}
\lim_{b\rightarrow 0} \int^\infty_{-\infty} dn \frac{1}{b} F(X) f(n) = f(0),
\ee
for all continuous $f(n)$ with compact support which implies that $F(X)$ is normalized as 
\be
\label{deltanorm}
\int^\infty_{-\infty} dX\,\, F(X) =1. 
\ee
Formally, each nascent delta function 
gives rise to a distribution of the form 
\be
\label{nasc}
\tilde{f}_{1/b} = \partial_n \Theta(n,b) = 
\frac{1}{b}F(n/b),
\ee 
and the sequence 
$\tilde{f}_1, \tilde{f}_2, \ldots $ converges
to the delta function distribution $\delta(n)$ if the limit defined on the LHS on \eqref{normDelta}
exists. 
 Some popular choices of the nascent delta function\footnote{See for example \cite{Lighthill} for a deeper discussion.}
$F(X)$
appearing in related literature are (i) $\frac{1}{\sqrt{\pi }} e^{- X^2 }$ (Gaussian distribution),
(ii) $\frac{1}{\pi (1+X^2)}$ (Cauchy distribution), (iii) $\frac{\sin(X)}{\pi X}$ (sinc function), etc. 
For the purpose of our work here, we adopt some choice of $F(X)$ that is even in $n$:
\footnote{We note that the symmetry property \eqref{parity} and the ansatz for the step function in \eqref{classical} imply that we are taking $\Theta(0) = \frac{1}{2}$.} 
\be
\label{parity}
F(X) = F(-X), \,\,\, \Theta_X \left( X \right) = -\Theta_X \left( -X \right),
\ee
with $F(X)$ being infinitely differentiable,
so that it is compatible with our use of its antiderivative $\Theta_X(X)$ to describe functions of varying degrees of smoothness. 

Taking $F(X)$ to be an even function inherits the nature of $\delta(n)$ being an even distribution,
and retains its original physical attribute of being symmetrical on either side of $\Sigma$. If we regard
the nascent delta function as a probability density function describing how energy-mass is localized in $\Sigma$ (idealized by the $\delta$-distribution), then \eqref{parity} translates to $n=0$ being the mean of the energy-mass distribution, which is expected on physical grounds.\footnote{See e.g. \cite{Lighthill,Saichev} for a more extensive discussion of various applications of even nascent delta functions.} As we shall point out shortly in Section \ref{GdRC}, the parity assignment \eqref{parity} also leads to a natural generalization of the derivatives of the $\delta$-function when they are integrated against non-smooth functions. 
\footnote{In general, any locally integrable function that can be normalized following \eqref{deltanorm}
can qualify as a nascent delta function. For the proof, we refer the reader to the seminal texts of L.Schwartz \cite{Schwartz}, Jones \cite{Jones}, Gel'fand and Shilov \cite{Shilov}, and a more accessible version in \cite{Rivas}. 
Our assumption of an even nascent delta function can in principle be relaxed, but does not alter
the conceptual basis of our approach in this paper.}

Parametrizing the step function in this manner implies that for a function $g(n)$ which may not be smooth at (and after) a certain order of its derivative at $n=0$, we can consider it as the limit of a sequence of functions as follows.  
\be
\label{gendiscont}
g(n) = g_1 (n) + \Theta (n) (g_2 (n) - g_1(n)) = \overline{g}(n) + \lim_{b\rightarrow 0} \Theta_X(X) [g(n)],
\ee
where $g_1(n), g_2(n)$ are infinitely differentiable functions of $n$ each analytically extending $g(n)$ beyond $n=0$ and we have defined 
$$
\overline{g}(n) = \frac{1}{2} \left( g_1 (n) + g_2 (n) \right), \quad
[g(n)] = g_2(n) - g_1(n), \quad g_2 (0) = g(0^+), \quad  g_1 (0) = g(0^-).
$$
For example, if $g(n)$ is of class $C^1$ with its second-order derivative being discontinuous at $n=0$, then $g_1 (0)= g_2 (0), g'_1 (0)= g'_2 (0), g''_1 (0) \neq g''_2 (0)$, etc. One can perform ordinary 
derivatives on $g(n)$ before finally taking the $b\rightarrow 0$ limit. 

In \eqref{integration}, 
the integrand is generally a rather complicated function of various
non-smooth curvature quantities, and it is crucial that we apply
\eqref{gendiscont} for all terms consistently before evaluating the
integral.  Formally, this procedure yields a sequence of regular
distributions (which is essentially a complicated function of the
nascent delta function and other functions of the metric tensor
analytic at $\Sigma$) that ensure the convergence of
\eqref{integration}.

We will find that this manner of expressing the step function (and
hence the delta functions and their derivatives) lends us a powerful
language for simplifying and organizing various types of singular
terms consistently in the integrand of \eqref{integration}.  It also
necessitates a more precise prescription of the infinitesimal
integration procedure in \eqref{integration} , one which takes into
account the relative scale separation between the sequence parameter
$b$ and the infinitesimal width $\epsilon$ of the surface $\Sigma$. We
now address this pivotal point that arises when we apply these
considerations to our problem --- the relative scaling of the
thin-width parameter $\epsilon$ and the nascent delta function
parameter $b$.

Recall that in deriving the junction equations, we perform an integral
with limits parametrizing the width of the infinitesimally thin
$\Sigma$, and hence we send the integral limits to zero after
integrating. The integral preserves only terms which accompany a delta
function singularity. Consider again \eqref{normDelta} with an
infinitely differentiable $f(n)$, but with the integral limits of
\eqref{integration}. We still expect to recover $f(0)$ on the RHS.
Expanding $f(n)$ around $n=0$ to obtain
\be
\label{weak2}
\lim_{\epsilon \rightarrow 0}
\lim_{b\rightarrow 0}
\int^{\epsilon}_{-\epsilon} dn
\,\,\delta_b (n) f(n)  
=
\lim_{\epsilon \rightarrow 0}
\lim_{b\rightarrow 0}
\int^{\epsilon/b}_{-\epsilon/b} dy
\, F(y) \left(  
f(0) + b y f'(0) + \frac{1}{2} b^2 y^2  f''(0) + \ldots
\right),
\ee
we find that we recover \eqref{normDelta} upon taking the double scaling limit
\be
\label{doublescale}
\epsilon \rightarrow 0, \qquad b/\epsilon \rightarrow 0.
\ee
We note that on the contrary, if we specify $b$ to vanish such that $b/\epsilon \rightarrow \mathcal{O}(1)$ 
or the converse $\epsilon/b \rightarrow 0$, then we do not recover $f(0)$. The physical 
reasoning behind \eqref{doublescale} is simple and intuitive: the thin-width limit must be kept away from/taken after the limit of the sequence of nascent delta functions, or otherwise we are not genuinely integrating a
singular delta function across $\Sigma$. 

Recall that in the field equations, we seek a solution to 
\be
\label{Gfield2}
\tilde{G}_{\mu \nu} = R_{\mu \nu} - \frac{1}{2} R g_{\mu \nu} + \ldots = 8 \pi \left( 
S_{\mu \nu} \delta(n) + \ldots
\right),
\ee
where the ellipses refer to all regular components of the energy-momentum tensor and possibly a linear combination of higher-order derivatives of the delta function, all of which do not survive the integral of \eqref{integration}. 
A solution to \eqref{Gfield2} must thus integrate to give $S_{\mu \nu}$
in the following manner. 
\be
\label{Gfield3}
\lim_{\epsilon \rightarrow 0}
\lim_{b\rightarrow 0}
\int^{\epsilon}_{-\epsilon} dn \left(
 R_{\mu \nu} - \frac{1}{2} R g_{\mu \nu} + \ldots \right) 
= 8 \pi  S_{\mu \nu},
\ee
where the limit in $b$ is taken for the sequence of distributions defined by 
\eqref{nasc}. We have to ensure that this sequence of integrals converges which generically requires 
additional conditions to be imposed on the extrinsic curvature
for the integral in \eqref{Gfield3} to be well-defined.

This double scaling limit also preserves the fact that we can obtain a well-defined operational meaning
for the derivatives of the delta functions when they are integrated against smooth functions, i.e.
that we have, in a weak distributional sense, 
\be
\label{deltaderiv}
\delta^{(k)} [g(n) ] \sim (-1)^k g^{(k)} (0),
\ee
for a $g(n)$ analytic at $n=0$. This formula needs to be refined generally for a non-smooth $g(n)$  --- an important yet tricky point
that is relevant for us since, as we shall see, the equations of motion of higher-derivative gravitational theories
typically contain terms which are products of non-smooth functions and a number of delta functions each equipped with some order of derivative.

\subsection{Generalized distributions and regularity constraints}
\label{GdRC}

In this section, we present a well-defined procedure that extends the usual notion of distribution for the delta function (and its derivatives) to one that could be applied when they are integrated against non-smooth functions. Essentially for our purpose here, this will ultimately turn out to furnish a simple and clear method to identify and classify singular terms arising from the junction equations, leading to regularity constraints which we can impose on the extrinsic curvature to eliminate singular terms that render \eqref{integration} to be ill-defined.

We begin with the basic examples of the $\delta(n), \delta'(n)$.

\subsubsection{Warm-up: more about $\delta (n), \delta'(n)$}

Let us consider a discontinuous test function for the delta function which we express as
$f(n) = \overline{f} (n) + \Theta_X (X) [f(n)] $ where
$\overline{f}(n) = \frac{1}{2} \left( f_1 (n) + f_2 (n) \right), 
[f(n)] = f_2(n) - f_1(n)$, with both $f_{1,2}(n)$ being smooth functions that extend $f(n)$ across $\Sigma$. 
Within the infinitesimal domain width, we can expand $f(n)$ about the origin. In the notations introduced in the previous section, we have
\be
\label{deltaexample}
\lim_{\epsilon\rightarrow 0} \lim_{b \rightarrow 0} \int^\epsilon_{-\epsilon} dn \,\, f(n) \delta_b (n) =
\lim_{\epsilon\rightarrow 0} \lim_{b \rightarrow 0} 
\sum_{l=0}\frac{b^l}{l!} \int^{\epsilon/b}_{-\epsilon/b}     dX\,\, X^l F(X) \left( \overline{f}^{(l)} (0) + \Theta_X (X) [f^{(l)}(0)] \right)
\ee
The only non-vanishing term is $l=0$, and since $F(X), \Theta_X (X)$ are even and odd in $X$ respectively, we obtain 
\be
\lim_{\epsilon\rightarrow 0}  \lim_{b\rightarrow 0} 
\int^\epsilon_{-\epsilon} dn \,\, f(n) \delta_b (n) = \overline{f} (0),
\ee
a result which is naturally intuitive and reduces correctly to the expected one in the continuous limit.  
Now we can extend this calculation to derivatives of the delta function, for example, $\delta'(n)$. 
Consider thus the integral 
\be
\label{deltapexample}
\lim_{\epsilon\rightarrow 0} \lim_{b \rightarrow 0} \int^\epsilon_{-\epsilon} dn \,\, f(n) \delta'_b (n) =
\lim_{\epsilon\rightarrow 0} \lim_{b \rightarrow 0} \sum_{l=0} \frac{b^{l-1}}{l!}
\int^{\epsilon/b}_{-\epsilon/b} dX \,\, X^l F'(X) \left( \overline{f}^{(l)} (0) +\Theta_X(X) [f^{(l)} (0)] \right).
\ee
For all $l > 1$ this vanishes. For $l=1$, since $F'(X), \Theta_X (X)$ are both odd in $X$, this term reads
\be
\label{deltapp}
\int^\infty_{-\infty}  dX\, X F'(X) \overline{f}'(0) = -\overline{f}'(0),
\ee
where we have integrated by parts. We are left with the $l=0$ term which is singular since
\be
\lim_{b\rightarrow 0} \frac{1}{b} \int^\infty_{-\infty} dX\,\, F'(X) \left(   \overline{f} (0) + \Theta_X (X) [f(0)] \right) = \lim_{b\rightarrow 0} \frac{[f(0)]}{b} \int^\infty_{-\infty} dX\,\, F'(X) \Theta_X(X).
\ee
The integral is then well-defined only if we impose from the outset the continuity of $f$ at $n=0$,\footnote{See for example Section 1.3 of \cite{Strichartz} for some brief comments on smoothness conditions of test functions for distributions.}
\be
\label{conti}
f_2(0) = f_1(0),
\ee
leading to 
\be
\label{derivNS1}
\lim_{\epsilon \rightarrow 0} \int^\epsilon_{-\epsilon} dn\,\, f(n) \delta' (n) = - \overline{f}'(0).
\ee
The continuity condition in \eqref{conti} for $\delta'(n)$ is the simplest example of what
we shall allude to as \emph{regularity constraints} that we would need to solve for, and impose when we encounter some integrand in \eqref{integration} that is typically a product of some nascent delta functions each possibly equipped with some order of derivative. Subject to \eqref{conti}, we can understand 
$\delta'(n)$ to yield (minus) the average of the first derivative of $f(n)$ evaluated at $n=0$ should the function be non-differentiable at $n=0$. In a similar vein, one can prove that more generally, we have 
\be
\label{derivNS}
\lim_{\epsilon \rightarrow 0} \int^\epsilon_{-\epsilon} dn\,\, f(n) \delta^{(k)} (n) = (-1)^k \overline{f}^{(k)}(0), \,\,\,
k \geq 1,
\ee
which is an intuitive generalization of the case where the test functions $f(n)$ are smooth. 
We note that the simple form of the RHS of \eqref{derivNS} follows by virtue of the symmetry property of the nascent delta function in \eqref{parity}
which, if relaxed, leads to a more complicated relation. For example, if $F$ has no definite parity, then 
generally instead of \eqref{derivNS1} we have
$
-\overline{f}'(0) - [f'(0)]\int^\infty_{-\infty} dX F^2(X) X
$, which depends on the precise form of $F$.

Usually, when $\delta(n), \delta'(n)$ are regarded as distributions on some open subset of $\mathbb{R}$, the class of test functions are smooth, compactly supported functions. As shown above, for test functions which are non-smooth,
$\delta (n), \delta'(n)$ can still be regarded as distributions with point support at $n=0$, but
for $\delta'(n)$ to be a sensible distribution, we need to restrict the class of test functions to be at least continuous at $n=0$.

\subsubsection{Regularity constraints for products of nascent delta
  functions
    \& their derivatives}

In the Gaussian normal chart, integrating the bulk equations of motion
(of a general higher-derivative gravitational theory) across $\Sigma$
is reduced to evaluating a set of one-dimensional integrals, with the
integrand being some complicated product of various derivatives of the
extrinsic curvature. Since generally, we take the extrinsic curvature
to be not necessarily continuous at $\Sigma$ in response to a
delta-singular energy source, this implies that in the absence of some
regularizing constraints, such an integral is generically singular.

In the same vein by which we have studied the basic examples of $\delta(n), \delta'(n)$, 
we now explain how we can solve for regularity constraints for products of non-smooth functions, delta functions and their derivatives. As we have seen in the previous examples, it is useful to describe a non-smooth function with the ansatz \eqref{gendiscont} which, in turn, contains smooth functions that can be
expanded around $n=0$ within the thin-width domain of integral.

This implies that we can break down the integral in \eqref{integration} into a linear combination of integrals of the following form.
\be
\label{prodN}
I(l, \vec{k})  = \lim_{\epsilon \rightarrow 0} \int^\epsilon_{-\epsilon} dn\,\, n^l \partial^{k_1}_n \Theta (n)
\partial^{k_2}_n \Theta (n) \ldots \partial^{k_j}_n \Theta (n),
\ee
where $l$ and the indices $\vec{k} = \{ k_1,k_2, \ldots, k_j \}$ are non-negative integers,
with each $k_i$ indicating the order of derivative. Together with the scaling limit \eqref{doublescale}, 
we can express \eqref{prodN} as 
\be
\label{prodN2}
I(l, \vec{k} ) = \lim_{b \rightarrow 0} b^{l+1- \sum^j_{m=1} k_m } \int^\infty_{-\infty} dX\,\, X^l 
\Theta_X^{(k_1 )} (X) \ldots \Theta_X^{(k_j )} (X).
\ee
Such a term vanishes for 
$l \geq \sum_m k_m$, remains finite for $l+1-\sum_m k_m = 0$,
and diverges for
\be
\label{diverge}
l+1 - \sum^j_{m=1} k_m <0, \qquad l= j - \sum^j_{m=1} k_m\,\,\,\,\,(\text{mod}\,\, 2),
\ee
with the second condition in \eqref{diverge} being due to the fact
that the integral $I(l, \vec{k} )$
vanishes by virtue of
$\Theta_X (X) = -\Theta_X(-X)$ for $l=1+j - \sum_m k_m$ \,\,\,(mod 2). 

After summing up the linear combination of $I(l,\vec{k} )$ defining
\eqref{integration}, 
for various singular terms which diverge as $b^{s}$ for some negative
index $s$, we can now sum them up and set the overall coefficient to
vanish. This naturally leads to a regularity constraint that we have
to impose separately for \emph{each order of singularity} labelled by
each distinct $s$. A caveat is that the definite integral in
\eqref{prodN2} is generally representation-dependent. Since we would
like to eliminate all singular integrals for any choice of nascent
delta function, we should then impose a stronger condition: that for
each family of integrals labelled by the same $\vec{k}$, every
singular $l$ satisfying \eqref{diverge} then leads to a regularity
constraint. This procedure then renders the integral in
\eqref{integration} to be well-defined for any choice of nascent delta
function. In the following, we elaborate on several illustrative
examples.

We begin with a basic example by taking $\vec{k}_0 = \{ 1,1,1 \}$, with
\be
\label{sexample}
I(l,\vec{k}_0) = \lim_{\epsilon \rightarrow 0} \int^\epsilon_{-\epsilon} dn\,\, n^l \, \left( \Theta'(n) \right)^3
= \lim_{b \rightarrow 0}  b^{l-2}   \int^\infty_{-\infty} dX\,\, X^l \, F^3 (X),
\ee
which diverges for $l=0$, vanishes for $l=1$ and $l \geq 3$, and for $l=2$, it evaluates to 
$
 \int^\infty_{-\infty} dX\,\, X^2\, F^3 (X)
$.
If we replace $n^l$ in the integrand of \eqref{sexample} by a function $\phi(n)$ that is analytic at $n=0$, 
then this implies that the integral diverges unless $\phi(0) = 0$, in which case we have 
\be
\label{sexample2}
I(\phi , \vec{k}_0 ) \equiv  \lim_{\epsilon \rightarrow 0} \int^\epsilon_{-\epsilon} dn\,\, \phi(n) \, \left( \Theta'(n) \right)^3
= \frac{\phi''(0)}{2} \int^\infty_{-\infty} dX\,\, X^2\, F^3 (X).
\ee
In this case, $\phi(0) =0$ would be what we call as a \emph{regularity constraint} which stipulates how fast $\phi(n)$ should grow near the origin for the integral to converge. The integral \eqref{sexample2} also demonstrates
that the naive product of three $\delta$-functions can be understood as a proper distribution, provided we restrict the space of test functions to be those which grow at least as $\phi (n) \sim n + \mathcal{O}(n^2)$ near the origin. 
\footnote{In our context, the integral limits are different from the typical ones ($\mathbb{R}^n$) used, but if desired, they can be extended to $\mathbb{R}$ provided they decay sufficiently fast enough at infinity for the $F(X)$ chosen. Note that such a distribution has point-support, and is equivalent to $\delta''(n)$ up to a normalization constant (which is $\frac{1}{2} \int^\infty_{-\infty} dX\,\, X^2\, F^3 (X)$), albeit with a different space of test functions. This agrees with the well-known fact that every distribution with point support is a finite linear combination of $\delta$-function and its derivatives (see e.g. \cite{Strichartz} for a semi-formal proof).} This serves as a simple example 
of how the use of nascent delta functions resolves the `ambiguity' that may arise in interpreting products of $\delta$-functions as articulated in \cite{Reina}.

As an another example, consider the sum 
$$ 
\sum_{l} C_l I(l, \vec{k} ),
$$
with a fixed $\vec{k}$, the set of regularity constraints are simply
\be
\label{exReg}
C_l=0,\qquad   \forall\, l \,\,\,\text{satisfying} \,\,\, \eqref{diverge}.
\ee 
In \eqref{integration}, as we shall see through explicit examples in later sections, it involves a linear combination of integrals of the form $I(l,\vec{k} )$ with a finite set of $\vec{k}$. The entire set of regularity constraints is then the union of all \eqref{exReg} associated with each $\vec{k}$. 

It is also instructive to check how \eqref{diverge} applies to the previous basic examples of 
$\delta(n), \delta'(n)$ in \eqref{deltaexample} and \eqref{deltapexample}. We note that 
for the two terms on the RHS of \eqref{deltaexample}, they correspond to $I(l, \{ 1 \})$ and $I(l, \{ 0,1 \})$
in the notation of \eqref{prodN}. With $\sum_m k_m =1$ for each term, there is no solution to \eqref{diverge} and thus no regularity constraint is needed. For \eqref{deltapexample}, each of the two terms on the RHS are of the form
$I(l, \{ 2 \})$ and $ I(l, \{ 0,2 \})$ respectively, with $\sum_m k_m =2$ for each term. From \eqref{diverge}, we thus see
that there is a regularity constraint needed for $I(l, \{0,2 \} )$ (where $j=2$) associated with $l=0$, and this is simply \eqref{conti}.

After imposing the necessary regularity constraints in \eqref{integration} following the above approach, we are left with a linear combination of well-defined integrals of the form
\be
\label{finiteI}
I(l, \vec{k} ) = I \left(\sum^j_{m=1} k_m -1 , \vec{k} \right)  =  \int^\infty_{-\infty} dX\,\, X^{(\sum^j_{m=1} k_m -1)}
\Theta_X^{(k_1 )} (X) \ldots \Theta_X^{(k_j )} (X).
\ee
Since the integrand of \eqref{finiteI} is odd for an even $j$, such a term only contributes to the junction equations
for odd values of $j$. For example, in \eqref{deltaexample}, the eventual expression corresponds to the first term within the RHS bracket which is of the form \eqref{finiteI} with $l=0,j=1,\sum_m k_m =1$,
whereas for \eqref{deltapexample}, we note that \eqref{deltapp} is of the form \eqref{finiteI}
with $l=1, j=1, \sum_m k_m = 2$. As another example, for the sum $ \sum_{l} C_l I(l, \vec{k} )$,
the only finite term (after imposing \eqref{exReg}) is 
\be
\label{finiteRep}
C_r I\left(r , \vec{k} \right),\qquad r=\sum^j_{m=1} k_m -1,
\ee
which is non-vanishing only if $j$ is odd. A caveat is that, like the definite integral in \eqref{diverge}, an expression like \eqref{finiteI} is 
generally sensitive to the choice of the nascent delta function, apart from terms
like 
\be
\label{FiniteInpJ}
\int dX\,\, \Theta'_X, \qquad \int dX\,\, X^p \Theta^{(p+1)}_X, \qquad \text{or} \qquad \int dX\,\, \left( \Theta_X \right)^p \Theta'_X,
\ee
which are independent of the choice. For example, consider
the integral
\be
\label{repEx}
I(2, \{1,1,1 \} ) =  \int^\infty_{-\infty} dX\,\, X^2 \left( \Theta'_X (X) \right)^3.
\ee
If we pick the nascent delta function to be the Gaussian $F(X) = \frac{1}{\sqrt{\pi}} e^{-X^2}$, then 
\eqref{repEx} evaluates to $1/(6\pi \sqrt{3} )$, whereas a choice of $F(X) = \frac{\sin (X)}{\pi X}$ yields
$1/(2\pi^2)$ instead.  It may seem like such junction terms are `regularization-dependent'. This feature is obviously absent in ordinary GR, and here we see that
the higher-order nature of the field equations may probe the form of the nascent delta function.
To write down consistent junction conditions that are insensitive to the choice of nascent delta functions, we could additionally set the coefficient $C_r$ in \eqref{finiteRep} to vanish each time it appears in \eqref{integration}. We are then left with junction terms arising from universal terms like those in 
\eqref{FiniteInpJ}. 

In most of our working examples for the rest of the paper, these representation-dependent terms turn out not to feature much. The only setting where it arises non-trivially in this work is the case of $R^3$ theory for which we find the terms 
$$
\int dX\, \Theta'^2_X \Theta_X X,\qquad \int dX\, \Theta''_X \Theta'_X \Theta_X X^2,
$$
to be manifest in the junction equations. As we shall demonstrate later, these terms would be absent if we further set $[K']=0$ as a regularity constraint for the $R^3$ theory's junction equations. In the general case, our
derivation procedure described above allows us to solve for the regularity constraints that will yield
the final junction conditions to be representation-independent if desired. Nonetheless, it is noteworthy 
that the convergence of the integral \eqref{integration} is compatible with the presence of these terms which, if allowed, implies that the specification of junction conditions is only complete with a choice of nascent delta function.

To summarize, we can now state explicitly how to read off the regularity constraints and junction conditions for \eqref{integration}. 
After parametrizing each discontinuous geometric quantity and its derivatives by $\Theta_X$, 
expanding the LHS of \eqref{integration} would yield integrand terms typically of the form 
\be
\label{GenericInt}
I_{\vec{k}} \equiv \phi (n) \Theta^{(k_1)}_X \Theta^{(k_2)}_X \ldots \Theta^{(k_j)}_X,
\ee
for some vector index $\vec{k}$, with $\phi (n)$ analytic at $\Sigma$, then the regularity constraints associated with $I_{\vec{k}}$ are 
\be
\label{genRegC}
\phi^l (0)  = 0,\,\,\,\forall\,\, l \leq \sum_{m=1}^j k_m -2, \,\,\, l = j - \sum_{m=1}^j k_m \,\,\, (\text{mod}\,\, 2),
\ee
with the junction term induced by integrating $I_{\vec{k}}$ across $\Sigma$ being 
\be
\label{junctionterm}
J_{\vec{k}} = \frac{1}{((\sum_m k_m) -1)!} \phi^{((\sum_mk_m) -1)}(0) \left(
\int^\infty_{-\infty} dX\, X^{(\sum_m k_m) -1} \Theta^{(k_1)}_X \ldots \Theta^{(k_j)}_X
\right).
\ee
The final junction condition arising from \eqref{integration} is then the sum of 
all the junction terms, each of the form \eqref{junctionterm}, subject to us imposing
all the regularity constraints, each of the form \eqref{genRegC}.

\subsection{The relation to Hadamard Regularization}
\label{Hadamard}

In this Section, we point out a relation between the regularity constraints
and Hadamard regularization \cite{Hadamard} ---  a well-studied procedure to regularize divergent integrals 
typically encountered in the theory of singular integral operators, and also commonly invoked when one handles
distributions defined by divergent integrals (for an emphasis in the theory of distributions, see for example \cite{Ram}).

Let us first briefly review the notion of Hadamard regularization with a simple example. Consider the (divergent) integral
$$
I_H = \int^\infty_0 dx\, \frac{\phi(x)}{x^{3/2}} = \lim_{\epsilon \rightarrow 0} \int^\infty_\epsilon dx\, \frac{\phi(x)}{x^{3/2}},
$$
where $\phi(x)$ is regular and continuous at the origin. From the mean value theorem, we can write 
$\phi (x) = \phi (0) + x \phi' (tx) \equiv \phi(0) + x \varphi(x), \,\,\, 0< t<1$ and thus we can express the above integral as
\be
\label{Hadexample}
I_H = \lim_{\epsilon \rightarrow 0} \frac{2\phi (0)}{\epsilon} + \lim_{\epsilon \rightarrow 0} \int^\infty_\epsilon dx\,\, 
\frac{\varphi(x)}{\sqrt{x}}.
\ee
The second integral in \eqref{Hadexample} converges and we define it to be Hadamard finite part of $I_H$, 
writing
\be
\label{Hexample}
FP \int^\infty_0 dx\,\,  \frac{\phi(x)}{x^{3/2}} = \lim_{\epsilon \rightarrow 0} \int^\infty_\epsilon dx \frac{\phi (x)}{x^{3/2}} 
- \lim_{\epsilon \rightarrow 0 } \frac{2\phi (0)}{\sqrt{\epsilon}}.
\ee
We have thus extracted the singular piece in the integral, with the first term of RHS of \eqref{Hexample} being 
the residual finite integral. 
The above example can be generalized to a similar regularization of the divergent integral
$$
I_D = \int^\delta_{-\delta}dx\,\,\, f(x) \phi (x), \qquad \delta >0,
$$
where we take $\phi (x)$ to be analytic at the origin, and $f(x) |x|^m$ has an algebraic singularity of order $m$ at $x=0$, i.e.
$m$ is some smallest positive integer such that $f(x)|x|^m$ is locally integrable.

We can extend this notion to $f(x)$ being a distribution, the Hadamard regularization of which is then defined as 
\be
\label{hadamardT}
FP \langle f, \phi \rangle = \int^\delta_{-\delta} dx\,\, f(x) \left(
\phi (x) -
\left(
\phi (0) + \phi'(0) x + \ldots + \frac{1}{m!} \phi^{(m-1)}(0) x^{m-1}
\right)
\right) \Theta \left( 1 - \frac{|x|}{\epsilon} \right).
\ee
We now examine how \eqref{hadamardT} is relevant for our context. 
From \eqref{prodN} and \eqref{prodN2}, we see that the sequence (in parameter $b$)
$$
\Delta_{\vec{k}} (n,b) \equiv \Theta^{(k_1)} (n,b) \ldots \Theta^{(k_j)} (n,b),
$$ 
converges to a distribution $\Delta_{\vec{k}} (n)$ that is of order of singularity 
$\left( \sum_{m=1}^j k_m \right) -1$, and further, upon integrating it against some function $\phi (n)$ that is analytic at $n=0$,
the Hadamard-regularized distribution reads
\bea
\label{Hcontext}
FP \langle  \Delta_{\vec{k}} (n), \phi (n) \rangle &=& 
\lim_{\epsilon \rightarrow 0}
\int^\epsilon_{-\epsilon} dn\,\, \Delta_{\vec{k}} (n) \cr
&&
\times \Bigg(
\phi (n)
- \left(
  \phi (0) + \phi'(0) n + \ldots + \frac{1}{(\sum_m k_m-2)!} \phi^{(\sum_m k_m-2)}(0)
 n^{\sum_m k_m-2}
\right)
\Bigg), \nonumber \\
\eea
where now $\delta = \epsilon = 0^+$ for the relevant integral domain. Comparing
\eqref{Hcontext} against 
\eqref{genRegC} and \eqref{junctionterm}, and 
recalling that $I(l,\vec{k} )$ vanishes
for $l=1+j - \sum_m k_m \,\,(\text{mod}\,\, 2)$,
it is then clear that:
\begin{itemize}
\item \emph{in the absence of the regularity constraints, the
  junction condition is nothing but the Hadamard-finite part of the
  LHS of \eqref{integration}},

\item imposing the regularity constraints then ensures that the integral
  in \eqref{integration} converges, being trivially equivalent to its
  Hadamard regularization.
\end{itemize}

Although it is nice to have recognized that our derivation of the
junction conditions (and regularity constraints) admits a natural
interpretation in terms of Hadamard regularization, we stress that our
method is independently consistent, and can be understood and
implemented without alluding to the latter.

\section{Generalized junction conditions for gravitational theories with quadratic terms}
\label{quadraticSec}

In this Section, we derive the generalized junction conditions for the following class of gravitational
theories with action terms quadratic in (various contractions of) the Riemann tensor. 
\be
\label{quadra}
\mathcal{L}_{quad} = \frac{1}{16 \pi } \left( R + \beta_1 R^2 + \beta_2 R_{uv} R^{uv} + 
\beta_3 R_{\alpha \beta \mu \nu} R^{\alpha \beta \mu \nu} \right).
\ee
We integrate the equations of motion $ \tilde{G}_{\mu \nu} = 8\pi T_{\mu \nu}$ across the infinitesimally thin surface $\int^\epsilon_{-\epsilon} dn\,\, \tilde{G}_{\mu \nu}$, identify the regularity constraints and derive the final 
explicit covariant form of the generalized junction condition. 
The equations of motion read
\bea
&&G_{\alpha \beta} + 2\beta_1 R R_{\alpha \beta} -4 \beta_3 R_{\alpha \mu} R^\mu_\beta
+2 \beta_3 R_{\alpha \rho \mu \nu} {R_\beta}^{\rho \mu \nu} + 
(2\beta_2 + 4 \beta_3) R_{\alpha \mu \beta \nu} R^{\mu \nu} \cr
&&- 2(\beta_1 + \frac{1}{2} \beta_2 + \beta_3)\nabla_\alpha \nabla_\beta R  
 +(\beta_2 + 4\beta_3) \Box R_{\alpha \beta}  - \frac{1}{2} g_{\alpha \beta} \bigg(
-(4\beta_1 + \beta_2) \Box R \cr
&&+ \beta_1 R^2 + \beta_2 R_{\mu \nu} R^{\mu \nu} + \beta_3 
R_{\rho \sigma \mu \nu}R^{\rho \sigma \mu \nu}
\bigg) = 8\pi  T_{\alpha \beta}.
\eea

In the following we consider $\tilde{G}_{ij}, \tilde{G}_{in}, \tilde{G}_{nn}$ separately. This Section is accompanied
by the Appendix \ref{AppA} which collects several useful identities that we developed for evaluating the integrals easily. 
We will find that in contrast to the case in GR, apart from bracketed quantities, the junction terms also involve
averaged quantities across $\Sigma$. In particular, the expression
$
\frac{1}{3} \left(  \overline{f g } + 2\overline{f} \overline{g} 
\right)[h] \equiv \myov{ f g } [h]
$
turns out to occur frequently (see \eqref{useful} in Appendix \ref{AppA} ).

\subsection{Junction terms from integrating $\tilde{G}_{ij}$}
In the following, we present explicitly the result of integrating $\tilde{G}_{ij}$ 
across $\Sigma$ for each term in the
equation of motion.
Below, the hatted expressions refer to intrinsic quantities, whereas
bracketed ellipses $( \ldots )$ refer to terms which do not contribute to \eqref{integration}.
For definiteness, we will focus on the case of $\Sigma$ being timelike unless
explicitly stated otherwise.

\begin{itemize}

\item 
For $g_{ij} \Box R = g_{ij} \left(
\partial^2_n R + K \partial_n R + g^{al} \left( 
\partial_l \partial_a R - \Gamma^k_{la} \partial_k R
\right) \right)
$, the Gauss-Codazzi equations enable us to express the first two terms as
$$
g_{ij} \left( \partial^2_n \left(  \hat{R} - 2 \partial_n K - K^2 - K^{ab} K_{ab} \right) 
+ K \partial_n \left(  \hat{R} - 2 \partial_n K - K^2 - K^{ab} K_{ab} \right) \right),
$$
whereas the remaining terms are 
$$
g_{ij} g^{al} \left( 
\partial_l \partial_a (-2\partial_n K) - \Gamma^k_{al} \partial_k (-2\partial_n K)
\right) + \left(  \ldots  \right).
$$
The singular terms arise from $-2 g_{ij} K \partial^2_n K$ and $4 K_{ij} \partial^2_n K$, the latter being
derived after an integration by parts. They sum up to be
$$
2 \left( g_{ij} [K]^2 -2 [K_{ij}][K] \right) \int dn\,\,\, (\Theta')^2.
$$
The remaining finite terms sum up to read 
\bea
&&-g_{ij} \left[  
2 D^2 K + 2K'' + 4KK' + K^3 + KK^{ab} K_{ab} + (K^{ab} K_{ab})'
\right] \cr
&&+8 \myov{K_{ij} K} [K] + 4 \myov{K_{ij} K^{ab} } [K_{ab}] + g_{ij}[K] \left( \myov{K^2} 
+ \myov{K^{ab}K_{ab}} \right) \cr
\label{GijBoxR}
&&+4\left(
g_{ij} \overline{K'} [K] + \overline{K_{ij}} [K'] - \overline{K'_{ij}} [K]
\right) + 8 \left( [K_{ij}] [K]^2 \right)\int^\infty_{0} dX\, X F^2(X),
\eea
where $D^2$ is the Laplacian defined on $\Sigma$, and we use the superscript prime
to denote $\partial_n = n^\alpha \nabla_\alpha$.


\item 
For $g_{ij} R^2$, the Gauss-Codazzi relations enable us to express it as
$$
g_{ij} \left( \hat{R} - 2 \partial_n K - K^2 - K^{ab} K_{ab} \right)^2.
$$
The singular term arises from $4 g_{ij} \partial_nK \partial_n K$, and reads
$$
4 g_{ij} [K]^2 \int dn\,\,\, (\Theta')^2,
$$
whereas the finite terms sum up to read
\be
\label{GijRSq}
g_{ij} \left(
-4[K] (\hat{R} - \myov{K^2} - \myov{K^{ab} K_{ab} } ) + 8 \overline{K'} [K] 
\right)
+ 16[K_{ij}][K]^2 \int^\infty_{0} dX\,  X F^2(X).
\ee


\item 
For $g_{ij} R_{\mu \nu} R^{\mu \nu} = g_{ij} (R_{ab} R^{ab} + R_{nn} R^{nn} ) + \left( \ldots \right)$, 
the Gauss-Codazzi relations enable us to express it as
\bea
&& g_{ij} \bigg( g^{mr} g^{ls} \left( -\partial_n K_{ml} + 2 K^a_m K_{la} - KK_{ml} + \hat{R}_{ml} \right)
\left( -\partial_n K_{rs} + 2 K^a_r K_{sa} - KK_{rs} + \hat{R}_{rs} \right) \cr
&& \left(  \partial_n K + K^{ab} K_{ab} \right)\left(  \partial_n K + K^{cd} K_{cd} \right) \bigg).
\eea
The singular term comes from the term
\be
g_{ij} g^{ml} g^{rs} \partial_n K_{ml} \partial_n K_{rs} = 
g_{ij} \partial_n K^{ab} \partial_n K_{ab} + 4 g_{ij} K^{ma} K_m^b \partial_n K_{ab},
\ee
and also we have another singular term in 
$$
g_{ij} \partial_n K \partial_n K,
$$
which yields the following overall singular term 
\be
g_{ij} \left(   [K^{ab} ] [K_{ab} ] + [K]^2  \right) \int dn\,\,\, (\Theta') ^2.
\ee
The finite terms sum up to read
\bea
\label{GijRuvSq}
&&g_{ij} \left(
2 \myov{KK^{ab} } [K_{ab}] - 2 \hat{R}^{ab} [K_{ab}] + 
\overline{K^{ab'}} [K_{ab}] + \overline{K_{ab}' } [K^{ab}]+2 \overline{K'} [K] 
+2 \myov{K^{ab} K_{ab}} [K]
\right) \cr
&& +4[K_{ij}] \left(   [K^{ab} ] [K_{ab} ] + [K]^2  \right) \int^\infty_{0} dX\,  X F^2(X).
\eea

\item
For $g_{ij} R_{\alpha \beta \mu \nu} R^{\alpha \beta \mu \nu} = 4 g_{ij} R_{ancn} g^{al} g^{cm}
R_{lnmn} + \left( \ldots \right)$, the Gauss-Codazzi relations enable us to express it as
\be
4g_{ij} g^{al} g^{mc} \left( -\partial_n K_{ac} + K_{ab} K^b_c \right)
\left( -\partial_n K_{ml} + K_{ms} K^s_l \right).
\ee
The singular term arises from $4g_{ij} g^{al} g^{mc} \partial_n K_{ac} \partial_n K_{ml}$ 
which yields
\be
4g_{ij} [K^{lm} ] [K_{lm} ] \int dn\,\,\, (\Theta')^2.
\ee
The finite terms sum up to read
\bea
\label{GijRabcdSq}
&&g_{ij} \left(
-8 \myov{K_l^m K^{ln} } [K_{mn}] + 4 \overline{K^{mn'}} [K_{mn}] + 4 \overline{K_{mn}'} [K^{mn}] 
+ 16 \myov{K^{mr} K_r^l } [K_{ml}]
\right) \cr
&&+16 [K_{ij}] [K^{lm} ] [K_{lm} ] \int^\infty_{0} dX\, X F^2(X).
\eea


\item
For $RR_{ij}$, the Gauss-Codazzi relations enable us to express it as
\be
\left(\hat{R} - 2 \partial_n K - K^2 - K^{ab} K_{ab} \right) \left( -\partial_n K_{ij} + 2K^a_j K_{ai}
-K K_{ij} + \hat{R}_{ij} \right).
\ee 
The singular term arises from $2 \partial_n K \partial_n K_{ij}$ and reads
\be
2 [K] [K_{ij} ] \int dn \,\,\, (\Theta')^2,
\ee
whereas the finite terms sum up to read 
\be
\label{RRij}
2 \overline{K'} [K_{ij} ] + 2 \overline{K_{ij}'} [K] - 2 \left( 2 \myov{ K_i^m K_{mj}   }
-\myov{KK_{ij}} + \hat{R}_{ij}
\right)[K]
- \left( 
\hat{R} - \myov{K^2} - \myov{K^{ab}K_{ab}}
\right) [K_{ij}].
\ee


\item
For $R_{i\mu} R^\mu_j = R_{im} R^m_j + \left(  \ldots  \right)$, the Gauss-Codazzi relations enable us to express it as
\be
g^{ml} \left( - \partial_n K_{im} + 2 K^a_i K_{am} - KK_{im} + \hat{R}_{im} \right)
\left( - \partial_n K_{jl} + 2 K^a_j K_{al} - KK_{jl} + \hat{R}_{jl} \right).
\ee
The singular term arises from $g^{ml} \partial_n K_{im} \partial_n K_{jl}$ and reads
$$
[K^a_i ][K_{ja} ] \int dn\,\,\, (\Theta')^2,
$$
whereas the finite terms sum up to read
\be
\label{RiuRuj}
\myov{K'_{l(i} } [K_{j)}^l ] -
\left( 2 \myov{K_m^l K_{l(j} } - \myov{ KK_{m(j} } + \hat{R}_{m(j} \right) [K_{i)}^m ]. 
\ee

\item 
For $R_{i\mu \alpha \beta} {R_j}^{\mu \alpha \beta} = 2 R_{inrn} R_{jnsn} g^{rs} + \left(  \ldots \right)$, the
Gauss-Codazzi relations enable us to express it as 
\be
2 g^{rs}   \left( - \partial_n K_{ir} + K_{im} K^m_r \right)\left( - \partial_n K_{js} + K_{jm} K^m_s \right).
\ee
The singular term arises from $2g^{rs} \partial_n K_{ir} \partial_n K_{js}$ and reads 
$$
2[K_i^s ][K_{js} ] \int dn \,\,\, (\Theta')^2,
$$
whereas the finite terms read
\be
\label{RiuabSq}
2 \overline{K'_{l (i }} [K_{j)}^l ] - 2 \myov{ K^l_m K_{l(i}    } [ K_{j)}^m ].
\ee

\item
For $R_{i\mu j\nu}R^{\mu \nu} = R_{iajb}R^{ab} +R_{injn} R^{nn} + \left(  \ldots  \right)$, the Gauss-Codazzi relations
enable us to express it as 
\bea
&&\left( \hat{R}_{iajb} - K_{ij} K_{ab} + K_{ib} K_{aj} \right) g^{al} g^{bm} 
\left( -\partial_n K_{lm} + 2 K^r_l K_{rm} - KK_{lm} + \hat{R}_{lm} \right)  \cr
&&
+ \left(  -\partial_n K_{ij} + K_{ia} K^a_j \right) \left( - \partial_n K
- K^{lm} K_{lm} \right).
\eea
The singular term arises from $\partial_n K_{ij} \partial_n K$ and reads
$$
[K_{ij} ][K] \int dn\,\,\, (\Theta')^2,
$$
whereas the finite terms sum up to read
\be
\label{RiujvRuv}
[K_{ab}] \left(
-\myov{K_i^b  K_j^a } - \hat{R}_{iajb} + \myov{K_{ij} K^{ab}}
\right)
+ \overline{K'_{ij} } [K] + \overline{K'} [K_{ij} ] - \myov{ K_{im} K^m_j  } [K] +
\myov{K^{ab} K_{ab}} [K_{ij}].
\ee

\item 
For $\nabla_i \nabla_j R$, the Gauss-Codazzi equation enables us to express it as
\be
\partial_i \partial_j (-2 \partial_n K) + K_{ij} \partial_n (-2 \partial_n K - K^2 - K^{ab} K_{ab} ) 
-\Gamma^k_{ij} \partial_k ( -2\partial_n K) + \left(  \ldots  \right)
\ee
The singular term comes from $-2 K_{ij} \partial^2_n K$ and reads
$$
2[K_{ij}][K] \int dn\,\, (\Theta')^2.
$$
The finite terms sum up to read
\be
\label{DiDjR}
-2 \left[  D_i D_j K \right] - \left[  K_{ij} (K^2 + K^{ab}K_{ab} ) \right] + 
[K_{ij}]( \myov{K^2} + \myov{K^{ab} K_{ab}} ) -2 \overline{K_{ij}} [K'] + 2 \overline{K'_{ij} } [K],
\ee
where $D_i$ is the affine connection defined on $\Sigma$. 
\item 
For $\Box R_{ij}$, it is useful to lay out explicitly various terms 
\bea
\Box R_{ij} &=& \partial_n ( \partial_n R_{ij} - \Gamma^l_{n(i} R_{j)l} ) 
- K^l_{(i} \left(  
 R'_{j)l} - R_{j)m} \Gamma^m_{nl} - \Gamma^m_{j)n} R_{lm}
\right) \cr
&&+ g^{rl} \left(
\partial_l (\nabla_r R_{ij} ) - \Gamma^k_{lr} \nabla_k R_{ij} -
\Gamma^k_{l(i} \nabla_r R_{j)k} - 
\Gamma^n_{l(i} \nabla_r R_{j)n} - \Gamma^n_{rl} \nabla_n R_{ij}   
\right). \;
\eea
The singular terms arise from $- K^l_{(i}  R'_{j)l} - \Gamma^n_{rl}
\nabla_n R_{ij}$ and they sum up to read
$$
\left(  -2[K^l_i K_{lj}] + [K][K_{ij}]  \right) \int dn\,\,\, (\Theta')^2 .
$$
The finite terms sum up to read
\bea
&&
-[D^2 K_{ij} ]+ 
\left[ 
2KK^l_i K_{lj} - K^2 K_{ij}
\right]
+[K] \myov{K K_{ij} }
+ \myov{KK^m_{(i} } [K_{j)m} ]
- 2\myov{K^m_{(i} K_{j)m}  } [K] \cr
\label{BoxRij}
&+&
\myov{K^m_l K^l_{(j}} [K_{i)m}] +
\overline{K^l_{(i}}[ K_{j)l}' ] - \overline{K^{'l}_{(i} }[ K_{j)l} ] +
\overline{K'} [K_{ij} ] - \overline{K} [K'_{ij}].
\eea

\end{itemize}

\subsection{Junction terms from integrating $\tilde{G}_{in}$}
In the following, we present explicitly the result of integrating 
\bea
\tilde{G}_{in} &\equiv& G_{in} + 2 \beta_1 R R_{in} - 4 \beta_3 R_{i\mu} R^\mu_n + 2\beta_3
R_{i\rho \mu \nu} {R_n}^{\rho \mu \nu} 
+ (2\beta_2 +4\beta_3) R_{i\mu n \nu } R^{\mu \nu} \cr
&&\,\,\,-2 ( \beta_1 + \frac{1}{2} \beta_2 + \beta_3 ) \nabla_i \nabla_n R + 
(\beta_2 + 4 \beta_3 ) \Box R_{in},
\eea
across $\Sigma$ for each term in the
equation of motion. There are no singular terms. In the following, we display the result of integrating each term across $\Sigma$ after invoking Gauss-Codazzi relations. 

\begin{itemize}
\item
From $R R_{in}$, we have 
\be
-2 \left( \overline{D^a K_{ia}} - \overline{D_i K} \right) [K].
\ee
\item
From $R_{i\mu} R^\mu_n$, we have 
\be
-\left( \overline{D^a K_{ka}} - \overline{D_k K} \right) [K_i^k ] -\left( \overline{D^a K_{ia}} - \overline{D_i K} \right)[K].
\ee
\item
From $R_{i\rho \mu \nu} {R_n}^{\rho \mu \nu} = 2R_{ikna} {R_n}^{kna} + \left( \ldots \right) $, we have
\be
-2[K^{ak}]  \left( \overline{D_k K_{ia}} - \overline{D_i K_{ak}} \right).
\ee
\item
From $R_{i\mu n \nu} R^{\mu \nu} = R_{ianb} R^{ab} + R_{innl}R^{nl}$, we have
\be
-[K^{ab}]\left( \overline{D_a K_{bi}} - \overline{D_i K_{ba}} \right) 
+ [ K_{il} ] \left( \overline{D^a K^l_a} - \overline{D^l K} \right).
\ee 
\item 
For $\nabla_i \nabla_n R$, we first note that 
\be
\nabla_i \nabla_n R = \partial_i \partial_n \left(  \hat{R} - 2 \partial_n K - K^2 - K^{ab}K_{ab}
\right) - K^j_i \partial_j \left(  \hat{R} - 2 \partial_n K - K^2 - K^{ab}K_{ab}
\right).
\ee
After integrating across $\Sigma$, we have
\be
-D_i \left[
2K'+ K^2  + K^{ab}K_{ab}
\right] + 2 \overline{K^j_i} [D_j K].
\ee 
\item
For $\Box R_{in} = \left( \nabla^k \nabla_k + \nabla^n \nabla_n \right) R_{in}$, we first 
note that 
\be
\nabla^n \nabla_n R_{in} = \partial_n (\nabla_n R_{in} ) - K^k_i (\partial_n R_{kn} - K^l_k R_{ln} ),
\ee
which, upon integrated across $\Sigma$, yields 
\be
n^\alpha [\nabla_\alpha \left( D^l K_{il} - D_i K \right)] - \overline{K^k_i} [D^l K_{kl} - D_k K].
\ee
For $\nabla^k \nabla_k R_{in}$, we note that 
\be
\nabla^k \nabla_k R_{in} = g^{kl} \left( \partial_l (\nabla_k R_{in} ) - \Gamma^j_{lk} \nabla_j R_{in} 
-\Gamma^n_{lk} \nabla_n R_{in} - \Gamma^j_{li} \nabla_k R_{jn} - \Gamma^n_{li} \nabla_k R_{nn} - 
\Gamma^j_{ln} \nabla_k R_{ij} \right),
\ee
and identify the terms which contribute to the integral to be 
\bea
&&\nabla_k R_{in} = - \Gamma^n_{ki} R_{nn} - \Gamma^a_{kn} R_{ia} + \left(  \ldots
\right), \quad
\nabla_j R_{in} = - \Gamma^n_{ji} R_{nn} - \Gamma^a_{jn} R_{ia} + \left(  \ldots
\right), \cr
&&\nabla_n R_{in} = \partial_n R_{in} + \left( \ldots  \right), \quad
\nabla_k R_{nn} = \partial_k R_{nn} + \left(  \ldots  \right), \cr
&& \nabla_k R_{ij} = \partial_k R_{ij} - \Gamma^l_{k(i} R_{j)l} + \left( \ldots \right),
\quad \nabla_k R_{jn} = - \Gamma^n_{kj} R_{nn} - \Gamma^a_{kn} R_{ja} + \left( \ldots
\right). \nonumber
\eea
Some straightforward (but lengthy) algebra then gives the
finite terms to be 
\be
-\overline{D^k K_{ki}} [K] - 2 \overline{K^l_i} [D_l K] - \overline{K} [\nabla_i K] + 
\overline{D^k K^a_k} [K_{ia} ] + 2 \overline{K^{ab}} [ D_a K_{ib} ] +
\overline{K} [D^a K_{ia} ].
\ee
\end{itemize}
A useful consistency check lies in taking the Gauss-Bonnet limit 
$
\beta_1 = \beta_3 =-\frac{1}{4} \beta_2.
$
where we find that the various junction terms sum up to vanish, in
accordance with the result reported in earlier literature obtained by
boundary variation of the surface term.

\subsection{Junction terms from integrating $\tilde{G}_{nn}$}
In the following, we present explicitly the result of integrating $\tilde{G}_{nn}$ 
across $\Sigma$ for each term in the
equation of motion. There are no singular terms. In the following, we display the result of integrating each term across $\Sigma$ after invoking Gauss-Codazzi relations. 

\begin{itemize}

\item 
For $RR_{nn}$, the Gauss-Codazzi relations imply that we can write it as 
\be
\left(\hat{R} - 2 \partial_n K - K^2 - K^{ab} K_{ab} \right) 
\left( - \partial_n K - K^{ab} K_{ab} \right),
\ee
from which we can read off the singular term to be 
$$
2 [K]^2 \int dn\,\,\, (\Theta')^2,
$$
and the finite terms to be 
\be
3 \myov{K^{ab}K_{ab}} [K] - [K] \left( \hat{R} - \myov{K^2} \right) + 4 \overline{K'} [K].
\ee

\item 
For $R_{n\mu} R^\mu_n$, the Gauss-Codazzi relations imply that we can write it as 
\be
(\partial_n K + K^{ab} K_{ab} )(\partial_n K + K^{ab} K_{ab} ),
\ee
from which we can read off the singular term to be 
$$
[K]^2 \int dn\,\,\, (\Theta')^2,
$$
and the finite terms to be 
\be
2[K] \left(  \overline{K'} + \myov{K^{ab}K_{ab}} \right).
\ee

\item 
For $R_{n \rho \mu \nu} {R_n}^{\rho \mu \nu}$, the Gauss-Codazzi relations imply that we can write it as
\be
2(-\partial_n K_{kl} + K_{ka} K^a_l ) g^{kr} g^{ls} (-\partial_n K_{rs} + K_{rb} K^b_s ),
\ee
from which we can read off the singular term to be
$$
2[K^{ab}][K_{ab}] \int dn\,\,\, (\Theta')^2,
$$
and the finite terms to be
\be
-4[K_{kl}] \myov{K^k_b K^{bl}} + 4 \overline{K'_{kl}}[K^{kl}].
\ee

\item
For $R_{n\mu n \nu} R^{\mu \nu}$, the Gauss-Codazzi relations imply that we can write it as 
\be
\left( -\partial_n K_{ij} + K_{ia} K^a_j \right) g^{ir} g^{js} \left( -\partial_n K_{rs} + 2 K^b_r K_{sb} - K K_{rs}
+ \hat{R}_{rs} \right),
\ee
from which we can read off the singular term to be 
\be
[K_{ab}][K^{ab}] \int dn\,\, (\Theta')^2,
\ee
and the finite terms to be 
\be
-[K_{ij}] \left( 
2 \myov{ K^{ib} K^j_b   } - \myov{KK^{ij}} + \hat{R}^{ij}
\right)
-[K_{rs}] \myov{K^r_a K^{as}} + 2 \overline{K'_{ij}} [K^{ij}].
\ee

\item
  For $\nabla_n \nabla_n R$, after invoking the Gauss-Codazzi relations, we have 
the finite terms
$$
[-2 \partial^2_n K - \partial_n (K^2) - \partial_n (K^{ab} K_{ab})  ].
$$

\item
For $\Box R_{nn} = -[\partial_n (\partial_n K + K^{ab} K_{ab} ) ] + g^{ab} \nabla_a \nabla_b R_{nn}$,
we first expand the second term to read
$$
g^{kl} \left(  
\partial_l \partial_k (-\partial_n K + \ldots ) - \Gamma^i_{lk} \partial_i (-\partial_n K + \ldots )
\right) - K \partial_n (\partial_n K + K^{ab}K_{ab} ) + 2 K^{ik} ( K_{ki} \partial_n K - K^j_k \partial_n K_{ij} + \ldots) 
$$
after which it is easier to read off the singular term to be
$$
[K]^2 \int dn\,\,\, (\Theta')^2,
$$
and the finite terms to be 
\be
-[KK'] - 2 \myov{KK_{ab}} [K^{ab}] + 2 \myov{K^{ab}K_{ab}} [K] - 2 \myov{K^{ik}K^j_k} [K_{ij}] - [\nabla^2_\Sigma K] - [\partial_n (K' + K^{ab}K_{ab} ) ] .
\ee

\item 
From $g_{nn} \Box R$, the singular term reads 
$$
2[K]^2 \int dn\,\,\, (\Theta')^2,
$$
whereas the finite term reads
\be
-\left[    
2\nabla^2_\Sigma K + 2K'' + 4KK' + K^3 + KK^{ab}K_{ab} + (K^{ab} K_{ab} )'
\right] 
+[K] (\myov{K^2} + \myov{K^{ab}K_{ab}} )+ 4 \overline{K}' [K].
\ee

\item 
From $g_{nn} R^2$, the singular term reads
$$
4[K]^2 \int dn \,\,\, (\Theta')^2,
$$
whereas the finite term reads
\be
-4[K] \left( 
\hat{R} - \myov{K^2} - \myov{K^{ab}K_{ab}}
\right)
+ 8 \overline{K'}[K].
\ee

\item
From $g_{nn} R_{\mu \nu} R^{\mu \nu}$, the singular term reads
$$
\left(   
[K^{ab}] [K_{ab}] + [K]^2
\right) \int dn\,\,\, (\Theta')^2,
$$
whereas the finite term reads
\be
2 \myov{K K^{ab}} [K_{ab}] - 2 \hat{R}^{ab} [K_{ab}] + \overline{K^{'ab}} [K_{ab}] + \overline{K'_{ab}}[K^{ab}]
+ 2 \overline{K'} [K] + 2 \myov{K^{ab}K_{ab}} [K].
\ee

\item 
From $g_{nn} R_{\alpha \beta \mu \nu} R^{\alpha \beta \mu \nu}$, we have the singular term
$$
4 [K_{ab}][K^{ab}] \int dn\,\,\, (\Theta')^2,
$$
whereas the finite term reads 
\be
8 \myov{K^\mu_\alpha K^{\alpha \nu}} [K_{\mu \nu}] + 4 \overline{K^{'\mu \nu}}
[K_{\mu \nu}]+ 4 \overline{K'_{\mu \nu}} [K^{\mu \nu}].
\ee

\end{itemize}

In the Gauss-Bonnet limit, we find that all terms from
integrating $\tilde{G}_{nn}$ across $\Sigma$,
both singular and finite, vanish identically.


\subsection{Regularity constraints}

We now examine the conditions under which the integration is
well-defined by summing up all singular terms arising from the
integration in all the field equations. As shown above, in this family
of theories, every such term is associated with the divergent term
\be
\lim_{b\rightarrow 0} \lim_{\epsilon \rightarrow 0} \int^\epsilon_{-\epsilon} dn\,\,
(\partial_n \Theta (n,b) )^2 =
\lim_{b\rightarrow 0} \frac{1}{b} \int^\infty_{-\infty}  dX F^2(X).
\ee
In the following, we gather the coefficients from each set of
components of the field equations 
which are various functions of jumps of the extrinsic curvature.

From $\tilde{G}_{ij}$, the various singular terms sum up to read
\bea
I^{(sing)}_{ij} &=& \left( 4\beta_3 - \beta_2 - 8\beta_1 \right) [K][K_{ij}] -2 \left( \beta_2 + 
4 \beta_3 \right)[K^l_i][K_{lj}] \cr
\label{constraintsR}
&& + g_{ij} \left(
(2\beta_1 + \frac{1}{2} \beta_2 ) [K]^2 -
( \frac{1}{2} \beta_2 + 2 \beta_3 ) [K^{ab} ][K_{ab}]
\right),
\eea
whereas from $\tilde{G}_{nn}$, the various singular terms sum up to be
\be
\label{constraintsnn}
I^{(sing)}_{nn} = \frac{3}{2} \left( [K]^2 (4\beta_1 + \beta_2) + [K^{ab}][K_{ab}] (\beta_2 + 4 \beta_3 ) \right).
\ee
Since there are no other singular terms from $\tilde{G}_{in} = 0$, we proceed to set both
\eqref{constraintsR} and \eqref{constraintsnn} to vanish. From \eqref{constraintsR}, we can take its trace 
to obtain 
\be
\left(  4 \beta_3 + (2d-8)\beta_1 + (\frac{d}{2}-1 ) \beta_2 \right)[K]^2
- 
(\beta_2 + 4 \beta_3) (2 + \frac{d}{2} ) [K^{ab}][K_{ab}] = 0.
\ee
We can classify solutions to $I^{(sing)}_{ij} = I^{(sing)}_{nn} =0$ as follows:
\begin{enumerate}[(I)]
\item
We first search for 
points in the moduli space such that there is no additional constraints to
be imposed on 
the extrinsic curvature.  From \eqref{constraintsR}, we can read off
\be
\beta_2 + 4\beta_3 = 0, \qquad 4\beta_3 - \beta_2 - 8 \beta_1 = 0,
\ee
which also solves \eqref{constraintsnn} and leads to 
\be
\label{GBline}
\beta_2 = -4\beta_3 = -4\beta_1.
\ee
This is precisly the combination for the Gauss-Bonnet theory! Recall that the
Einstein-Gauss-Bonnet theory is defined with the following addition
\be
\alpha \left( R^2 - 4 R_{\alpha \beta} R^{\alpha \beta} + R^{\alpha \beta \mu \nu}
R_{\alpha \beta \mu \nu} \right),
\ee
to the Einstein-Hilbert action, with $\alpha$ being some constant
parameter (that could be $\sim l^2_s$ where $l_s$ is the string length
if there is a string-theoretic origin \cite{Myers}).
In \cite{Reina},
the authors imposed the smoothness condition that terms naively
containing product of delta-functions should be forbidden in the
action from the outset. This implies that the extrinsic curvature has
to be continuous and it was argued that the junction conditions are
identical to GR for the theory with an added Gauss-Bonnet term in the
action. Further, since $[K_{ab}]=0$, no thin-shell singular sources
should be permitted. However, we find that in the related past
literature
\footnote{For example, in \cite{Grumiller}, a similar result
  was obtained through a similar derivation formulated in terms of
  differential forms and in the Gaussian chart. Using a specific
  example (4D cosmological brane in a 5D spacetime with negative
  cosmological constant) where a $\mathbb{Z}_2$ symmetry was further
  imposed, they showed how this junction condition can be equivalently
  derived by integrating over $\Sigma$ starting from the bulk
  equations and assuming a delta-singular source (eqn. (28) of
  \cite{Grumiller}).
In \cite{Dolezel}, a similar derivation was made,
  and the authors showed how in the Gaussian chart, we can simply use
  the Gauss-Codazzi relations to read off the extra junction condition
  term in \eqref{GB}
  (term in $\beta_1$) starting from the bulk field
  equations. In equation
  (B8) of \cite{Dolezel},
  one can find a formula for this extra
  terms in Gaussian coordinates and we have checked that it is
  equivalent to its expression in differential forms as defined in
  eqn. (12) of \cite{Grumiller}.
},
a consistent junction condition for
Einstein-Gauss-Bonnet theory has been presented in a few papers (that
is different from the above conclusion). In the next Section, we will
show that our results recover those of \cite{Myers, Davis, Gravanis, Dolezel}
in the topological limit. This serves as a stringent
consistency check for many of our equations, and 
demonstrates definitively that the Gauss-Bonnet theory
does have non-trivial junction conditions.

\item
Since imposing $[K_{ij}]=0$ naturally removes all singular terms, we look for less stringent conditions on the extrinsic curvature.
From the form of \eqref{constraintsR} and \eqref{constraintsnn}, we find the following class of solutions
\be
\label{traceless}
[K]=0, \qquad \beta_2 + 4 \beta_3 = 0,
\ee
corresponding to a family of theories for which we can set just the trace of the extrinsic curvature 
instead of all its components to vanish.  
In the space of the couplings $(\beta_1, \beta_2, \beta_3)$, this is a plane containing the 
Gauss-Bonnet `line' (\eqref{GBline}).

\item
Finally, we have the trivial solution 
\be
[K_{ij}] = 0,
\ee
which is accompanied by no other constraints on the coupling parameters. There are still non-trivial junction equations to write down even in this case, since as we have seen earlier, in general, these equations sometimes involve normal derivatives of the extrinsic curvature, i.e. while the first (normal) derivative of the metric has to be continuous, the higher-order ones need not be.

\end{enumerate}

We have focussed on the case where $\Sigma$ is timelike for definiteness. Nonetheless, this derivation can be repeated in an identical fashion for a spacelike $\Sigma$ since the difference lies in a few signs
to be switched on in the Gauss-Codazzi relations. In particular we find that the regularity constraints for the spacelike case remain the same, and the classification of junction conditions presented above is also preserved. Explicitly, restoring $\xi$ (the sign of $n^2$) we find
that \eqref{constraintsR} and \eqref{constraintsnn} generalize to 
\bea
I^{(sing)}_{ij} &=& \left( 4\beta_3 - \beta_2 - 8\beta_1 \right) [K][K_{ij}] -2 \left( \beta_2 + 4 \beta_3 \right)[K^l_i][K_{lj}] \cr
\label{constraintsR2}
&& + g_{ij} \left(
(2\beta_1 + \frac{1}{2} \beta_2 ) [K]^2 - \left( \frac{1}{2} \beta_2 + 2 \beta_3 \right) [K^{ab} ][K_{ab}]
\right)\cr
&&+ ( \xi -1 )(4 \beta_1 + \beta_2 ) [K] \left( [K] g_{ij} - [K_{ij}] \right).
\eea
There are no singular terms identically from $\tilde{G}_{in}$, 
whereas for $\tilde{G}_{nn}$, the various singular terms sum up to be
\bea
\label{constraintsnn2}
I^{(sing)}_{nn} &=& \frac{3\xi}{2} \left( [K]^2 (4\beta_1 + \beta_2) + [K^{ab}][K_{ab}] (\beta_2 + 4 \beta_3 ) \right) \cr
&&+ (1-\xi) [K]^2 \left( 
 2\beta_1 + \frac{3}{2} \beta_2 + 4 \beta_3
\right).
\eea
It is then straightforward to show that setting $I^{(sing)}_{ij}$ and $I^{(sing)}_{nn}$ to vanish
gives the same classification of junction conditions as described earlier.

In the following section, we simplify and present the final form of the junction conditions for Class II and III
theories with their respective regularity constraints implied. It is worthwhile to note that 
the finite terms multiplied to the representation-dependent
factor $\int^\infty_{-\infty} dX\, \Theta X F^2(X)$ sum up to zero in the junction equations
since from integrating $\tilde{G}_{ij}$ across $\Sigma$, we find that these terms assemble to read
\be
\label{repd}
2\left((4\beta_1 + \beta_2)[K]^2 - (\beta_2 + 4 \beta_3) [K^{ab}][K_{ab}]\right)[K_{ij}]\times 
\int^\infty_{0} dX\,  X F^2(X),
\ee
and one can check that for each of the above classes of theories, upon imposing their respective smoothness conditions, \eqref{repd} vanishes exactly. 
The final junction equations
are thus insensitive to the choice of the nascent delta function for all three families of theories.


\subsection{A summary of results}

Having solved for the appropriate regularity constraints, we can now
impose them on the set of general finite junction terms derived
earlier, and obtain the generalized Darmois-Israel junction
conditions. For a generic choice of coupling parameters the resulting
junction equations can be rather elaborate. In the following, we
present the explicit junction equations for each of the three classes
of theories defined in the previous section, including the ordinary
Darmois-Israel junction terms. On the RHS of the each equation, the
boundary energy-momentum tensor is defined as the singular part of the
bulk energy-momentum tensor localized within $\Sigma$, denoted by
$S_{\alpha \beta} = \lim_{\epsilon \rightarrow 0}
\int^\epsilon_{-\epsilon} T_{\alpha \beta}$.

For Case (I), we find that our results reduce nicely to those for
Gauss-Bonnet obtained independently by the method of boundary
variation. This topological theory is defined by the line of couplings
$\beta_1 = \beta_3 = -\frac{1}{4}\beta_2$ and we find that the
junction conditions simplify to read
\be
\label{GBexplicit}
-2 \beta_1 \left[  
  3J_{ij}
 -  J h_{ij}
\right]-4\beta_1 [K^{cd}] 
P_{icdj}
+ [K]h_{ij} - [K_{ij}] = 8 \pi  S_{ij},
\ee
where 
\bea
J_{ij} &\equiv& \frac{1}{3} \left( 2K
K_{ic} K^c_j + K_{cd} K^{cd} K_{ij} - 2 K_{ic} K^{cd} K_{dj} - K^2 K_{ij} \right),
\quad J \equiv J^k_k \cr
P_{icdj} &\equiv & 
\frac{1}{2} \hat{R} g_{i[d} g_{j]c} + \hat{R}_{icdj} + \hat{R}_{c[d} g_{j]i} - \hat{R}_{i [d} g_{j]c}.
\eea
We note that the bracketed rank-4 curvature tensor in the second line
turns out to be the divergence-free component of the intrinsic Riemann
tensor. There are no junction terms arising from integrating
$\tilde{G}_{in},
\tilde{G}_{nn}$. We
also note
that \eqref{GBexplicit} is equivalent to the junction condition
derived in earlier literature \cite{Myers, Davis, Gravanis, Dolezel}
by taking the boundary variation of the Gauss-Bonnet surface
term. Thus, this gives a strong consistency check of our general
derivation which is noticeably absent in other previous proposals for
junction conditions in quadratic gravity \cite{Reina,Berezin}.  Since
this is an important point, in Appendix \ref{GBsec}, we present a
detailed proof of how our equations reduce to \eqref{GBexplicit}.

We should mention that a previous work in \cite{Reina}, which proposes junction conditions for quadratic gravity, fails this consistency 
test. The authors argued one has to take the extrinsic curvature to be continuous at $\Sigma$ even for such the Gauss-Bonnet case, and its junction conditions should read simply as $ [K_{ij}]=0$. This
contradicts the result obtained by either the boundary variation of the Gauss-Bonnet surface term, or integrating the equations of motion across $\Sigma$. In contrast, our derivation elucidates how various junction terms involving
both averaged and jump quantities assemble nicely in the Gauss-Bonnet limit to yield \eqref{GBexplicit}, consistent
with what we expect from the action principle.

For Case (II) where we take $[K]=0$, parametrizing this class of theories by $\{ \beta_1, \beta_3 \}$, we have 
the junction equations
\bea
\label{CaseIIa}
&&2(\beta_1 - \beta_3) \Bigg( -g_{ij} 
\left[ 
2K'' + 4K K' + K K^{ab}K_{ab} + n^\alpha \nabla_\alpha ( \text{Tr} (K^2) )
\right] + 4 \myov{K_{ij} K^{ab} } [K_{ab}] \cr
&& 
- [K_{ij}] \overline{R} +\overline{K_{ij}} [ K^{ab}K_{ab} + 6 K' ]
\Bigg)  -2\beta_3 \Bigg[ J_{ij} - \frac{1}{3} \left(  
\left( 3K K_{cd} K^{cd} - 2 K^{cd}K_{ac} K_d^a \right) h_{ij}
\right) \cr
&&+
2 K^{cd} \left(   
\frac{1}{2} \hat{R} g_{i[d} g_{j]c} + \hat{R}_{icdj} + \hat{R}_{c[d} g_{j]i} - \hat{R}_{i [d} g_{j]c}
\right)  - K_{ij} \Bigg]= 8 \pi  S_{ij},
\eea
\be
\label{CaseIIb}
2(\beta_1 - \beta_3) h^\alpha_i  \nabla_\alpha [2K' + K^{ab}K_{ab}] = 8 \pi S_{in},
\ee
\be
\label{CaseIIc}
-2(\beta_1 - \beta_3) \left[  
2KK' + K K^{ab} K_{ab}
\right] = 8 \pi  S_{nn},
\ee
where $K' = n^\alpha \nabla_\alpha K, K'' = n^\alpha n^\beta \nabla_\alpha \nabla_\beta K$, restoring covariance in notation. 
In the $\beta_1 \rightarrow  \beta_3$ limit, we recover the Gauss-Bonnet junction equations
with the additional constraint $[K]=0$. In the $\beta_3=0$ limit, we recover the simplest
example of $\mathcal{F}(R)$-type gravitational theories of which Lagrangian is an analytic function of 
the Ricci scalar $R$.

For Case (III) where $[K_{ij}] =0$, we have the set of equations 
\bea
-h_{ij} (4\beta_1 + \beta_2) \left( [K''] + 2K[K'] + \frac{1}{2} [ (\text{Tr}(K^2) )' ] \right) &
+ &4 (3 \beta_1 + \beta_2 + \beta_3) K_{ij} [K'] \cr
+(\beta_2 + 4 \beta_3) \left( K^l_{(i}[K'_{j)l}] - K[K'_{ij}]\right) & =& 8\pi S_{ij}, \cr
4(\beta_1 + \frac{1}{2}\beta_2 + \beta_3) h^\alpha_i \nabla_\alpha [K'] + (\beta_2 + 4 \beta_3) n^\alpha [\nabla_\alpha R_{in}] &= &
8\pi  S_{in}, \\
-(4\beta_1 + \beta_2) [KK'] - \frac{1}{2} (\beta_2 + 4 \beta_3) [n^\alpha \nabla_\alpha(\text{Tr}(K^2))] &=& 
8\pi S_{nn}.
\eea
 As we see here, away from the Gauss-Bonnet limit, it could have non-vanishing components in orthogonal directions. Note that
we expect all terms to vanish in the Gauss-Bonnet limit in which setting $[K_{ij}]$ to vanish implies the absence of junction conditions.


\section{Junction conditions for other examples of higher-derivative
 theories}
\label{OtherSec}

\subsection{$\mathcal{F}(R)$ theories}
A class of gravitational theories considered in literature on modified gravity are $\mathcal{F}(R)$ theories which refer to a Lagrangian that is an analytic function of the Ricci scalar. The equations of motion read
\be
\partial_R \mathcal{F}(R) R_{\mu \nu} - \frac{1}{2} \mathcal{F}(R) g_{\mu \nu} + \left( 
g_{\mu \nu} \Box - \nabla_\mu \nabla_\nu
\right) \partial_R \mathcal{F}(R) = 8 \pi  T_{\mu \nu}.
\ee

\subsubsection{More about junction conditions of  $R^2$ theory}
\label{411}
In the previous section, we covered a simple example of such theories: the $\mathcal{F} (R) = R + \beta_1 R^2$ Lagrangian which belongs
to the class of theories with quadratic curvature invariants that admit the regularity condition $[K]=0$. 
Setting $\beta_3  = 0$ in equations \eqref{CaseIIa}---\eqref{CaseIIc}, we obtain the junction conditions to be
\bea
\label{R21}
&&-[K_{ij}] +2 \beta_1  \bigg( -g_{ij} 
\left[ 
2K'' + 4K K' + K K^{ab}K_{ab} + ( \text{Tr} (K^2) )'
\right] \cr
&&\qquad + 4 \myov{K_{ij} K^{ab} } [K_{ab}] - [K_{ij}] \overline{R} +\overline{K_{ij}} [ K^{ab}K_{ab} + 6 K' ]
\bigg) 
= 8 \pi S_{ij}, \\
\label{R22}
&&2 \beta_1 \nabla_i [2K' + K^{ab}K_{ab}] = 8 \pi  S_{in},\\
\label{R23}
&&
-2 \beta_1  \left[  
2KK' + K K^{ab} K_{ab}
\right] = 8 \pi S_{nn}.
\eea
Before we briefly comment on some typical features of the junction conditions for a general $\mathcal{F}(R)$
theory, let us review
in detail
some aspects of the junction equations in the $R^2$ theory which would serve to
highlight certain useful points. The equations of motion for the $R^2$ theory reads
\be
\label{EOMR2}
G_{\mu \nu} + \beta_1 \left(
2R R_{\mu \nu} - \frac{1}{2}R^2 g_{\mu \nu} + 2(g_{\mu \nu} \Box - \nabla_\mu \nabla_\nu )
R \right)= 8 \pi T_{\mu \nu}.
\ee
Recall that $R = \hat{R} - 2 \partial_n K - K^2 - K^{ab}K_{ab}$. Setting $[K]=0$ thus ensures
that $R$ is non-singular. From inspection, there is no singular term arising from integrating 
\eqref{EOMR2} across $\Sigma$. The $R^2 g_{\mu \nu}$ term contains no delta-singularity and hence
it integrates to zero. Consider the term $2 g_{\mu \nu} \Box R$, and take
the indices to be parallel to $\Sigma$. The terms which could be non-vanishing after integrating are
$g_{ij} \partial^2_n R + g_{ij} g^{kl} \Gamma^n_{kl} \partial_n R = 
g_{ij} \partial^2_n R + g_{ij} K \partial_n R = \partial_n (g_{ij} \partial_n R)- \partial_n g_{ij} \partial_n R +
g_{ij} K \partial_n R$. Integrating $ \partial_n (g_{ij} \partial_n R)$ across $\Sigma$ yields
$[g_{ij} \partial_n R ]$. The other two terms require some work. As we discussed in Section 3, we can proceed by expanding all terms using the Gauss-Codazzi relations and then integrating the terms ensuring
that the various discontinuities present in each term and its derivatives are handled appropriately (e.g. using the integral identities in Appendix \ref{AppA}). We find
\bea
\label{rr1}
-\int dn\,\, \partial_n g_{ij} \partial_n R &=& 
2\overline{K_{ij}} \left( 2 [K'] + [K^{ab} K_{ab}] \right) + \frac{1}{3}[K_{ij}][K^{ab}][K_{ab}] = 
-2\overline{K_{ij}} [R] + \frac{1}{3}[K_{ij}][K^{ab}][K_{ab}],
 \cr
\int dn\,\, g_{ij} K \partial_n R &=& -g_{ij} K (2[K'] + [K^{ab} K_{ab}] ) = g_{ij}K[R].
\eea
The remaining terms in \eqref{EOMR2} integrate to yield
\bea
\label{rr2}
\int dn\,\, RR_{ij} &=& 2\overline{K'} [K_{ij}] - \left(\hat{R} - K^2  - \myov{K_{ab} K^{ab}} \right) [K_{ij}]
= - \overline{R} [K_{ij}] - \frac{1}{6} [K^{ab} ][K_{ab}][K_{ij}], \cr 
\int dn\,\, \nabla_i \nabla_j R &=& \overline{K_{ij}} [R] - \frac{1}{6} [K^{ab} ][K_{ab}][K_{ij}].
\eea
Summing up all terms, we find the junction condition
\be
\label{R21p}
-[K_{ij}] + 2\beta_1 \left(
g_{ij} [n^\alpha \partial_\alpha R ] - \overline{R} [K_{ij}] -3 \overline{K_{ij}} [R] + [R]  Kg_{ij} 
+ \frac{1}{3}[K_{ij}][K^{ab}][K_{ab}] 
\right)= 8 \pi  S_{ij},
\ee
which is equivalent to \eqref{R21} but with some expressions succintly expressed in terms of the Ricci scalar. We can also express \eqref{R22} and \eqref{R23} as 
\be
\label{NNNIR2}
2\beta_1 \nabla_i [R] = - 8 \pi S_{in}, \qquad
2\beta_1 K [R] = 8 \pi  S_{nn}.
\ee
Incidentally, in \cite{SenoFR} and \cite{Sasaki}, the authors attempted to study the junction conditions for
$\mathcal{F} (R)$ gravity and obtained different results in general. In \cite{Sasaki}, the junction conditions were
derived under the assumption of a continuous Ricci scalar. 

To facilitate comparison of our results against those of \cite{SenoFR,Sasaki}, 
consider the case where we impose $R$ to be continuous at $\Sigma$. 
This renders the junction conditions in directions not parallel to $\Sigma$ trivial,
whereas \eqref{R21} or \eqref{R21p} reduce to 
\be
\label{R21pp}
2\beta_1 \left(
g_{ij} [n^\alpha \partial_\alpha R ] - R [K_{ij}] 
+ \frac{1}{3}[K_{ij}][K^{ab}][K_{ab}] 
\right)= 8 \pi  S_{ij}.
\ee
The first two terms on the LHS of \eqref{R21pp} are naively expected from integrating
$2 g_{ij} \partial^2_n R$ and $2RR_{ij}$ respectively after using Gauss-Codazzi relations, and were
also obtained in \cite{SenoFR} (see Appendix, eqn. (A17)) and \cite{Sasaki} (see eqn. (3.11)). 
But the third term requires a more careful evaluation of the integrals as we have pinpointed 
in each of the integrals in \eqref{rr1} and \eqref{rr2}. 

We note that the appearance of the `anomalous' term
$[K_{ij}][K^{ab}][K_{ab}] $ in \eqref{rr1} and \eqref{rr2} has been
missed in \cite{SenoFR,Sasaki}. In \cite{Sasaki}, the authors also
employed the technique of integrating the equations of motion across
$\Sigma$ but they missed this term due to essentially an incorrect
ansatz for the Ricci scalar as we shall shortly elaborate upon. In
\cite{SenoFR}, a different method for deriving the junction conditions
was proposed which, as mentioned earlier, appears to be afflicted with
not having a consistent Gauss-Bonnet limit when it is applied to
quadratic gravity.


A common feature that characterizes both \cite{SenoFR} and
\cite{Sasaki} is an improper treatment of $\Theta^2 (n)$ term that is
present in the Ricci scalar when the extrinsic curvature is
discontinuous.  
As a concrete illustration of its consequence on junction terms, let
us revisit \eqref{rr2}, this time sparing more details. Acting on the
Ricci scalar with covariant derivatives, we have
\be \label{ddR}
\nabla_i \nabla_j
R = K_{ij} \partial_n R + \partial_i \partial_j R - \Gamma^k_{ij}
\partial_k R.
\ee
Recall \eq{T2one} that
\be
R = R^+ \Theta + R^- (1 - \Theta) - [K_{ab}][ K^{ab}] (\Theta^2 -
\Theta),
\ee
where $\Theta (n) = \frac{1}{2} + \Theta_X(n)$ is the step function adopted in
\cite{Mars,SenoFR}.
This implies 
only the first term on the RHS of \eq{ddR} is singular with
\be
\partial_n R = [R] \delta - [K_{ab}][K^{ab}] (2\Theta -1) \delta +
\ldots.
\ee
Integrating across the surface, we have
\be
\label{T2tw}
\int dn \,\,\, \nabla_i \nabla_j R = \int dn \,\, K_{ij} [R] \delta -
\int dn\,\, [K_{ab}] [K^{ab}]K_{ij} (2\Theta - 1)\delta.
\ee
Substituting
\be
K_{ij} = K^-_{ij} + [K_{ij}] \Theta, \,\,\,\,\, \int
dn\,\, (2\Theta -1) \delta = 0, \,\,\,\,\, \int dn \,\, \Theta
(2\Theta -1 ) \delta = \frac{1}{6},
\ee
into \eqref{T2tw}, we finally have
\be
\label{T2th}
\int dn \,\,\, \nabla_i \nabla_j R = \overline{K_{ij}} [R] -
\frac{1}{6} [K_{ij}] [K_{ab}][K^{ab}],
\ee
which is of course
\eqref{rr2}. It should be clear from this explicit computation that
one will miss the term $- \frac{1}{6} [K_{ij}] [K_{ab}][K^{ab}]$ if we
have taken $\Theta^2 - \Theta = 0$ right from the outset
erroneously. This error would then lead to the absence of the second
term in \eqref{T2tw} and only the first term would then remain in
\eqref{T2th}, as was the case in \cite{SenoFR,Sasaki}.  Fundamentally,
our use of nascent delta functions enables us to compute the integral
of products of the step function, delta function and their derivatives
in a well-defined and unambiguous manner, as we discussed in a more
generic fashion earlier in Section \ref{GdRC}.

\subsubsection{Junction conditions for $R^3$ theory}

In the following, we derive the junction conditions explicitly for the gravitational theory with additional 
$\beta R^3$ term in the action with the equation of motion 
\be
\label{R3EOM}
G_{\mu \nu}+ \beta\left(   3R^2 R_{\mu \nu} - \frac{1}{2}R^3 g_{\mu \nu} + 3(g_{\mu \nu} \Box - \nabla_\mu \nabla_\nu ) R^2 \right)
= 8 \pi  T_{\mu \nu},
\ee
where
$g_{\mu \nu} \Box - \nabla_\mu \nabla_\nu  
= g_{\mu \nu} \left( \partial^2_n + K \partial_n + D^2 \right) - \left( K_{\mu \nu}
\partial_n +
D_\mu D_\nu \right)$, 
and 
$D_\alpha$ is the covariant derivative associated with $\Sigma$.  We
impose the constraint that $R$ be non-singular, thus taking $$[K]=0,$$ 
 as our regularity condition. From inspection, since the highest
order $n$-derivative is a $\partial^2_n$ acting on $R$ which would
yield at most a delta-like singularity, this suffices to be our only
regularity condition.

This time, since we only have the Ricci scalar and no other curvature
invariants, we express the Ricci scalar and its derivative explicitly
in terms of the nascent delta functions, their derivatives and
functions of the extrinsic curvature right from the outset.  We first
note that we can express the Ricci scalar and its derivative as
\bea
\label{Rgc}
R &=& \left( \hat{R} - 2 \overline{K'} - \overline{K}^2 - \overline{K^{ab}}
\;
\overline{K_{ab}} \right) + 
\Theta_X ( -2[K'] - 2[K]\overline{K} - 2[K^{ab}]\overline{K_{ab}} ) \cr
&&\;\; + 
\Theta^2_X (-[K]^2 - [K^{ab}][K_{ab}] ) - 2\Theta' [K]   \cr
&\equiv& R_{(1)} +
\Theta_X [R] + \Theta^2_X R_{(2)}- 2\Theta' [K]  ,\;\; \\
\label{RgcDer}
\partial_n R &=& \Theta' [2K'-R] - 2 \Theta'' [K] + 2\Theta_X \Theta'  R_{(2)}
+ \left( \ldots \right),
\eea
where the ellipses represent finite quantities which vanish in the
double scaling limit when we integrate \eqref{R3EOM} across
$\Sigma$. We note that in \eqref{R3EOM}, the terms with the covariant
derivative intrinsic to $\Sigma$ do not generate non-vanishing terms
in the junction equation, and we can evaluate the normal derivatives
by the usual chain rule $\partial_n f = R' \partial_R f$, etc. Another
pertinent point is that while we impose $[K]=0$ as our
regularity condition, we should keep factors of $[K]$ in all our
expressions first before evaluating the integral since as a function
of $n$, $[K] \neq 0$ and various terms in \eqref{R3EOM} are
generalized distributions which may probe the non-vanishing
derivatives of $[K]$ at $n=0$ in the integral.

We begin by considering the indices to be those of $\Sigma$, with
$\{\mu ,\nu\} = \{i,j\}$.  Using $R_{ij} = - \partial_n K_{ij} +
2K^a_j K_{ai} - K K_{ij} + \hat{R}_{ij}$, and substituting \eqref{Rgc}
and \eqref{RgcDer} into \eqref{R3EOM}, we organize all terms according
to the order of singularity defined by various powers and derivatives
of $\Theta'$. In \eqref{R3EOM} only $R^2 R_{\mu \nu}$
and $ \left( g_{\mu \nu} \Box - \nabla_\mu \nabla_\nu \right) R^2$
terms give junction terms, which we collate below. In the following,
all curvature quantities are understood to be evaluated at $n=0$.

\begin{enumerate}[(i)]

\item \underline{From the $3R^2 R_{ij}$ term:}

\begin{itemize}
\item $\Theta'$: 
\be
- 3 \int dn\,\, \Theta' [K_{ij}] \left(  
R^2_{(1)} + \Theta^2_X ([R]^2+ 2 R_{(1)} R_{(2)}) + \Theta^4_X R^2_{(2)} 
\right),
\ee

\item $(\Theta' )^2$: 
\be
 \int dn\,\, n \Theta'^2 \Theta_X (12[K'][R][K_{ij}]),
\ee

\item $(\Theta' )^3$:  
\be
 \int dn\,\, n^2 \Theta'^3 (-12[K_{ij}] [K']^2).
\ee
\end{itemize}

\item \underline{From the $3\left( g_{ij}
  \Box - \nabla_i \nabla_j \right) R^2$ term:}

We first note that after integration by parts,
\be
\int dn\,\,\,3\left( g_{ij} \Box - \nabla_i \nabla_j \right) R^2
=6 [g_{ij} R \partial_n R] + 6 \left(
\int dn\,\, \left( g_{ij} K - 3 K_{ij} \right) R \partial_n R 
\right).
\ee
For a more compact notation at this point, we introduce $c_1= g_{ij} K - 3
\overline{K_{ij}}$.
Organizing all terms according to their order of singularity, we have 

\begin{itemize}
\item $\Theta'$: 
\bea
&&6\int dn\,\,\,  \Theta' \Bigg(    ([R]-2[K'])c_1 R_{(1)} \cr
&&+ 
\Theta^2_X (  ([R]-2[K'])(c_1R_{(2)} + c_2 [R] ) + 2 R_{(2)} ( c_1[R] + c_2 R_{(1)} ) )
+2\Theta^4 c_2 R^2_{(2)}
\Bigg), \nonumber
\eea

\item $(\Theta' )^2$: 
\be
6\int dn \,\,\, n( \Theta')^2 \Theta_X 
\left(
(-2[K'])(-3[K_{ij}]([R]-2[K']) + 2c_1 R_{(2)} )
\right),
\ee

\item $\Theta'' $: 
\be
6\int dn\,\, \Theta'' n (-2[K']) \left( c_1 R_{(1)} + \Theta^2_X (c_1 R_{(2)}
- 3[K_{ij}][R])  \right),
\ee

\item $\Theta' \Theta'' $: 
\be
6 \int dn\,\, n^2 \Theta_X \Theta' \Theta'' (-12) [K_{ij}] [K']^2.
\ee

\end{itemize}

\end{enumerate}

The junction equation is then the sum of all above terms. We note that there
are representation-dependent integrals which are not all independent as they
satisfy the following two constraints.
\bea
&&\int dn\,\, \left(  
(\Theta' )^3n^2 + 2 \Theta'' \Theta' \Theta_X n^2 + 2\Theta'^2 \Theta_X n 
\right) = 0, \cr
&& \int dn \,\, \left(
\Theta'' \Theta^2_X n + 2\Theta'^2\Theta_X n
\right) = -\frac{1}{12}, \nonumber
\eea
which straightforwardly follows by integrating $\partial_n ( \Theta'^2 \Theta_X n^2 ),
\partial_n (\Theta^2_X \Theta' n )$. After some simplification, we find that the
representation-dependent terms remain and they sum up to be 
\be
\mathcal{T}_{rep} \equiv 24 [K'][K_{ij}] \left([K_{ab} K^{ab}] (\int dn\,\, \Theta'^2 \Theta_X n ) -2[K'] 
(\int dn\,\, \Theta'' \Theta' \Theta_X n^2)
 \right).
\ee
Gathering all terms, we find the junction equation in the $ij$-directions to be 
\bea
\label{GijR3a}
8 \pi S_{ij} &=& 
\beta \Bigg(6[g_{ij} R R']-3 R^2_{(1)} [K_{ij}] - \frac{7}{4}[K_{ij}] \left(   2 R_{(1)} R_{(2)} + [R]^2 \right) 
- \frac{39}{80}[K_{ij}] R^2_{(2)}  \cr
&+& 6c_1[R] ( R_{(1)} + \frac{1}{4}  R_{(2)} ) 
+ \mathcal{T}_{rep} \Bigg) - [K_{ij}].
\eea
We can also express \eqref{GijR3a} in terms of the mean 
$\overline{R} = \hat{R} - 2 \overline{K'} - K^2 - \overline{K^{ab}K_{ab} }$,
and difference of the Ricci scalar $[R]$
across $\Sigma$, using the identity
$$
\overline{K^{ab}K_{ab} } - \overline{K^{ab}} \overline{K_{ab}} = \frac{1}{4}[K^{ab}][K_{ab}] \equiv -\frac{1}{4} R_{(2)}. 
$$
When $[K]=0$, we have $\overline{R} = R_{(1)} + \frac{1}{4} R_{(2)}$.
Then, in terms of $\overline{R}$, \eqref{GijR3a} can also be written as 
\bea
\label{GijR3b}
8 \pi  S_{ij} &=& 
\beta \left( 6[g_{ij} R R']-3 \overline{R}^2 [K_{ij}] + [K_{ij}]R_{(2)}
(-2\overline{R}+ \frac{1}{5}R_{(2)} )  
+ 6c_1[R] \overline{R}
-\frac{7}{4}[K_{ij}][R]^2
+ \mathcal{T}_{rep} 
\right)- [K_{ij}], \cr
c_1&=& g_{ij} K - 3 \overline{K}_{ij}, \qquad R_{(2)} = - [K^{ab}][K_{ab}].
\eea
The first two terms on the RHS would be what we naively expect from
integrating $3g_{ij} \partial^2_n R^2 + 3R^2 R_{ij}$, whereas the remaining terms
arise from a more careful evaluation.

For the other directions not completely parallel to $\Sigma$, we need to integrate
\bea
\label{GnnR3}
&&\partial_R \mathcal{F} R_{nn} - \frac{1}{2}\mathcal{F} + (\Box - \nabla_n
\nabla_n )\partial_R \mathcal{F}, \\
\label{GinR3}
&& \partial_R \mathcal{F} R_{in} - \nabla_i \nabla_n \partial_R \mathcal{F},
\eea
across $\Sigma$. 
For \eqref{GinR3}, we can integrate straightforwardly and obtain an explicit
expression
for a general $\mathcal{F}(R)$. Since $R_{in} = D^k K_{ik} - D_i K$ is non-singular
and so is $\partial_R \mathcal{F}$,
the only term that could contain a delta-singularity is 
\be
-\nabla_i \nabla_n (\partial_R \mathcal{F}) = \partial_n \left( - \partial_i
\partial_R \mathcal{F}) + K^j_i \partial_j \partial_R \mathcal{F} \right).
\ee
This integrates to $-[\partial_i R \partial^2_R \mathcal{F} ]$, and so the
junction condition reads simply as
\be
\label{NIR3}
8 \pi S_{in} = - [\partial_i R \partial^2_R \mathcal{F} ] =-6\beta [R\partial_i R ] .
\ee
For \eqref{GnnR3}, since $\partial_R \mathcal{F} R_{nn} - \frac{1}{2} \mathcal{F}(R)$ is non-singular,
it vanishes upon integration and we are left with 
\be
g^{kl} \nabla_k \nabla_l \partial_R \mathcal{F} = -g^{kl} \Gamma^n_{kl} R' \partial^2_R \mathcal{F} + \left( \ldots \right)
= K \partial^2_R \mathcal{F} \partial_n R + \left( \ldots \right)
\ee
to consider, with the ellipses representating other terms that would not survive the integral. 
If $R$ is continuous, then $\partial_n R$ is no longer singular, so clearly this junction condition probes
the discontinuity of $R$. 
For the $R^3$ theory, we need to integrate
\bea
&&\int dn\,\,6( \overline{K} + \Theta_X [K] ) \left( R_{(1)} + \Theta_X [R] + \Theta^2_X R_{(2)} - 2 \Theta' [K] \right) \cr
&&\qquad \times \left(
(R'_{(1)} + \Theta_X [R'] + \Theta^2_X R'_{(2)} ) + 
\Theta' ([R] - 2[K'] + 2 \Theta_X R_{(2)} ) 
-2\Theta''[K]
\right).
\eea
Only terms containing derivatives of $\Theta'$ survive the integration. In the following, we 
organize various contributions as we have done earlier for the junction equation in the $ij$-directions.
\begin{itemize}
\item $\Theta'$:  
$$6([R] -2 [K'] ) K R_{(1)} + \frac{1}{2} \left( 2 R_{(2)} K [R] + 
([R]-2[K'] )K R_{(2)} \right),
$$

\item $(\Theta')^2$: 
$$
-24K [K'] R_{(2)} \int dn\,\, (\Theta')^2 \Theta_X \,n,
$$

\item $\Theta''$: 
$$
-12[K']K R_{(1)} \int dn\, \Theta'' n - 12 K[K']R_{(2)} \int dn\,\, \Theta'' \Theta^2_X n.
$$

\end{itemize}
Note that the term containing $\Theta' \Theta''$ integrates to zero since $[K]=0$. 
After some algebra, and invoking the relation $ \int dn \,\, \left(
\Theta'' \Theta^2_X n + 2\Theta'^2\Theta_X n
\right) = -\frac{1}{12}$, we obtain the junction condition 
\be
\label{NNR3}
8 \pi  S_{nn} = 6 \beta [R]K \overline{R}.
\ee
In summary, for the $\mathcal{F}(R) = R +\beta R^3$ theory, we have the regularity condition $[K]=0$, and the junction conditions
specified by \eqref{GijR3b}, \eqref{NIR3} and \eqref{NNR3}.

\subsubsection{Some general comments}

Having obtained the explicit junction equations for the $R^2$ and $R^3$ theory via a rather intricate
integration procedure, let us briefly comments on some features that we can deduce for the general $\mathcal{F}(R)$ case without alluding to some specific form of $\mathcal{F}$. 

\begin{itemize}

\item \textbf{\underline{$[K]=0$ as the regularity condition}}

We have seen that $[K]=0$ is an appropriate regularity condition for the $R^2$ and $R^3$ theories,
and it is easy to see that it is valid for a general analytic $\mathcal{F}$. If $[K]=0$, then $R$ is non-singular, so 
$\partial_R \mathcal{F} R_{\mu \nu}$ cannot generate divergent terms upon integration, since there is at most a delta-singularity carried by $R_{\mu \nu}$. The term $\mathcal{F} g_{\mu \nu}$ vanishes after integration leaving us with only $(g_{\mu \nu} \Box - \nabla_\mu \nabla_\nu) \partial_R \mathcal{F}$. Consider the normal derivatives --- after an integration by parts, we are left with only $\partial_n$ acting on $\partial_R \mathcal{F}$ which generate
at most a delta-singularity that yields finite quantities after integration. This applies to the junction equations in all directions. Hence, $[K]=0$ is a valid regularity condition generally. 

When we discuss the case of quadratic gravity earlier, we obtained the general equations of which solutions give all possible regularity constraints. For the $R^2$ theory, we found no regularity condition
apart from $[K]=0$ and similarly, one can show that this is the case for the $R^3$ theory as well.

\item \textbf{\underline{On taking $R$ to be continuous: }}

Another constraint that we can impose on top of $[K]=0$ is the continuity of $R$ at $n=0$. This implies
that we take 
\be
\label{reg}
2[K'] + [K^{ab} K_{ab}] = 0, \qquad [K]=0.
\ee
As we observed earlier, the junction equations in the orthogonal directions would be trivial in this limit
leaving only those parallel to $\Sigma$. In the specific cases of $R^2$ and $R^3$ theories, this is evident
in equations \eqref{NNNIR2}, \eqref{NNR3} and \eqref{NIR3}. 
Thus, if we further take $[R]=0$, the discontinuity in the extrinsic curvature
can be physically supported purely by a singular source that only has non-vanishing components parallel to $\Sigma$. For the $R^2$ theory, the junction equation in the case of a continuous $R$ can be found in \eqref{R21pp}
whereas for the $R^3$ theory, we have explicitly
\bea
\label{IJR3}
8 \pi  S_{ij} &=& \beta \Bigg( 6 g_{ij} R [R'] - 3 R^2 [K_{ij}] + [K_{ij}] R_{(2)} (-2R + \frac{1}{5} R_{(2)} ) \cr
&&\,\,\, + 12 [K'][K_{ij}] R_{(2)} \int dn \,\, (\Theta')^3 n^2 \Bigg) - [K_{ij}].
\eea
In both \eqref{R21pp} and \eqref{IJR3}, we see that although the first two terms on the RHS may be naively
expected from integrating $g_{ij} \partial^2_n  \partial_R \mathcal{F}$ and  $(\partial_R \mathcal{F}  )R_{ij}$ respectively, there are non-trivial terms which arise from the intricacies of the integral.

\item \textbf{\underline{On representation-dependent terms and $[K']$}}

In \eqref{GijR3b}, we note the appearance of $\mathcal{T}_{rep}$ which is sensitive to the
choice of the nascent delta function. We note that such terms always come with at least a factor of $[K']$. To see this, from \eqref{Rgc} and \eqref{RgcDer}, we see that the $\Theta', \Theta''$ terms in $R, \partial_nR$ 
are each multiplied to a factor of $[K]$. Now the representation-dependent terms can always be traced
to a product of them. Let $m$ be the total order of derivatives defined as the sum of the order of derivative on each $\Theta_X$ in some product. Any such term is generically representation-dependent,
with the junction term arising from integrating them against the $(m-1)^{\text{th}}$ derivative of the 
coefficient. These derivatives must act only on factors of $[K]$ as otherwise, the term will vanish since we have imposed $[K]=0$ at $n=0$. Thus, we always have $[K']$ as part of the overall coefficient of any
representation-dependent term. Although the actual form of $\mathcal{T}_{rep}$ depends on 
what $\mathcal{F}(R)$ is as a function of $R$, this implies that universally across $\mathcal{F}(R)$-theories, setting
\be
\label{repInd}
[K']=0,
\ee
implies the absence of these representation-dependent terms. We stress however that 
such a constraint is not necessary for the regularity of the junction equations.

\item \textbf{\underline{A restrictive set of regularity constraints}}

We consider a set of regularity constraints at $n=0$ which allows us to explicitly derive the 
appropriate junction equations for a generic analytic $\mathcal{F}(R)$. This amounts to simply taking \emph{all} 
components of 
$R$ to be continuous at $\Sigma$. Since $R = \hat{R} - 2 \partial_nK - K^2 - K^{ab}K_{ab}$, 
this implies that we impose 
\be
\label{regFR}
[K_{ij} ] = 0, \qquad [K' ] =0.
\ee
From \eqref{reg} and \eqref{repInd}, this set of regularity constraints can also be understood as the one that leads to 
an absence of representation-dependent terms and 
a singular energy source that doesn't have components orthogonal to $\Sigma$.
 
Keeping only terms involving normal derivatives, the equations of motion simplify to read
\bea
&&- ( \partial_R \mathcal{F}) \partial_n K_{ij} + g_{ij} \partial_n \left(   
\partial_n R (\partial^2_R \mathcal{F} )\right) 
+ g_{ij} R' K( \partial^2_R \mathcal{F}) + K_{ij} R' (\partial^2_R \mathcal{F}) + \left( \ldots \right) = 8 \pi  T_{ij},
\cr \cr
&& -\partial^2_R \mathcal{F} \partial_n \partial_i R - \partial_n R \, \partial_i R \, \partial^3_R \mathcal{F} -2 (\partial^2_R \mathcal{F}) K^m_i \partial_m \partial_n K + \left( \ldots \right) = 8 \pi T_{in}, \cr \cr
&&-(\partial_R \mathcal{F}) \partial_n K + K \partial_n R \, (\partial^2_R \mathcal{F}) + \left( \ldots \right)
 = 8 \pi  T_{nn}, \qquad  
\eea
where we use the Gaussian normal chart in which $\Gamma^n_{ij} = - K_{ij}, \Gamma^i_{nj} = K^i_j$, 
the Gauss-Codazzi relations and note that the components $\tilde{G}_{ni}=0$ identically. 
Integrating across $\Sigma$ in the double scaling limit then yields the junction conditions
to be
\be
 g_{ij} (\partial^2_R \mathcal{F}) \left(  
2[K''] + (K^{ab} K_{ab} )'
\right) = -8 \pi S_{ij}, 
\ee
with no other junction conditions arising from other components $\tilde{G}_{\alpha \beta}$. 
Restoring covariance, we note that
$[K''] = n^\alpha n^\beta [ \nabla_\alpha \nabla_\beta K]$ and similarly
$(K^{ab} K_{ab} )' = n^\alpha \nabla_\alpha (K^{ab} K_{ab} )$.

Again, we wish to emphasize that the smoothness conditions $[K_{ij}] = 0, [\partial_n K] = 0$ that arise from a continuous Ricci scalar are not the least restrictive ones. As we have seen in the previous 
examples of $R^2, R^3$ theories, one could just impose $[K]=0$, leading to a more complicated set of junction conditions. In particular, 
there are non-trivial ones which generically require the singular source at $\Sigma$ to have 
non-vanishing orthogonal components.

\end{itemize}

\subsection{Low-energy effective action from toroidal compactification of the Heterotic String}

As an illustration of how our method could be applied straightforwardly to the presence of matter couplings in
higher-derivative gravitational theories, we consider a simple example motivated by string phenomenology - the low energy effective action arising from a particular compactification
of a ten-dimensional string theory. This effective action involves two scalar fields coupled to the Riemann tensor in a certain manner. 

To first-order in $\alpha'$ expansion, suppressing all gauge fields for simplicity, the 10D effective action
of the Heterotic Superstring reads (see e.g. Chapter 12 of \cite{Polchinski})
\be
S = \frac{g^2_s}{\kappa^{(10)}} \int d^{10} x \sqrt{|g|} e^{-2\phi}  \left( R  + 
4 \partial^\mu \phi \partial_\mu \phi - \frac{1}{12}H^2 + \frac{\alpha'}{8} R_{\mu \nu \rho \sigma}
R^{\mu \nu\rho \sigma} \right),
\ee
where $\kappa^{(10)} = 16 \pi G^{(10)}$, $g_s$ is the string coupling and the Riemann-squared term 
is required for supersymmetry to first order in $\alpha'$ (essentially this follows from the
Chern-Simons terms in the 3-form field strength). It was shown in \cite{Cano} that upon 
compactification on a $T^6$, up to leading order in the string coupling, the effective 
$4D$ action can be simplified to read 
\be
\label{Het}
S_{eff} = \frac{1}{\kappa} \int d^4x \sqrt{|g|} \left( R - \frac{1}{2} \left(\partial^\mu \phi \partial_\mu \phi 
+ \partial^\mu \varphi \partial_\mu \varphi \right) - \frac{\alpha'}{8}\phi
 \left( R^2 -4 R^{\mu \nu} R_{\mu \nu} + R_{\mu \nu \alpha \beta} R^{\mu \nu \alpha \beta} 
\right)
+ 
\frac{\alpha'}{8}\varphi R_{\mu \nu \rho \sigma} \tilde{R}^{\mu \nu \rho \sigma} \right),
\ee
with $\kappa \sim \kappa^{(10)}/\text{Vol}(T^6)$, $\tilde{R}_{\mu \nu \alpha \beta} = \frac{1}{2} 
\epsilon_{\mu \nu \rho \sigma} R^{\rho \sigma}_{\alpha \beta} R^{\mu \nu \alpha \beta}$,
and $\phi, \varphi$ can be interpreted as
the dilaton and axion fields respectively. We see that the effective action is a sum 
of the Gauss-Bonnet and Chern-Simons terms, each coupled to a scalar field. 
Since our method works for any coefficients of the interaction terms, in the following, 
we derive the junction conditions for the above theory with the factors of $\pm \frac{\alpha'}{8}$
being replaced by $\beta_{2,1}$ respectively which are arbitrary constants in units of $\alpha'$. 

The equations of motion read 
\bea
\label{EOMHet1}
&&R_{\mu \nu} - \frac{1}{2} R g_{\mu \nu} - \beta_1 g_{\nu \lambda} 
\delta^{\lambda \sigma \alpha \beta}_{\mu \rho \gamma \delta} {R^{\gamma \delta}}_{\alpha \beta} 
\nabla^\rho \nabla_\sigma \phi + 2 \beta_2 \nabla^\rho \nabla^\sigma \left( 
\tilde{R}_{\rho (\mu \nu) \sigma } \varphi
\right) +\beta_1 \phi H_{\mu \nu}  \cr
&& \qquad \qquad \qquad + \frac{1}{2} \left(  \partial_\mu \phi \partial_\nu \phi - \frac{1}{2} g_{\mu \nu} (\partial 
\phi )^2 \right) + \frac{1}{2} \left(  \partial_\mu \varphi \partial_\nu \varphi - \frac{1}{2} g_{\mu \nu} (\partial 
\varphi )^2 \right) =  8 \pi  T_{\mu \nu},\\
\label{EOMHet2}
&& \nabla^2 \phi = - \beta_1 \left( 
R^2 -4 R^{\mu \nu} R_{\mu \nu} + R_{\mu \nu \alpha \beta} R^{\mu \nu \alpha \beta} 
\right),\\
\label{EOMHet3}
&&\nabla^2 \varphi = - \beta_2 R_{\mu \nu \rho \sigma} \tilde{R}^{\mu \nu \rho \sigma},
\eea
where $T_{\mu \nu}$ represents the external energy-momentum tensor field,
and $$H_{\mu \nu} = 2 \left(RR_{\mu \nu} - 2 R_{\alpha \mu} R^\alpha_\nu + R_{\mu \alpha \beta \gamma}
{R_\nu}^{\alpha \beta \gamma} - 2 R_{\mu \alpha \nu \beta}R^{\alpha \beta} \right)
-\frac{1}{2} g_{\mu \nu} \left( R^2 - 4 R_{\alpha \beta} R^{\alpha \beta} + R^{\rho \sigma \alpha \beta}
R_{\rho \sigma \alpha \beta}\right)$$ is the second Lovelock tensor. For the term in $\beta_2$, it is useful to invoke Bianchi identities
to rewrite it as 
\be
\beta_2 \nabla^\rho \nabla^\sigma \left( 
\tilde{R}_{\rho (nn) \sigma } \varphi
\right) =  \beta_2 \nabla_k \left(  \partial^m \varphi \, \epsilon_{mefn} {R_n}^{kef} \right).
\ee
before proceeding to derive the junction equations below.

We first examine the singular terms in \eqref{EOMHet1} --- \eqref{EOMHet3}. 
Let us begin with the matter fields' equations of motion. Since there is no term more singular than the delta function in the 
Gauss-Bonnet nor the Chern-Simons term, this implies that integrating 
\eqref{EOMHet2}, \eqref{EOMHet3} across $\Sigma$ will yield a finite quantity. 
In the Gaussian normal chart, $\nabla^2 = \partial^2_n + \nabla^2_{\Sigma}$. 
Hence, we 
have the continuity constraints for the scalar fields:
\be
[\phi] = 0, \qquad [\varphi] = 0,
\ee
since otherwise, the action of $\partial^2_n$ on the fields would lead to a singular integral across $\Sigma$. To preserve as much generality as possible, we do not assume however that their normal derivatives are continuous at $\Sigma$. We can integrate \eqref{EOMHet2} and \eqref{EOMHet3} across $\Sigma$
obtain 
\be
[\phi' ] = -4\hat{R} [K] + 8 \hat{R}^{ab} [K_{ab}] - 4 [ KK^{ab} K_{ab} ] + \frac{4}{3} [K^3] + \frac{8}{3} 
[K^m_l K^{ln} K_{mn}],
\ee 
\be
[\varphi' ] = 8 \beta_2 \epsilon_{ijk} D^j \overline{K^{kb}}  K^i_{b}.
\ee
Let us now study if \eqref{EOMHet1} yields any non-trivial 
regularity constraints. 
Consider first $\tilde{G}_{ij}$, suppressing the Gauss-Bonnet and scalar field terms,
\be
\label{smoo1}
\tilde{G}_{ij} = - \beta_1 g_{j \lambda} \delta^{\lambda \sigma \alpha \beta}_{i \rho \gamma \delta} 
R^{\gamma \delta}_{\alpha \beta} \nabla^\rho \nabla_\sigma \phi +  
\beta_2 \nabla_k ( \partial^m \varphi \epsilon_{mef(j} {R_{i)}}^{kef} ) + \ldots,
\ee
where the ellipses refer to finite terms in \eqref{integration}. 
For a $1/b$-type singularity, we need terms of the form $R^{na}_{nb}$ and $\partial^2_n \phi$ to be present simultaneously or a term that goes as $\nabla_n R^{na}_{nb}$. By observation, this cannot
arise from the dilaton term where for the axion interaction term, consider the term where
we take the indices $k=f=n$ in \eqref{smoo1}. After some algebra, we can simplify this term 
to read $-2\beta_2 \partial^m \varphi \epsilon_{me(j} \partial^2_n K^e_{i)}$. But integrating this term
across $\Sigma$ yields 
$$-{\epsilon^{me}}_j [\partial_m \varphi] [K_{ei}] \lim_{b \rightarrow 0 }\frac{1}{b} \int dX F^2(X) + \ldots$$ 
where we suppress the manifestly finite quantities. Since $\varphi$ and thus $\partial_m \varphi$ is continuous across $\Sigma$, we find no singularity here. Finally, we note that
as shown in the previous section, there is no singular term descending from $H_{\mu \nu}$ (no additional regularity constraints required for pure Einstein-Gauss-Bonnet theory). 
Thus, from inspection,
we can see that there is no singular term from integrating $\tilde{G}_{ij}$ across $\Sigma$. 

Similar arguments apply for the other components of the field equations, explicitly, 
\bea
\label{gnn1}
\tilde{G}_{nn} &=& - \beta_1 g_{n n} \delta^{n \sigma \alpha \beta}_{n \rho \gamma \delta} 
R^{\gamma \delta}_{\alpha \beta} \nabla^\rho \nabla_\sigma \phi +  
2\beta_2 \nabla_k ( \partial^m \varphi \epsilon_{mef(n} {R_{n)}}^{kef} ) + \ldots \\ \cr
\label{gnn2}
\tilde{G}_{in} &=& - \beta_1 g_{ik} \delta^{k \sigma \alpha \beta}_{n \rho \gamma \delta} 
R^{\gamma \delta}_{\alpha \beta} \nabla^\rho \nabla_\sigma \phi +  
\beta_2 \nabla_k ( \partial^m \varphi \epsilon_{mef(i} {R_{n)}}^{kef} ) + \ldots
\eea
By inspection, one can again draw the conclusion that no singular terms remain after integrating
\eqref{gnn1} and \eqref{gnn2}
across $\Sigma$ by checking the absence of terms of the form $\nabla_n R^{na}_{nb}$
or products of $R^{na}_{nb}$ and $\partial^2_n \phi$. Altogether, the above considerations
reveal that just like the pure Gauss-Bonnet theory, there is no additional
regularity constraints that we need to impose here. This fits intuitively well with the fact that 
the interaction terms mixing the otherwise free scalar fields and the graviton happen to be linear in the Gauss-Bonnet and Chern-Simons terms. 

We now proceed to derive the junction equations. From the form of \eqref{gnn1}, we find
that there are no terms which carry delta-like singularities so integrating \eqref{gnn1} across $\Sigma$
cannot give any junction condition. For \eqref{gnn2}, the dilaton interaction term could not give rise
to any junction term since the generalized Kronecker delta symbol already contains an $n$-index, and 
introducing some pair of $n$-indices (for either $R^{na}_{nb}$ or $\partial^2_n \phi$ ) would annihilate the term by symmetry of the symbol. 

For the axion interaction term in \eqref{gnn2} which reads
\be
\label{GinAxion1}
-\beta_2 \left( 
\nabla_a \left(\partial^\beta \varphi \epsilon_{\beta \mu \nu i} {R_n}^{a \mu \nu}\right)
+ \nabla_\alpha \left(  \partial^b \varphi \epsilon_{buvn} {R_i}^{\alpha u v} \right)
\right).
\ee
For the second term in \eqref{GinAxion1}, the only non-vanishing term arises from taking 
the dummy index $\alpha$ to be $n$, generating the following junction term after integration:
$$
\beta_2 \epsilon_{b u v n}\partial^b \varphi 
\left[  {R_i}^{nuv} \right].
$$
For the first term in \eqref{GinAxion1}, it is helpful to first expand all terms that come with the 
covariant derivative. 
\bea
&& \partial^\beta \varphi \epsilon_{\beta \mu \nu i} 
\left( 
 \partial_a {R_n}^{a\mu \nu} + \Gamma^a_{al} {R_n}^{l \mu \nu} + \Gamma^\mu_{ak} {R_n}^{ak\nu} 
+ \Gamma^\nu_{ak} {R_n}^{a\mu k} + \Gamma^\mu_{an} {R_n}^{an \nu} + \Gamma^\nu_{an} {R_n}^{a\mu n}
\right) \cr
&=& 2 \partial^\beta \varphi  \epsilon_{\beta n v i} \left( \partial_a {R_n}^{anv} + \Gamma^a_{al} {R_n}^{lnv} \right) 
+ 2 \epsilon_{nuvi} \varphi' K^u_a {R_n}^{anv} + 2 \epsilon_{buni} \partial^b \varphi \Gamma^u_{ak} {R_n}^{akn} + \left(
\ldots \right), \nonumber \\
\eea
where we have only displayed terms that will survive the integral.
Invoking the Gauss-Codazzi relation
${R^n}_{snv} = -\partial_n K_{sv} + K_{bv} K^b_s
$,
and assembling all terms, we integrate to obtain
the junction condition for $\tilde{G}_{in}$ to be 
\be
\label{GinHet}
8 \pi  S_{in} = \beta_2 \left(
\epsilon_{b u v n}\partial^b \varphi 
\left[  {R_i}^{nuv} \right] + 2 \epsilon_{nmri}
\left( \partial^m \varphi [D_a K^{ar} ] - \myov{\varphi' K^m_a}
[K^{ar}]  \right)
\right).
\ee
We now consider the junction condition for $\tilde{G}_{ij}$. Apart from the
Gauss-Bonnet term (multiplied to $\phi$) which integrates to yield $\phi J^{(GB)}_{ij}$
where $J^{(GB)}_{ij}$ is the junction equation
of the pure Gauss-Bonnet theory (see
\eq{GBexplicit}),
the dilaton and axion interaction terms
contribute to the junction condition as well. 

Let us begin with the 
dilaton interaction term which could contribute to the junction equation with
the following two
terms from \eqref{smoo1}:
\be
\label{Gijdil}
\tilde{G}_{ij} = - \beta_1 g_{jl} \delta^{lnab}_{ingd} R^{gd}_{ab}\nabla^n \nabla_n \phi  
- \beta_1 g_{jl} \delta^{lsnb}_{irnd} R^{nd}_{nb}\nabla^r \nabla_s \phi
+ \left( \ldots \right)
\ee
For the first term, expanding the generalized Kronecker symbol, we obtain
\be
\beta_1 \left(  g_{ji} R^{gd}_{gd} - g_{jg} R^{gd}_{id} + g_{jd} R^{gd}_{ig} - g_{jd} R^{gd}_{gi}
+g_{jg} R^{gd}_{di} - g_{ij} R^{gd}_{dg} \right) \nabla^n \nabla_n \phi = -4 \beta_1 G_{ij} \nabla^n \nabla_n \phi
\ee
Integrating across $\Sigma$ then gives
\be
-4\beta_1 \myov{G_{ij}} [\partial_n \phi ] \equiv 
-4\beta_1 \left(
-\overline{K'_{ij}} + 2 \myov{K^a_j K_{ai} } - \myov{KK_{ij}} + \hat{R}_{ij} - 
\frac{1}{2}g_{ij} \left( \hat{R} - 2 \overline{K'} - \myov{K^2} - \myov{K^{ab}K_{ab}} \right)
\right)[\partial_n \phi ]
\ee
For the second term in \eqref{Gijdil}, we first note that 
\be
g_{jl} \delta^{lab}_{igd} \nabla^g \nabla_a \phi \, K^d_b = 
g_{ji} \nabla^2_\Sigma \phi K - \nabla_j \nabla_i \phi K + \nabla^m \nabla_i \phi K_{jm} - 
\nabla^2_\Sigma \phi K_{ji} + \nabla_j \nabla_d \phi K^d_i - g_{ji} \nabla^g \nabla_d \phi K^d_g
\ee
after expanding the generalized Kronecker symbol. Integrating the second term of \eqref{Gijdil} across
$\Sigma$ then gives, after some algebra,
\be
\beta_1 \left(  
\nabla^2_{\Sigma} \phi ( g_{ij} [K] - [K_{ij}] ) + (D_m D_a \phi + \myov{K_{ma} \phi' }) 
\mathcal{C}^{mars}_{ij} [K_{rs}] 
\right),
\ee
where we have defined 
$$
\mathcal{C}^{mars}_{ij} \equiv \delta^a_i \delta^r_j g^{ms} + \delta^m_j \delta^s_i g^{ar} - \delta^m_j \delta^a_i g^{rs} - g^{ar} g^{ms} g_{ij}.
$$
For the axion interaction term, it is helpful to first write out explicitly three types of terms which
contribute to the junction equation. From the second term of \eqref{smoo1}, we have 
\be
\label{axionij}
\beta_2 \left(  
 2 \nabla_n ( \partial^m \varphi \epsilon_{men(j} {R_{i)}}^{nen} ) + \nabla_n ( \partial^n \varphi \epsilon_{nef(j}{R_{i)}}^{nef} ) + 2 \nabla_k ( \partial^m \varphi \epsilon_{mnf(j} {R_{i)}}^{knf} )
\right) + \left( \ldots \right)
\ee
Expanding the first term of \eqref{axionij} and using Gauss-Codazzi relation to express the Christoffel symbols in terms of the extrinsic curvature gives
\be
\nabla_n (\partial^m \varphi \epsilon_{men(j} {R_{i)}}^{nen} ) =  
\partial_n ( \partial^m \varphi \epsilon_{men(j} {R_{i)}}^{nen} ) + K^m_l  \partial^l \varphi  \epsilon_{men(j} {R_{i)}}^{nen} + \partial^m \varphi \epsilon_{men(j} \left( {R_{i)}}^{nln} K^e_l -  K^a_{i)} {R_a}^{nen} \right)
\ee
After integration this gives the junction terms
\be
[\partial^m \varphi \epsilon_{men(j} {R_{i)}}^{nen} ] + \overline{K^m_l} \partial^l \varphi \epsilon_{men(j} 
[K^e_{i)} ] + \overline{K^e_l} \partial^m \varphi \epsilon_{men(j} 
[K^l_{i)} ] - \partial^m \varphi \epsilon_{men(j} \overline{K^a_{i)}}
[K^e_{a} ]
\ee
The second term of \eqref{axionij} simply gives, after integration,
$$
\left[ \partial^n \varphi \epsilon_{nef(j} {R_{i)}}^{nef} \right],
$$
whereas the third term of \eqref{axionij} yields
$$
\partial^m \varphi \epsilon_{mnf(j} \left( - \overline{K}_{i)k} [K^{kf} ] + \overline{K} [K^f_{i)} ] \right),
$$
after we note that 
$
\nabla_k \left( \partial^m \varphi \epsilon_{mnf(j} {R_{i)}}^{knf} \right) = 
\partial^m \varphi \epsilon_{mnf(j} \left(  K_{i)k} {R_n}^{knf} - K {R_{i)}}^{nnf} + \ldots \right) 
$.
Assembling all terms together, we find the junction condition associated
with $\tilde{G}_{ij}$ to be
\bea
\label{HetJ}
8 \pi  S_{ij} &=& 
\beta_2 \left(  
\left[ \partial^n \varphi \epsilon_{nef(j} {R_{i)}}^{nef} \right] +
2 \partial^m \varphi \epsilon_{mnf(j} \big( - \overline{K}_{ki)} [K^{kf} ]
+ \overline{K} [K^f_{i)} ] \big)
\right)
\cr
&&
+2 \beta_2
\left([\partial^m \varphi \epsilon_{men(j} {R_{i)}}^{nen} ]
+ \partial^l \varphi \epsilon_{men(j} \overline{K^m_l}
[K^e_{i)} ] + \partial^m \varphi \epsilon_{men(j} \overline{K^e_l}
[K^l_{i)} ] \right)
\cr
&+& \beta_1 \left(
\nabla^2_\Sigma \phi ( g_{ij} [K] - [K_{ij}] ) + (D_m D_a \phi + \myov{K_{ma} \phi' }) 
\mathcal{C}^{mars}_{ij} [K_{rs}] - 4 \myov{G_{ij}} [ \phi' ] + \phi [J^{(GB)}_{ij}]
\right) \nonumber \\
&+&[K] h_{ij} - [K_{ij}].
\eea
Together with \eqref{GinHet}, \eqref{HetJ} specify the appropriate
junction condition for the low-energy effective theory defined by the
action \eqref{Het}. As a simple consistency check, we note that taking
the scalar fields to be constant give the junction conditions for the
pure Gauss-Bonnet theory (for a continuous axion, the pure
Chern-Simons gravity theory has a trivial junction condition). If we
assume a stronger constraint $[K_{ij}] =0$, then \eqref{HetJ}
simplifies to read 
\be
8 \pi  S_{ij} = -2 \beta_2  n_\alpha   \epsilon_{mln(j}  \left[  \nabla^\alpha K_{i)}^l \right]
\partial^m \varphi ,
\ee
where only the axion-coupling remains in the junction condition.

\subsection{Higher-dimensional Euler densities}

The Gauss-Bonnet term is non-topological beyond four dimensions, and is an example of the Euler characteristic that is the most general extension of the Einstein-Hilbert action that yield at most second-order field equations. The appropriate 
surface terms were derived some time ago and their variation with respect 
to the induced metric of $\Sigma$ gives the junction conditions - as was shown explicitly in the Gauss-Bonnet  case in \cite{Myers, Davis, Gravanis}. 

 The topological Euler density term for a $2m$-dimensional manifold is defined as  
\be
\label{Eulerform}
\mathcal{L}_m =\Omega^{a_1 b_1} \wedge \ldots \wedge \Omega^{a_m b_m} \wedge
\epsilon_{a_1 b_1 \ldots a_m b_m} =
 \frac{1}{2^m} \delta^{[ c_1 d_1 \ldots  c_m d_m ]}_{a_1 b_1 \ldots a_m b_m} 
{R^{a_1 b_1}}_{c_1 d_1} \ldots {R^{a_m b_m}}_{c_m d_m},
\ee
where the Kronecker $\delta$-function above is totally antisymmetric in both sets of indices, 
$\Omega$ is the curvature two-form, 
and we have
normalized it such that the anti-symmetrization symbol in \eqref{Eulerform} has no other normalization factor. 
Note that $\mathcal{L}_1, \mathcal{L}_2$ are the Ricci scalar and Gauss-Bonnet term respectively.
One can consider extending \eqref{Eulerform} to other dimensions apart from $2m$. 
For
dimensions less than $2m$, it simply vanishes whereas for dimensions higher, it will be non-topological. 

Now the Euler-Lagrange equations of motion of \eqref{Eulerform} read
\be
\label{EOMeuler}
\tilde{G}_{ij} = -\frac{1}{2^{m+1}} g_{i\mu} \delta^{[\mu a_1 \ldots a_{2m}]}_{j b_1 \ldots b_{2m}}
{R_{a_1 a_2}}^{b_1 b_2} \ldots {R_{a_{2m-1} a_{2m}}  }^{b_{2m-1} b_{2m}}.
\ee
This follows from a well-posed variational principle if appropriate surface terms can be added at $\Sigma$
such that setting $\delta g_{ij} = 0$ at $\Sigma$ is sufficient for the vanishing of the action variation
and no terms of the form $\nabla_k \delta g_{ij}$ survive. In the following, we briefly review how such a surface term was derived in \cite{Myers} in the language of differential forms. 
One begins by defining a Chern-Simons form $Q_m$ such that 
\bea
\label{CS1}
\mathcal{L}_m (\omega) - \mathcal{L}_m (\omega_0) &=& dQ_m (\omega, \omega_0),
\\
\label{CS2}
Q_m &=& m \int^1_0 ds\,\, \theta^{a_1 b_1} \wedge \Omega^{a_2 b_2}_s \wedge \ldots \wedge \Omega^{a_m b_m}_s \wedge \epsilon_{a_1 b_1 \ldots a_m b_m},
\eea
where $\omega$ is the connection one-form, $\omega_0$ is the conection one-form defined on $\Sigma$,
$\theta = \omega- \omega_0$ is the extrinsic curvature/second fundamental form, and finally $\Omega_s = d\omega_s + \omega_s \wedge \omega_s$ is the curvature two-form defined with $\omega_s = \omega - s \theta$. In \cite{Myers}, it was shown that 
taking the variation of \eqref{CS1} implies
$
\delta_\omega \int_M  \mathcal{L}_m = \int_M d(\delta_\omega Q_m),
$
which leads naturally to the surface term in the action 
\be
I_m = \int_M \mathcal{L}_m - \int_{\partial M} Q_m.
\ee
In coordinate form, the Chern-Simons form that enacts the surface term can be written as \cite{Myers}
\be
Q_m = \int^1_0 ds\,\, \delta^{\mu_1 \ldots \mu_{2m-1}}_{\nu_1 \ldots \nu_{2m-1}} K^{\nu_1}_{\mu_1} \times
\left(   
\frac{1}{2} R^{\nu_2 \nu_3}_{\mu_2 \mu_3} - s^2 K^{\nu_2}_{\mu_2} K^{\nu_3}_{\mu_3}
\right)\ldots
\left(   
\frac{1}{2} R^{\nu_{2m-2} \nu_{2m-1}}_{\mu_{2m-2} \mu_{2m-1}} - s^2 K^{\nu_{2m-2}}_{\mu_{2m-2}} 
K^{\nu_{2m-1}}_{\mu_{2m-1}}
\right).
\ee
In principle, one could take the variation of $Q_m$ with respect to the induced metric to 
obtain the junction condtions. Yet even for the Gauss-Bonnet case, this can be a rather 
elaborate calculation as shown in \cite{Davis, Gravanis}. 

In the following, we will integrate the equations of motion $\tilde{G}_{ij}$ across $\Sigma$ and obtain the junction conditions for the Euler density term valid in
dimension $> 2m$.
We first check that there are no singular terms and hence no regularity constraints to impose. 
This is manifest in the form of \eqref{EOMeuler}. 
Consider again the Gauss-Bonnet term ($m=2$) as an example. The singular terms can only 
arise from a product of two Riemann tensors each of which carries two  `$n$'  indices. 
This leads to the antisymmetric delta function of the form 
$$
\delta^{[\mu\,\, n \,\, \alpha_2 \,\, n \,\, \alpha_4 ]}_{\,\,\,j \,\, n \,\, \beta_2 \,\,n \,\, \beta_4},
$$
which is identically zero. Similarly for $m>2$, the singular terms arise from delta functions of the form
$$
\delta^{[\mu\,\, n \,\, \alpha_2 \,\, n \,\, \alpha_4 \,\, n \,\, \ldots ]}_{\,\,\,j \,\, n \,\, \beta_2 \,\,n \,\, \beta_4 \,\, n \,\, \ldots}.
$$
The permutations among the `$n$' indices come in pairs of $\pm$ signs and hence they sum to zero. 
A subtle point is that as mentioned earlier, in general, there are different classes of divergent terms
defined by the number of Riemann tensors with a pair of `$n$' indices. Each class vanishes separately by the same reason. 

Using \eqref{EOMeuler} as the new starting point, we now rederive the junction condition for the Gauss-Bonnet theory which gives us some intuition on how this generalizes for the higher Euler densities. 
The form and symmetry of the antisymmetric Kronecker $\delta$-function implies that we can 
pick one of the Riemann tensors to have two `$n$' indices and thus
carry the singular delta function (i.e. the term $\sim -\partial_n K_{ab}$), whereas 
the other one should carry all indices parallel to $\Sigma$.  Up to some degeneracy factor, we have 
the junction term 
\be
\label{Hgb1}
\sim -g_{ik} \delta^{[k\,a\,b\,n\,\alpha_4 ]}_{\,\, j\,c\,d\,n\,\beta_4}
\left( \hat{R}_{ab}^{cd} - \myov{K^c_a K^d_b} + \myov{K^c_b K^d_a} \right)
     [K^{\beta_4}_{\alpha_4} ],
\ee
where we have used the form of antisymmetric Kronecker $\delta$-function
to deduce that 
Riemann tensor with only one `$n$' index does not contribute. 
Note the 
exchange symmetry between the pairs of indices $(a,c), (b,d)$ and
$(\alpha_4, \beta_4)$ and that the 
term $[K^{\beta_4}_{\alpha_4} ]$ 
arises from $-\partial_n K^{\beta_4}_{\alpha_4} $. 
Taking into account the $m$-dependent coefficient of $\tilde{G}_{ij}$ and some
symmetry factors, we have the junction term 
\be
\label{SlickGB}
\frac{(-1)^2 (2 \times 4) }{2^{2+1}} g_{ik}
\delta^{[k\,a\,b\,n\,\alpha_4 ]}_{\,\, j\,c\,d\,n\,\beta_4}
\left[ \left( \hat{R}_{ab}^{cd} - \frac{2}{3} K^c_a K^d_b
  \right) K^{\beta_4}_{\alpha_4} \right].
\ee
One can show that this is identical to the junction condition we
derived earlier in \eqref{GBexplicit}.  It is apparently in a more
compact form due to the choice of expressing the Gauss-Bonnet
equations of motion using the antisymmetric Kronecker
$\delta$-function. But more importantly, it generalizes to the higher
Euler densities fairly straightforwardly.  After some algebra, we find
the junction condition for the theory defined by $I_{GR}+ \beta I_m,
m>2$
to be
\bea
&&\beta \frac{2m}{2^m} g_{i\mu} \delta^{[ \mu \, n \, a_2 \, a_3\, a_4\, \ldots a_{2m} ]}_{j \, n \, b_2 \, b_3 \, b_4 \, \ldots 
b_{2m}} 
\sum^{m-1}_{l=0} \frac{(-2)^l}{2l+1} \left( \hat{R}_{a_3 a_4}^{b_3 b_4} \ldots \hat{R}_{a_{2(m-l)-1} a_{2(m-l)}}^{b_{2(m-l)-1} b_{2(m-l)}}
K^{b_2}_{a_2}  
K^{b_{2(m-l)+1}}_{a_{2(m-l)+1}} \ldots
K^{b_{2m}}_{a_{2m}} \right) \cr
&&\qquad \qquad \qquad \qquad = 8\pi S_{ij} +[K_{ij}] - [K]h_{ij},
\eea
where $\beta$ is the coupling constant for $I_m$ added to the ordinary Einstein-Hilbert term in the action. There are no non-trivial junction conditions 
in directions non-parallel to $\Sigma$.

\section{Applications}
\label{AppSec}

\subsection{Thin-Shell wormholes in $R^2$ gravity}

As an application, we examine the energy conditions governing thin-shell wormholes
contructed by a cut-and-paste method (similar to the way we define $\Sigma$ as an identification
between two manifolds). For definiteness, let us again consider the $R + \beta R^2$ theory which has garnered much interest in recent literature since it was proposed in 1979 by Starobinsky and Gurovich in \cite{Starobinsky} as a natural model for cosmological inflation. We had earlier derived the junction conditions for such a theory in Section \ref{quadraticSec} and also in equations 
\eqref{R21p}, \eqref{NNNIR2}
which we reproduce here for reading convenience. 
\bea
\label{R21pCopy}
&&-[K_{ij}] + 2\beta \left(
g_{ij} [\partial_n R ] - \overline{R} [K_{ij}] - \overline{K_{ij}} [R] + [R] (-2 \overline{K_{ij}} + Kg_{ij} )
+ \frac{1}{3}[K_{ij}][K^{ab}][K_{ab}] 
\right)= 8 \pi  S_{ij}, \cr
\label{NNNIR2Copy}
&&2\beta \nabla_i [R] = - 8 \pi  S_{in}, \qquad
2\beta K [R] = 8 \pi  S_{nn},
\eea
together with the condition $[K]=0$.
From the equation of motion \eqref{EOMR2}, we see that any Ricci-flat geometry is valid as a vacuum solution. 
Thin-shell wormholes have been studied in this theory in a few works such as \cite{Eiroa,Lobo:2009},
\footnote{See, for example, \cite{Richarte1}---\cite{Richarte5} for thin-shell wormhole solutions in other gravitational theories beyond GR. }  
but unfortunately assuming an incorrect set of junction conditions. Consider the following
spherically symmetric ansatz
\be
ds^2 = -A(r) dt^2 + A(r)^{-1} dr^2 + r^2 ( d\theta^2 + \sin^2 \theta d\phi^2 ),
\ee
with $r>0, \theta \in[0,\pi], \phi \in [0, 2\pi)$. We can construct a simple model of a thin-shell wormhole 
by picking some radius $a$ and identifying two copies of the region $r \geq a$ with $\Sigma$
as the hypersurface $r=a$. Such a construction leads to geodesically complete wormhole
with $r=a$ being the throat of minimal radius. (In terms of the proper radial distance $l = \int^r_a dr\,A^{-1/2}$, the throat is defined by $l=0$.) We can also conveniently obtain a family of timelike $\Sigma$ by setting 
$$\Sigma: \{r=a(t)\}, \qquad n_\alpha = (-\dot{a}, 1, 0, 0)/\sqrt{A - \dot{A}^2/A}.$$ 
For such a thin-shell wormhole with $\Sigma$ as its throat, the components of the extrinsic
curvature read
\be
K^\theta_\theta = K^\phi_\phi = \pm \frac{1}{a} \sqrt{A(a) + a'^2}, \qquad
K^t_t = \pm \frac{\partial_r A (a) + 2a'' }{2 \sqrt{A(a) + a'^2} }
\ee
where $a' \equiv \frac{da}{d\tau}$, $\tau$ being the proper time on $\Sigma$ of which induced metric reads
\be
ds^2 = -\left( A - \frac{\dot{a}^2}{A}  \right) dt^2 + a^2 d\Omega^2 = -d\tau^2 + a^2(\tau) d\Omega^2.
\ee
For simplicity, let us now pick $A$ to correspond to a solution with constant Ricci scalar $R_0$ (so for example if $R_0 = 0$, then $A = 1- \frac{2M}{r}$). This implies that terms such as $[R], [R']$ vanish in the junction equations. 
For the theory with Lagrangian $R + \beta R^2$, 
the junction conditions imply the following for the singular source
$S^i_j = \text{diag} (-\sigma, P, P )$.
\bea
\label{sigma1}
\sigma &=& \frac{1}{2 \pi  a} \left( 1 + 2\beta R_0 - \frac{2\beta}{3}
       [K^{ab}][K_{ab}] \right) [K^t_t]
= 2 P, \\
\label{sigma2}
a''&=& -\frac{1}{2}\partial_r A - \frac{2}{a} \left(  
A + a'^2
\right),\\
\label{sigma3}
      [K^t_t] &=& -\frac{4\sqrt{A + a'^2}}{a}, \qquad
      [K^{ab}][K_{ab}] = \frac{3}{2} [K^t_t]^2, 
\eea
where \eqref{sigma2} and \eqref{sigma3} arise from the regularity constraint $[K]=0$. 

As a simple example, let us consider static solutions with $a(\tau) = a_0$, with $a_0$ being some positive constant radius parameter. From \eqref{sigma2}, we have
\be
\label{staticc}
a_0 \partial_r A (a_0) + 4 A (a_0) = 0. 
\ee
If we take $R_0 = 0$, we are inevitably led to the Schwarzschild ansatz
$A(r) = 1 - \frac{2M}{r}$ for which \eqref{staticc} implies that $a_0 = 3M/2$. 
Since this is unfortunately smaller than the Schwarzschild radius, we can't construct
a typical thin-shell wormhole in this manner with $r=a_0$ as the time-like throat hypersurface of minimal area. 

Suppose we take $R_0 >0$ and in particular that it arises from a positive cosmological constant,
with $R_0 = 4\Lambda$, then we are led to the Schwarzschild- de Sitter ansatz with $A(r) = 1 - \frac{2M}{r} - \frac{\Lambda r^2}{3}$. The regularity constraint in \eqref{staticc} translates to 
\be
\label{staticSch}
\Lambda a^3_0 - 2a_0 + 3M = 0. 
\ee
To obtain some finite region of positive $g_{tt}$ in the line element, we need to restrict $\Lambda$ to the domain $0<\Lambda M^2 < 1/9$
, where we have two horizons. 
The cosmological horizon $r_c$ and black hole event horizon $r_h$
can be analytically solved to read 
\be
r_c/M = 2 \sqrt{\frac{1}{M^2\Lambda}} \cos \left(
\frac{1}{3} \cos^{-1} (-3\sqrt{M^2\Lambda} )
\right), \qquad 
r_h/M = 2 \sqrt{\frac{1}{M^2\Lambda}} \cos \left(
\frac{1}{3} \cos^{-1} (-3\sqrt{M^2\Lambda} )
-\frac{2\pi}{3} \right).
\ee
On the other hand, solving for $a_0$ in \eqref{staticSch} in this domain yields
\be
\label{SchDe}
a_0/M = 2\sqrt{\frac{2}{3M^2\Lambda}} \cos \left(
\frac{1}{3} \cos^{-1} \left( - \frac{9}{4} \sqrt{\frac{3M^2\Lambda}{2}}   \right)
\right).
\ee
We find that \eqref{SchDe} falls nicely between the horizons, as depicted in Figure \ref{FigSchdS} below.

\begin{figure}[h!]
\centering
\includegraphics[width=130mm]{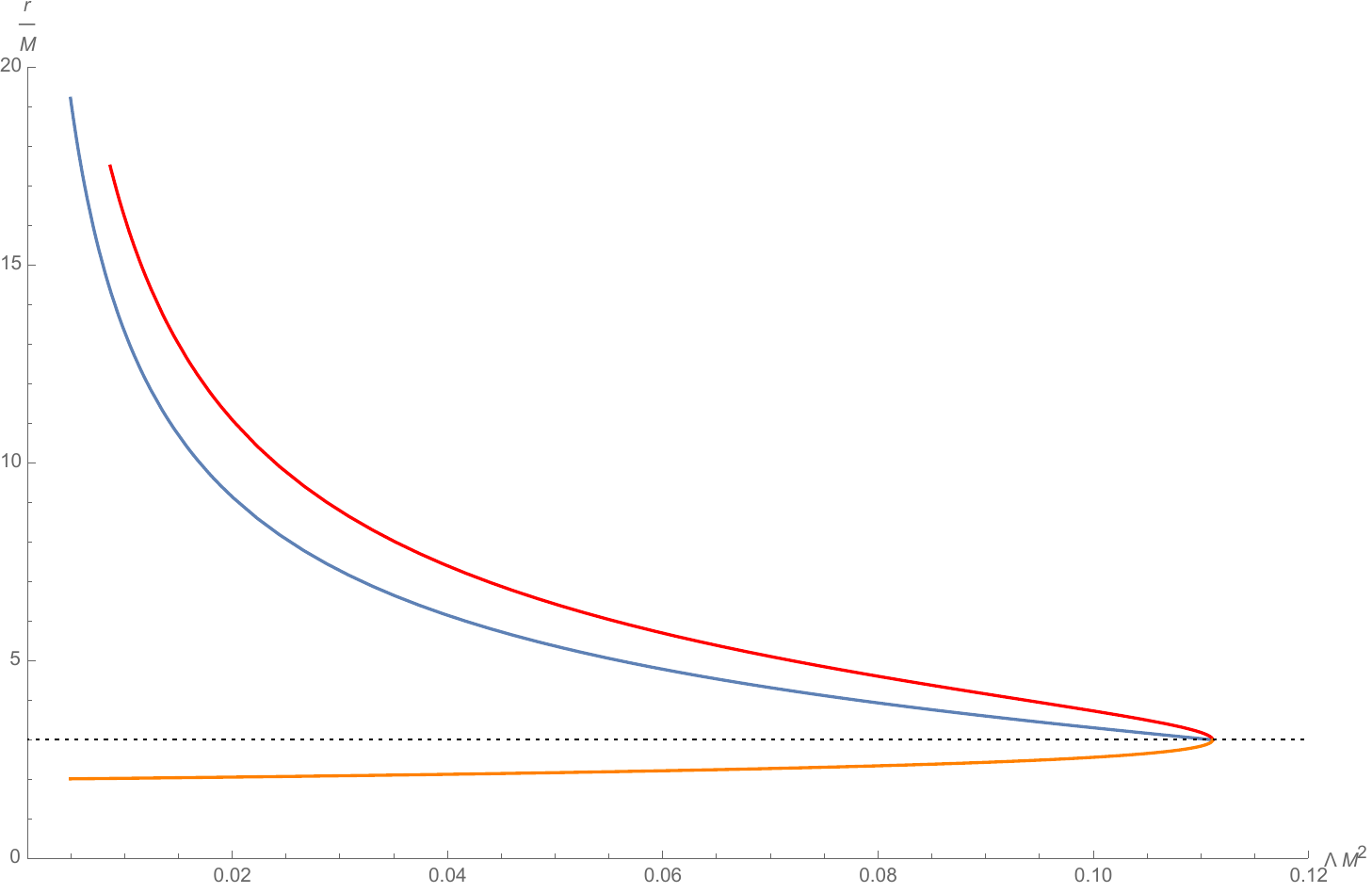}
\caption{In this Figure, the wormhole throat $a_0$ is represented by the
  solid blue line which 
  falls between the event horizon $r_h$ (orange) and the cosmological horizon
  $r_c$ (red) for all 
  $\Lambda M^2 < 1/9$. We note that $\Lambda M^2 = 1/9$ is the extremal limit
  where both horizons
  degenerate into one and all $r_c, r_h, a_0 \rightarrow 3M$. This point is
  excluded from our wormhole construction domain. Also, we note that the
  $\Lambda = 0$ limit is singular.
}
\label{FigSchdS}
\end{figure}
Thus, we see that contrary to the Schwarzschild case, the regularity constraint 
$[K]=0$ is compatible with the wormhole construction procedure of identifying
exteriors of the Schwarzschild-de Sitter spacetime.
For this class of thin-shell wormholes, we find that the energy density
$\sigma$ is 
unfortunately negative definite if we also adopt the unitarity constraint
$1 + 2\beta R_0>0$ (see for e.g. \cite{Eiroa}). 
For the weak energy condition to be obeyed, we require the coupling
parameter $\beta$ to satisfy
\be
\label{WEC}
1 + \beta \left( 2 R_0 - [K^t_t]^2  \right) \leq  0.
\ee
We find that $2R_0 - [K^t_t]^2 $ is positive definite for $\Lambda M^2 < 1/9$, and that 
there is no negative $\beta$ which satisfies both \eqref{WEC} and the unitarity condition 
$1 + 2\beta R_0>0$. This implies that the family of Schwarzschild-de Sitter wormholes constructed here
has to be supported by exotic matter at the throat. They are also unstable under radial perturbation.
\footnote{ 
In the following, we follow the linear stability analysis in \cite{Eiroa}. 
To see the instability of our solution under radial perturbation, we note that
the $[K]=0$ constraint can be 
expressed more suggestively as 
$$\partial_a U(a) + \frac{4}{a} U = - \partial_a A - \frac{4}{a} A(a), 
\,\,\, U \equiv a'^2,
$$
which can be integrated to yield 
$
 a'^2 = - A(a)  + \frac{a^4_0}{a^4} A(a_0) \equiv -V(a).
$
For the above wormholes, $V'(a_0) = 0$ and since the second derivative 
$V''(a_0) = -\frac{4M}{a^3_0} - \frac{2\Lambda}{3} - \frac{20}{a^2_0} \left(
1 - \frac{2m}{a_0} - \frac{\Lambda a^2_0}{3} \right)$ 
is negative definite, the geometry is unstable under radial perturbation. }

\subsection{Implications for stellar models}

In the absence of a singular source, the generalized junction equations reduce to a set of conditions for the geometry induced by a non-singular energy-momentum tensor that is possibly discontinuous at $\Sigma$.
In the ordinary Einstein theory, this simply translates to continuity in the extrinsic curvature, but taking $S_{\mu \nu} = 0$ in the generalized junction conditions typically implies more complicated smoothness conditions on the extrinsic curvature.

In this Section, we briefly discuss the form of the junction conditions 
when the source is non-singular,
and some implications for the $R+ \beta R^2$ theory. 
In this case, it turns out to be convenient to begin by first setting
$S_{in} = 0$, which implies that 
\be
[R] = R_0, 
\ee
where $R_0$ is some constant. In the following, we classify the junction conditions according to whether $R_0$ is zero. 
\begin{enumerate}[(I)]
\item \underline{$[R] = 0$}: 

From the vanishing of $S_{ij}$, we have   
$$
2\beta g_{ij}  n^\alpha [\nabla_\alpha R ] +
[K_{ij}] \left(
\frac{2\beta}{3} \left(   [K^{ab}][ K_{ab}] - 3R \right) -1
\right) = 0.
$$
Taking the trace implies that $n^\alpha [\nabla_\alpha R ] =0$ together with the junction conditions
\be
[K_{ij}] = 0, \qquad \text{or} \,\,\,\,\,\, [K^{ab}][K_{ab}] - 3R = \frac{3}{2\beta}.
\ee

\item \underline{$[R] = R_0 \neq 0$}:

In this case, $S_{nn} = 0$ implies that $K=0$ and from the vanishing of $S_{ij}$, we have 
$n^\alpha [\nabla_\alpha R ] =0$ together with the junction conditions
\be
 [K_{ij}] \left(
\frac{2\beta}{3} \left(   [K^{ab}][ K_{ab}] - 3\overline{R} \right) -1
\right) = 6\beta R_0 \overline{K}_{ij}.
\ee

\end{enumerate}

Generally, if $\Sigma$ is embedded in the bulk with the extrinsic and intrinsic curvature parametrically independent of $\beta$,
then the junction conditions for cases (I) and (II) reduce to $n^\alpha [\nabla_\alpha R ] =0$, and 
\be 
\label{IndJunction}
[R]=[K_{ij}]=0, \qquad  \text{or} \,\,\,\,\,\,\, [R]= R_0 \neq 0,\,\,\, K_{ij} = 0.
\ee 
As an application, let's apply \eqref{IndJunction} to a well-known family of line elements which model static stars with spherical symmetry. 
 In ordinary GR, this class of solutions is constructed
by matching a Schwarzschild exterior to a perfect fluid interior with metric of the form
\be
\label{lineAB}
ds^2_{int} = -A(r) dt^2 + B^{-1} (r) dr^2 + r^2 ( d\theta^2 + \sin^2 \theta d\phi^2 ),
\ee
where 
\be
\label{lineABAnsatz}
A(r) = e^{2\alpha (r)}, \,\,\,B(r) = 1- \frac{2m(r)}{r},\,\,\, \alpha'(r) = \frac{m(r)+4\pi r^3 p(r)}{r(r-2m(r))},
\ee
which solves the field equations with the energy-momentum tensor of a perfect fluid
\be
\label{GRt}
T_{\mu \nu} = (\rho (r) + p(r))U_\mu U_\nu + p(r) g_{\mu \nu}, \,\,\, U_{\mu} = (\sqrt{A}, 0, 0, 0),
\ee
with the Tolman-Oppenheimer-Volkoff equation
\be
p'(r) = - \frac{m(r)}{r^2} \rho (r) \left( 1 + \frac{p(r)}{\rho(r)} \right) \left( 1 + \frac{4\pi r^3 p(r)}{m} \right)
\left( 1 - \frac{2m(r)}{r} \right)^{-1}.
\ee
The Schwarzschild solution is still a vacuum solution in $\mathcal{F}(R)= R + \beta R^2$ theory,
and we also take the same interior solution as defined above but now this is sourced by a different energy-momentum which is the sum of \eqref{GRt} and extra terms arising from 
the field equations of the $\mathcal{F}(R)$ theory. 

The surface $\Sigma$ is the star's boundary defined by $r=R_s$ for some constant $R_s$ (which typically satisfies an appropriate Buchdahl bound). At $\Sigma$, imposing
metric continuity and $[K_{ij}]=0$ lead to the boundary conditions 
$$
m(R_s) = M, \,\,\,p (R_s) =0 ,
$$
where $M$ is the mass parameter of the Schwarzschild exterior. Note that for the metric of the form \eqref{lineAB},
the components of the extrinsic curvature are $K^t_t = \frac{1}{2} \frac{\sqrt{B}A'}{A}, K^\theta_\theta = K^\phi_\phi = \frac{\sqrt{B}}{R_s}$. Thus, matching it to a Schwarzschild exterior leads to $p(R_s)=0$ after using \eqref{lineABAnsatz}. Since $R = -8\pi T^\mu_\mu = 8 \pi (\rho-3p)$, the additional junction conditions $[R]=0, [\partial_r R]=0$ further impose the additional boundary conditions 
\be
\rho (R_s) = 0, \,\,\, p'(R_s) = 0, \,\,\, \rho'(R_s) = 0.
\ee
For a polytropic equation of state of the form $p \propto \rho^\gamma$ for some positive constant $\gamma$, these boundary conditions are only compatible with the case of radiative matter $p = \frac{1}{3}\rho$. These results were similarly
presented in \cite{Lobo:2009} albeit through a different set of junction conditions. 
For the Tolman-Oppenheimer-Volkoff stellar model above, our junction conditions lead to an identical final set of boundary conditions on the interior fluid's density and pressure.

Another well-studied stellar model that is also a cut-and-paste solution involving a Schwarzschild exterior is 
the Oppenheimer-Snyder solution where the interior is a closed FRW universe sourced by a pressureless dust. 
For this model of stellar formation, the matching surface $\Sigma$ is taken to preserve the $SO(3)$ isometry,
and is comoving with the FRW interior of which metric reads
\be
ds^2 = a^2 (\tau) \left( -d\tau^2 + dR^2 + \sin^2 R (d\theta^2 + \sin^2 \theta d\phi^2) \right).
\ee
The surface $\Sigma$ is defined as the sphere $R=R_c$ for some constant $R_c$, or in the coordinates of the Schwarzschild exterior, $r(\tau) = a(\tau) \sin (R_c)$, with the scale factor $a(\tau)$ satisfying the Friedmann equations for a pressureless dust.
It is straightforward to see that this solution is incompatible with the junction conditions \eqref{IndJunction} since
the Ricci scalar $R=8\pi \rho$. 

These simple examples appear to indicate that an embedding of GR solutions with discontinuous but non-singular sources 
is subject to rather stringent constraints associated with the generalized junction conditions in $F(R)$ theory. 
It is however important to note that we have examined only the simplest embedding, retaining the full GR metric and the geometry of $\Sigma$. The caveat is that this implies that 
the interior is sourced by an energy-momentum tensor of the form
$$
T_{\mu \nu} = T^{(GR)}_{\mu \nu} + \frac{\beta}{4\pi} \left(
(g_{\mu \nu} \Box - \nabla_\nu \nabla_\mu )R + RR_{\mu \nu} - \frac{1}{4}R^2 g_{\mu \nu}
\right).
$$
It would be interesting to consider stellar models where $T_{\mu \nu}$ inherits a more physically motivated form,
as well as other shapes of $\Sigma$ which may be dependent on various theory couplings in the $F(R)$ theory. This would probe a much wider landscape of solutions compatible with our generalized junction conditions.

\section{Concluding Remarks}
\label{Conclude}

We have presented a general method to derive the appropriate Darmois-Israel junction conditions for gravitational
theories with higher-order derivative terms by integrating the bulk equations of motion
across the infinitesimal width of the singular hypersurface $\Sigma$ as defined in \eqref{integration}. 
A salient feature of our work is the presence of regularity constraints which impose conditions on the extrinsic curvature
such that the integral in \eqref{integration} converges. Geometrically, they specify the conditions under which the embedding of $\Sigma$ into the bulk spacetime is compatible with the delta-singular source localized within $\Sigma$.

Our method fundamentally relies on defining the $\delta$-distribution
as the limit of a sequence of classical functions as expressed in
\eqref{classical} and \eqref{classical2}.  We found that the use of
delta-convergent sequences yields a powerful language for organizing
various terms with different orders of singularities appearing in the
integral \eqref{integration}, and is intimately related to the
procedure of Hadamard regularization commonly invoked in the theory of
distributions. Upon imposing the regularity constraints, the integral
in \eqref{integration} converges and is well-defined. Our method
passes a stringent consistency test (that is noticeably absent in
previous literature) : that the junction conditions for Gauss-Bonnet
gravity can be obtained as a suitable limit of those of quadratic
gravity when the coupling constants reduce to those of the 4D Euler
density term. This is a rigorous check of validity since the junction
conditions for Gauss-Bonnet gravity can also be independently derived
by boundary variation of a suitable surface term in the action.

As explicit examples of our approach, we demonstrated in detail how to
obtain the regularity constraints and junction conditions for (i)
quadratic gravity (ii) $\mathcal{F}(R)$ theories (iii) a 4D low-energy
effective action in string theory and (iv) Euler density action terms
which are higher-dimensional analogues of the 4D Gauss-Bonnet term. We
have expressed these generalized junction conditions explicitly as
functions of the extrinsic curvature tensor and its
derivatives. Generically, they also involve components that are
non-parallel to $\Sigma$, in contrast to the case in ordinary GR. To
our knowledge, all of these generalized junction conditions are novel
results. Although there have been past attempts to derive junction
conditions for quadratic gravity \cite{Reina} and $\mathcal{F}(R)$
theories \cite{Berezin,SenoFR,Sasaki}, their results or underlying
methodologies did not appear to demonstrate consistency with the
Gauss-Bonnet case. In this aspect, we hope that our work has also
clarified some of the ambiguities encountered in these previous
studies.

The details of our derivation procedure presented here should be
pedagogically useful towards adopting our methodology to derive
junction conditions for other more complicated gravitational theories,
including those with matter and gauge couplings. We should also
mention that although our method applies rather widely to
gravitational actions built out of curvature invariants, by
definition, it does not apply to topological boundary terms in the
action since they do not manifestly modify the bulk equations of
motion. In an upcoming work \cite{Tan}, we derive and examine the
generalized junction conditions for Chern-Pontryagin density terms by
boundary variation of suitable surface terms. This class of theories
includes, in particular, the (non-dynamical) `Chern-Simons gravity'
theory in 4D of which surface term was derived in \cite{Grumiller}.

We hope that these junction equations will furnish the essential first
steps towards exploring a potentially rich and phenomenologically
interesting landscape of classical solutions which have singular
hypersurfaces as their defining geometric feature. In this work, we
have briefly touched upon a couple of applications in the
$\mathcal{F}(R) = R+ \beta R^2$ theory, where we found a thin-shell
wormhole constructed by identifying the exterior regions of two
identical copies of Schwarzschild-de Sitter spacetime. We showed that
many stellar models in ordinary GR which are lifted directly to this
particular $\mathcal{F}(R)$ theory violate its regularity
constraints. It would be very interesting to carry out a more
extensive exploration of thin-shell and stellar geometries in many
gravitational theories beyond GR, now that we are freshly equipped
with the fundamental junction conditions to work with. In particular,
we note that thin-shell wormholes have been recently revisited as
black hole mimickers for LIGO events \cite{Cardoso}.

Another natural avenue for future work lies in extending our approach
to cover $\Sigma$ which is lightlike. In this case, there is no unique
definition of the extrinsic curvature once the induced metric on the
surface becomes degenerate, since the normal vector defined in the
setting of timelike/spacetlike $\Sigma$ is then tangent to $\Sigma$,
and naively we need another notion of a `transverse' vector. This
subtle point has been addressed in \cite{Barrabes} where a proposal
for junction conditions in the case of a null surface was
presented. It would be interesting to generalize the results of
\cite{Barrabes} to higher-derivative gravitational theories, and see
if some aspects of our method remain useful.

\vskip7mm
\section*{Acknowledgments}

We acknowledge support of this work by
the National Center of Theoretical Science
(NCTS) and the grant 107-2119-M-007-014-MY3 of the Ministry of Science and
Technology of Taiwan.


\appendix

\section{Some useful integral identities}
\label{AppA}

In this Appendix, we collect a set of integral identities (i)---(v)
which involve products of discontinuous functions and their
derivatives. This accompanies the detailed derivation in Section
\ref{quadraticSec} .  The scaling limit \eqref{doublescale} is
implicitly taken for each final expression throughout this
Section. When the argument of a function is omitted, it is understood
to be evaluated at $n=0$.
\bea
\label{I1}
\text{(i)} I_1 &=&  \int^\epsilon_{-\epsilon} dn \,\,\, f(n) \partial^2_n g(n) \cr
&=& \int^\epsilon_{-\epsilon} dn\left(f_1 (n) + \Theta (n,b) (f_2 (n) - f_1(n)) \right)
\partial^2_n \left(g_1 (n) + \Theta (n,b) (g_2 (n) - g_1(n)) \right) \cr
&=& 
\frac{1}{2}(f_1 (0) + f_2(0)) [g'_2 (0) - g'_1(0)] - \frac{1}{2}(g_1 (0) + g_2(0)) [f'_2 (0) - f'_1(0)]  \cr
&&- \int^\epsilon_{-\epsilon}dn\,\,\, (f_2(n) - f_1 (n)) (g_2(n) - g_1(n))\left(  \Theta'(n,b) \right)^2 \cr
&\equiv& \overline{f} [g'] - \overline{f}' [g] -
 \int^\epsilon_{-\epsilon}dn\,\,\, [f(n)][g(n)] \left(  \Theta'(n,b) \right)^2 \\
\label{I2}
\text{(ii)} I_2 & =& \int^\epsilon_{-\epsilon} dn \,\,\, \partial_n f(n) \partial_n g(n) \cr
&=& 
 \int^\epsilon_{-\epsilon} dn  \left(f_1 (n) + \Theta (n,b) (f_2 (n) - f_1(n)) \right)'
\left(g_1 (n) + \Theta (n,b) (g_2 (n) - g_1(n)) \right)' \cr
&=& \overline{g}' [f] + \overline{f}' [g] +  \int^\epsilon_{-\epsilon}dn\,\,\, [f(n)][g(n)] \left(  \Theta'(n,b) \right)^2.
\eea
In both $I_1, I_2$, the singular parts of the integral may only arise in the term 
\be
\label{Idiv}
I_{div} \equiv
 \int^\epsilon_{-\epsilon}dn\,\,\, [f(n)][g(n)] \left(  \Theta'(n,a) \right)^2
= \lim_{b\rightarrow 0} \frac{1}{b} [f(0)][g(0)] \int^\infty_{-\infty} dX \,\,\, F^2(X),
\ee
where we have expanded $[f(n)][g(n)]$ around the origin, and restored the vanishing limit symbol for
parameter $b$ to indicate the term's singular nature. This is a particular case of the general formula we develop in \eqref{prodN}, with 
$\sum_m k_m =2$ and thus $l=0$ is the only singular mode. In applying this integral identity
to the junction equations, we note that we have to collect all singular terms that similarly diverge as $1/b$ and set the coefficient (which is typically a function of the extrinsic curvature and its derivatives) to vanish. 
One can similarly simplify integrals of the form 
\be
\label{genDiv}
 \int^\epsilon_{-\epsilon} dn \,\,\, \partial_{k_1} f_1 \, \partial_{k_2} f_2 \, \ldots 
\partial_{k_r} f_r,
\ee
where $f_1, f_2, \ldots, f_r$ are discontinuous functions. After expanding various functions (apart
from $\delta_b (n)$ and its derivatives) about the origin, we then obtain a linear combination of 
\eqref{prodN}. Only a finite number of terms remain, including the singular terms. 
As an another example, let's consider 
$$
I_3 =   \int^\epsilon_{-\epsilon} f(n) \partial^3_n g(n).
$$
After some similar manipulations as in the previous examples, we obtain
\bea
\text{(iii)} I_3 &=& [f g'' ] - [f] \overline{g}'' - \overline{f}' [g'] + \overline{f}''[g] \cr
&&+  \int^\epsilon_{-\epsilon} dn \,\,\, (\Theta'(n,b))^2 \left(
2[f(n)][g'(n)] - [f'(n)][g(n)]
\right)
+ \Theta'(n,b) \Theta''(n,b) [f(n)][g(n)]  \cr
&=&[f g'' ] - [f] \overline{g}'' - \overline{f}' [g'] + \overline{f}''[g]  
-\lim_{b \rightarrow 0}
\frac{3}{2b}\left( [f][g'] - [f'][g] \right) \int^\infty_{-\infty} dX F^2 (X)
\eea
Again, we see that there is one singular term that diverges as $1/b$. 

Another useful formula that we will need is
\be
\label{useful}
\text{(iv)} I_4 = \int^\epsilon_{-\epsilon} dn\,\,\,  f (n) g (n) \partial_n h(n) = \frac{1}{3} \left(  \overline{f g } + 2\overline{f} \overline{g} 
\right)[h] \equiv \myov{f g} [h],
\ee
where the various functions are all discontinuous at $n=0$ in the limit $\epsilon \rightarrow 0$, and we have taken the liberty to introduce
a bold overline for notational simplicity since as we shall see, such combination of averaging over functions (of extrinsic curvature) occurs frequently in the junction equations for gravitational theories
with Lagrangian terms that are quadratic invariants of the Riemann tensor. 
To obtain \eqref{useful}, we simply note that
the LHS is equivalently
\be
\int^\epsilon_{-\epsilon} dn\,\,\, \left( f_1 (n) + \Theta (n) (f_2(n) -f_1(n))\right)\left( g_1 (n) + \Theta (n) 
(g_2(n) -g_1(n)) \right)
\Theta'(n) \left(
h_2 (n) - h_1 (n)
\right),
\ee
with all functions being continuous. 
Upon using the identity $\int dn\,\, \Theta' (n) \Theta^k(n) F(n) = \frac{1}{k+1} F(0)$, we are then led to 
\eqref{useful}.

Finally, another useful integral which we will encounter in deriving junction equations for gravitational
Lagrangians with quadratic invariants 
is
\bea
\label{repDep}
\text{(v)} I_5 &=& \int^\epsilon_{-\epsilon} dn \,\, g_{ij} \partial_n f \partial_n h
\\
&=& 
g_{ij} \overline{f'} [h] + \overline{h'} [f] g_{ij} + \lim_{b\rightarrow 0} \frac{1}{b} g_{ij} [h][f] \int^\infty_{-\infty}dX\,
F^2(X)  + 2[f][h][g'_{ij}] \int^\infty_{0} dX\, X F^2(X). \nonumber
\eea
We note that there is a singular term arising from this integral and another \emph{finite} term that
is depends on the choice of the nascent delta function. This can be interpreted as an (infinite) sum
of various moments of $\delta_b (X)$ after expanding $\Theta (bX,b)$ and a factor of $F(X)$ in $X$. 
Since the coefficient is of the same form as the singular term (apart from replacing $g_{ij}$ with its normal
derivative), we will find that this representation-dependent term naturally cancels out when we impose the 
regularity condition. 

To arrive at \eqref{repDep}, we can apply the same techniques that we used in proving the previous integral identities, taking into account that we take $g_{ij}$ to be generally non-differentiable. As a consistency check,
let us write the integrand in the LHS of \eqref{repDep} as 
\be
 g_{ij} \partial_n f \partial_n h = \partial_n (g_{ij} f) \partial_n h - (\partial_n g_{ij} ) f \partial_n h,
\ee
and similar to how we obtain $I_2$ in \eqref{I2}, it is straightforward to obtain
\bea
\int^\epsilon_{-\epsilon} dn\,\, \partial_n (g_{ij} f) \partial_n h &=& 
g_{ij} \overline{f'} [h] + \overline{h'} [f] g_{ij} + \lim_{b\rightarrow 0} \frac{1}{b} g_{ij} [h][f] \int^\infty_{-\infty} dX\,\, F^2(X) \cr&&+ 2 [f][h][g'_{ij}] \int^\infty_{0} dX\,  X F^2 (X) \cr
\label{consis}
&&\qquad + \left(
(\overline{g'_{ij}} -\frac{1}{2} [g'_{ij}] )[h] \overline{f} + \frac{1}{2}[h][g'_{ij}](\overline{f} -
\frac{1}{2}[f] ) + \frac{1}{3} [g'_{ij}][f][h]
\right). \nonumber \\
\eea
After some algebra, one can show that the last line (in brackets) of \eqref{consis} is precisely
$\int^\epsilon_{-\epsilon} dn \, f \partial_n g_{ij} \partial_n h$ after using the identity \eqref{useful},
and thus furnishing a nice consistency check between \eqref{repDep} and \eqref{useful}.


\section{From quadratic gravity to Gauss-Bonnet theory: some notes on the junction conditions}
\label{GBsec}
We have seen that for the Gauss-Bonnet theory, there are no additional constraints on the extrinsic
curvature that we have to impose for regularity. Let us now elaborate on a subtlety arising in this topological limit.

For the case of Gauss-Bonnet gravity, the junction conditions were derived in \cite{Dolezel} via a bulk integration across 
$\Sigma$.  
An alternate derivation was performed in \cite{Davis} via the boundary variation of an appropriate surface term that appeared earlier in \cite{Myers} --- a seminal work where surface terms for higher-dimensional topological Euler density terms 
were also derived. These surface terms are required for a well-defined action principle with Dirichlet conditions. As explained in 
\cite{Myers}, the Euler density terms $\chi_{2m}$ that one can define in every even dimension $2m$ are precisely the linear combination of curvature invariants that generate only second-order field equations, and thus, in principle, these exists 
appropriate surface terms for them. The simplest example would be the Ricci scalar $\chi_2$ being the Euler density in two dimensions. The surface term for $\chi_2$
is the Gibbons-Hawking-York term
$
S_{GHY} =  \frac{1}{8\pi } \int d^{d-1}x \,\, \sqrt{h} K
$.
This continues to hold for dimensions $>2$, and hence the junction conditions \eqref{DIJunction}, even when the theory itself is no longer topological in dimensions $>2$. Similarly, the Gauss-Bonnet term $\chi_{4}$ is the topological Euler density term in 4D, trivial in lower dimensions and non-topological in dimensions $>4$. Adding it to the Einstein-Hilbert action yields additional junction terms
which are third-order polynomials in the extrinsic curvature \cite{Dolezel,Davis}. 
For junction terms descending from a boundary variation of surface terms, 
since the normal vector to $\Sigma$ points in 
opposite directions when we take $\Sigma$ to bound the spacetime region on each side, this naturally leads
to junction terms which are expressible in terms of bracketed quantities like \eqref{DIJunction}.

As already argued in \cite{Myers,Madsen} and definitively shown in \cite{Smolic}, the equations of motion of a generic higher-derivative gravitational theory do not descend from a well-posed variational principle with Dirichlet conditions. Hence, the approach of obtaining their junction conditions by boundary variation of surface terms is not applicable, since an appropriate surface term does not exist for a generic higher-derivative theory. On this point, we note that a recent work \cite{Berezin} claimed to have derived the junction conditions for quadratic gravity by using the variational principle. However, in their derivation, they 
imposed a form (eqn. 62 in \cite{Berezin}) for the variation of the extrinsic curvature that is problematic, and does not follow consistently from its fundamental definition (for the interested reader, see e.g. \cite{Davis,Grumiller} for the correct expression for $\delta K_{ij}$.)

Nonetheless, the junction conditions for the Euler-density terms can serve as vigorous consistency checks for those belonging to gravitational theories of which action is constructed from some linear combination of curvature invariants including those defining $\chi_{2m}$. Varying the action with the Gauss-Bonnet term 
$$ \mathcal{L}_{GB} = R +  \beta_1 \left( R^2 - 4 R_{\alpha \beta} R^{\alpha \beta} + R^{\alpha \beta \mu \nu} R_{\alpha \beta \mu \nu} \right), 
$$
with respect to the metric yields the following the equation of motion
\bea
\label{GB-EOM}
\frac{\delta \mathcal{L}}{\delta g^{\alpha \beta}} &=&
G_{\alpha \beta} + 2\beta_1 \left( R R_{\alpha \beta} - 2 R_{\alpha \mu}
{R^\mu}_\beta - 2R^{\mu  \nu} R_{\alpha \mu \beta \nu} +
{R_\alpha}^{\mu \nu \chi} R_{\beta \mu \nu \chi} \right)
\nonumber\\
&& \quad -\frac{\beta_1}{2} g_{\alpha \beta}
\left( R^2 - 4 R_{\alpha \beta} R^{\alpha \beta}
+ R^{\alpha \beta \mu \nu} R_{\alpha \beta \mu \nu} \right).
\eea
We now proceed to verify that taking $\beta_2=-4\beta_3=-4\beta_1$
in $\tilde{G}_{ij}$ yields exactly
the same junction conditions derived from the method of boundary variation. 
This furnishes a strong
consistency check of many expressions in the previous Section (explicitly,
they are equations 
\eqref{GijRSq}, \eqref{GijRuvSq}, \eqref{GijRabcdSq}, \eqref{RRij}, \eqref{RiuRuj},
\eqref{RiuabSq} and \eqref{RiujvRuv}).

In the following, we focus on the part in \eqref{GB-EOM} that is coupled to
$\beta_1$ which we suppress for the workings below for notational simplicity. 
Consider first all terms involving the intrinsic curvature of $\Sigma$ which are various contractions of products of $\hat{R}_{abcd}$, the extrinsic curvature tensor and the metric tensor. After some algebra, we find that they sum up to read
\be
2 \left(  
-2 \hat{R}_{ij} [K] + 2 \hat{R}_{m(j} K^m_{i)} + 2 [K^{ab} ] \hat{R}_{iajb} - \hat{R} [K_{ij}] 
\right) -4g_{ij} \left(
\hat{R}_{cd} [K^{cd}] - \frac{1}{2} \hat{R} [K]
\right),
\ee
which can be written as
$-4 P_{icdj} [K^{cd}]$ with $P_{icdj}$ being
the divergence-free part of the Riemann curvature tensor of $\Sigma$ as defined in 
\eqref{GBexplicit} earlier.

Next, we consider all terms that contain a factor of the induced metric $g_{ij}$ which sum up to read
\be
\label{finiteGij}
-\frac{1}{2}g_{ij} \left(   
4[K] \myov{K^2} - 4 [K] \myov{K^{ab}K_{ab}} - 8 \myov{KK^{ab}}[K_{ab}]
+8 \myov{K^b_l K^{la}}[K_{ab}]
\right)
\ee
Expressing all quantities in \eqref{finiteGij} in terms of the ordinary averaging symbol, and using the fact that
\bea
[K^m_l K^{la} K_{ma}] &=& \overline{K^m_l K^{la}} [K_{ma}] + [K^m_l K^{la}] \overline{K_{ma}} = 
\overline{K^m_l K^{la}} [K_{ma}] + 2 \overline{K^m_l} \overline{K^{la}} [K_{ma}], \cr
[KK^{ab} K_{ab} ] &=& [K] \overline{K^{ab} K_{ab}} + [K^{ab} K_{ab}]\overline{K} = 
[K] \overline{K^{ab} K_{ab}} + 2 \overline{K^{ab}}\overline{K} [K_{ab}], \cr
\label{IdAvg}
[KK^{ab} K_{ab} ] &=& \overline{KK^{ab}} [K_{ab}] + [K K_{ab}]\overline{K_{ab}} = 
\overline{KK^{ab}} [K_{ab}] + \overline{K}\overline{K^{ab}} [K_{ab}]
+[K]\overline{K^{ab}} \overline{K_{ab}},
\eea
one can straightforwardly show that \eqref{finiteGij} can be expressed purely in terms of commutators and
read as
\be
-\frac{1}{2} g_{ij} \frac{4}{3} \left( [K^3] - 3[KK^{ab}K_{ab}] + 2 [K_l^m K^{la}K_{ma} ] \right)
\ee
Finally, let's consider all other terms. For those involving $K_{ij} K^2$, they arise from \eqref{RRij} and
can be simplified as 
\be
\label{KijKSq}
2 \left(
2 \myov{KK_{ij}} [K] + \myov{K^2}[K_{ij}] \right)= 
\frac{4}{3} \left(  \overline{KK_{ij}} + 2 \overline{K} \, \overline{K_{ij}} \right)[K] + 
\frac{2}{3} (\overline{K^2} + 2 \overline{K}^2 )[K_{ij}] 
\ee
By noting that 
\bea
[K_{ij} K^2 ] &=& \overline{K_{ij}K} [K] + [K_{ij} K] \overline{K} = \overline{K_{ij}K} [K] + \overline{K_{ij}} \, \overline{K} [K] + \overline{K}^2 [K_{ij}], \cr
&=& \overline{K^2} [K_{ij}] + [K^2 ] \overline{K_{ij}} = \overline{K^2} [K_{ij}] + 2 [K] \overline{K}\, \overline{K_{ij}}
\eea
some straightforward algebra then leads one to see that \eqref{KijKSq} can be written purely in terms of 
a bracket and reads
\be
2 \left(
2 \myov{KK_{ij}} [K] + \myov{K^2}[K_{ij}] \right)=2[K_{ij} K^2]
\ee
Next, we consider terms of the form $K_{ij} K^{ab} K_{ab}$ which arise from 
\eqref{RRij} and \eqref{RiujvRuv}. They sum up to read
\be
-2 \left( \myov{K^{ab} K_{ab} } [K_{ij}] + 2 \myov{K_{ij} K^{ab}} [K_{ab}] \right)
= -2 \left( [K_{ij} K_{ab} K^{ab} ] \right)
\ee
where we have invoked identities similar to the form of \eqref{IdAvg}. Thus, all terms 
containing a factor of $K_{ij}$ sum up to be a single bracket of the form
\be
2 \left[ K_{ij} (K^2 - K_{ab} K^{ab} ) \right]
\ee
Next, we consider terms of the form $K_{ja} K_{ib} K^{ab}$
which arise in \eqref{RiuRuj}, \eqref{RiuabSq} and \eqref{RiujvRuv}.
They sum up to 
read
\be
2 \left(  2 \myov{K^l_m K_{l(i}} [K_{j)}^m ] + [K^l_m] \myov{K_{l(j} K^m_{i)}}
\right) = 2 [K^l_m K_{l(j} K^m_{i)} ]
\ee
where again we have invoked identities similar to the form of \eqref{IdAvg}.
Finally, 
we are left with terms of the form $KK_{ia} K^a_j$. They arise in 
\eqref{RRij}, \eqref{RiuRuj} and \eqref{RiujvRuv}, and sum up to read
\be
-2 \left(  \myov{K_{m(j} K^m_{i)}} [K] + 2\myov{K K_{m(j}} [K^m_{i)}] \right)
=-2 [ K_{m(j} K^m_{i)} K ].
\ee
Gathering all terms together, and including the Gibbons-Hawking terms from
Einstein-Hilbert action, 
we find the junction condition to be
\be
\label{GB}
[K] h_{ij} - [K_{ij}] - 2\beta_1 \left[ (3J_{ij} -J h_{ij} + 2 P_{icdj} K^{cd}) \right]
= 8 \pi  S_{ij},
\ee
where $P_{icdj}$ is the divergence free part of the Riemann tensor and
$J_{ij} = \frac{1}{3} (2K
K_{ic} K^c_j + K_{cd} K^{cd} K_{ij} - 2 K_{ic} K^{cd} K_{dj} - K^2 K_{ij}) $.
This is identical
to the junction condition derived in earlier literature
\cite{Myers, Davis, Gravanis, Dolezel} via boundary variation.



\end{document}